\documentclass[aps,10pt,prd,tightenlines,twocolumn,twoside,showpacs,nofootinbib,superscriptaddress]{revtex4-2}
\usepackage{alltt}
\usepackage{amsmath}
\usepackage{amsthm}
\usepackage{amssymb}
\usepackage{bm}
\usepackage{booktabs}
\usepackage{cancel}
\usepackage{color}
\usepackage{csquotes}
\usepackage{enumerate}
\usepackage[marginal]{footmisc}

\usepackage{graphicx,bm}
\usepackage[justification=raggedright, singlelinecheck=false]{caption}
\usepackage{lipsum}
\usepackage{listings}
\usepackage{mathtools}
\usepackage{mathrsfs}
\usepackage{physics}
\usepackage{ragged2e}
\usepackage{relsize}
\usepackage{slashed}
\usepackage{soul}
\usepackage{spverbatim}
\usepackage{siunitx}
\usepackage{tabularx}
\usepackage{wrapfig}
\usepackage{xcolor}
\usepackage{diagbox}
\usepackage{multirow}
\usepackage{subcaption}

\usepackage[pdftex,hypertexnames=false,linktocpage=true]{hyperref}
\hypersetup{colorlinks=true,linkcolor=blue,anchorcolor=blue,citecolor=darkgreen,filecolor=blue,urlcolor=blue,bookmarksnumbered=true,
	pdfview=FitB
}

\colorlet{darkgreen}{green!60!black}
\colorlet{brightyellow}{yellow!75!red}
\colorlet{orange}{red!50!yellow}
\colorlet{darkblue}{blue!60!black}
\colorlet{darkred}{red!80!black}
\colorlet{greenblue}{green!50!blue}

\makeatletter

\newcommand{\Rmnum}[1]{\expandafter\@slowromancap\romannumeral #1@}
\makeatother

\def\imag{{\mathrm{i}}}

\begin{document}
\title{Basis light-front quantization for the $\Lambda_b$ and $\Sigma_b$ baryons}

\author{Lingdi~Meng}
\email{lingdimeng@impcas.ac.cn}
\affiliation{%
	Department of Physics, Hebei University, Baoding, 071002, China
}%
\affiliation{%
	Institute of Modern Physics, Chinese Academy of Sciences, Lanzhou, Gansu 730000, China
}%

\author{Tian-Cai~Peng}
\email{pengtc20@lzu.edu.cn} 
\affiliation{School of Physical Science and Technology, Lanzhou University, Lanzhou 730000, China}
\affiliation{Lanzhou Center for Theoretical Physics, Key Laboratory of Theoretical Physics of Gansu Province, Lanzhou University, Lanzhou 730000, China}
\affiliation{Key Laboratory of Quantum Theory and Applications of MoE, Lanzhou University, Lanzhou 730000, China}
\affiliation{Research Center for Hadron and CSR Physics, Lanzhou University and Institute of Modern Physics of CAS, Lanzhou 730000, China}

\author{Zhi~Hu}
\email{huzhi0826@gmail.com}
\affiliation{High Energy Accelerator Research Organization (KEK), Ibaraki 305-0801, Japan}

\author{Siqi~Xu}
\email{xsq234@impcas.ac.cn}
\affiliation{%
	Institute of Modern Physics, Chinese Academy of Sciences, Lanzhou, Gansu 730000, China
}
\affiliation{%
	School of Nuclear Science and Technology, University of Chinese Academy of Sciences, Beijing 100049, China
}
\affiliation{%
	Department of Physics and Astronomy, Iowa State University, Ames, Iowa 50011, USA
}

\author{Jiangshan~Lan}
\email{jiangshanlan@impcas.ac.cn}
\affiliation{%
	Institute of Modern Physics, Chinese Academy of Sciences, Lanzhou, Gansu 730000, China
}
\affiliation{%
	School of Nuclear Science and Technology, University of Chinese Academy of Sciences, Beijing 100049, China
}
\affiliation{%
	Advanced Energy Science and Technology Guangdong Laboratory, Huizhou, Guangdong 516000, China
}

\author{Chandan~Mondal}
\email[corresponding author: ]{mondal@impcas.ac.cn}
\affiliation{%
	Institute of Modern Physics, Chinese Academy of Sciences, Lanzhou, Gansu 730000, China
}
\affiliation{%
	School of Nuclear Science and Technology, University of Chinese Academy of Sciences, Beijing 100049, China
}

\author{Guo-Li~Wang}
\email{wgl@hbu.edu.cn}
\affiliation{%
	Department of Physics, Hebei University, Baoding, 071002, China
}%
\affiliation{%
	Key Laboratory of High-precision Computation and Application of Quantum Field Theory of Hebei Province, Baoding, China
}%

\author{Xingbo~Zhao}
\email{xbzhao@impcas.ac.cn}
\affiliation{%
	Institute of Modern Physics, Chinese Academy of Sciences, Lanzhou, Gansu 730000, China
}
\affiliation{%
	School of Nuclear Science and Technology, University of Chinese Academy of Sciences, Beijing 100049, China
}
\affiliation{%
	Advanced Energy Science and Technology Guangdong Laboratory, Huizhou, Guangdong 516000, China
}

\author{James~P.~Vary}
\email{jvary@iastate.edu}
\affiliation{%
	Department of Physics and Astronomy, Iowa State University, Ames, Iowa 50011, USA
}

\collaboration{BLFQ Collaboration}

\date{\today}

\begin{abstract}
	

Within the basis light-front quantization framework, we compute the masses and light-front wave functions of the $\Lambda_b$ baryon and its isospin triplet counterparts $\Sigma_b^+$, $\Sigma_b^0$, and $\Sigma_b^-$ using a light-front effective Hamiltonian in the leading Fock sector. These wave functions are obtained as eigenstates of the effective Hamiltonian, which incorporates the one-gluon exchange interaction with fixed coupling and a three-dimensional confinement potential. With the quark masses and the couplings as adjustable parameters, the computed masses are set within the experimental range. The resulting predictions for their electromagnetic properties align well with other theoretical calculations. Additionally, the parton distribution functions (PDFs) of these baryons are obtained for the first time, with gluon and sea quark distributions dynamically generated through QCD evolution of the valence quark PDFs.

\end{abstract}

\maketitle

\section{Introduction}

The $\Lambda_b$ baryon, the lowest-lying bottom baryon with $J^P = \frac{1}{2}^+$, was first observed by the CERN R415 Collaboration in 1981~\cite{qcd1}. Its mass has been measured as $5619.51 \pm 0.23$ MeV, with a mean lifetime of $(1466 \pm 10) \times 10^{-15}$ s~\cite{pdg}. Due to its very short lifetime, experimentally probing the internal structure of the $\Lambda_b$ baryon poses a significant challenge. Nevertheless, various decay modes of the $\Lambda_b$, including both semileptonic and baryonic decays, have been observed~\cite{pdg}. Understanding the decay mechanisms of bottom hadrons, such as the $\Lambda_b$, is crucial for testing the validity of the Standard Model and uncovering potential new physics, particularly in the context of $CP$ violation (CPV). While CPV has been observed in $B$ meson decays~\cite{Belle:2001zzw}, notable differences in the decays of $\Lambda_b$ baryons persist, presenting a puzzle in heavy flavor physics.

A comprehensive understanding of hadrons as bound states of quarks and gluons, particularly in terms of confinement, spectra, and internal structure, remains one of the central challenges of hadronic physics and quantum chromodynamics (QCD)~\cite{qcd1,qcd2}. Several theoretical approaches, such as QCD factorization~\cite{QCDF1,QCDF2,QCDF3} and perturbative QCD~\cite{PQCD1,PQCD2,cpv:fsyu}, have been employed to study $\Lambda_b$ decays. However, the exploration of its internal structure is still in its early stages and constitutes the focus of our work here.

Various nonperturbative approaches have been developed to predict aspects of hadron spectra and to illuminate the partonic structures of hadrons~\cite{Hagler:2009ni,Bashir:2012fs,Brodsky:2014yha,PhysRevC.81.035205}. Among these, Basis Light-front Quantization (BLFQ) has emerged as a promising Hamiltonian framework for solving the nonperturbative dynamics of QCD~\cite{PhysRevC.81.035205}. Within the Hamiltonian formalism, BLFQ offers a practical computational framework for addressing relativistic many-body bound-state problems in quantum field theories~\cite{PhysRevC.81.035205,honkanen2011,PhysRevD.89.116004,ZHAO201465,PhysRevD.91.105009,heavymeson,Lan:2019rba,sreeraj2022,sreeraj2023,Li:2013cga,PhysRevD.96.016022,LI2016118,PhysRevD.98.114038,PhysRevC.99.035206,lanjiangshan2019m,Mondal:2019jdg,xusiqi2021b,adhikari2021m,lanjiangshan2022m,liuyiping2022b,kuangzhongkui2022t,pengtiancai2022b,huzhi2022b,zhuzhimin2023m,zhuzhimin2023m1,xusiqi2023b,linbolang2023b,gross2023,meng2024}, including systems involving strange and charmed baryons~\cite{lambdac,zhuzhimin2023m}. Here, we extend these investigations to bottom baryons, focusing on the $\Lambda_b$ baryon and its isospin triplet.

Electromagnetic form factors (EMFFs) and parton distribution functions (PDFs) are essential tools for probing the internal structure of bound states. EMFFs describe the spatial distributions of electric charge and magnetization within a hadron, while PDFs capture the nonperturbative structure of the hadron by describing the longitudinal momentum and polarization distributions of its constituent quarks and gluons. PDFs are typically extracted from deep inelastic scattering (DIS) experiments and are expressed as functions of the light-front longitudinal momentum fraction ($x$) carried by the hadron's constituents.
At leading twist, the complete spin structure of spin-$\frac{1}{2}$ hadrons is characterized by three independent PDFs: the unpolarized $f_1(x)$, the helicity $g_1(x)$, and the transversity $h_1(x)$. Together, EMFFs and PDFs provide critical insights into the interplay of nonperturbative and perturbative QCD effects encoded in hadrons.
While numerous theoretical and experimental studies have investigated these observables in nucleons (see Ref.~\cite{xusiqi2021b} and references therein), information on bottom baryons remains scarce. A few theoretical predictions have been made regarding the magnetic moments~\cite{mgntmt36,EMS,mgntmt37,mgntmt45,mgntmt47,mgntmt50} and the EMFFs~\cite{EMFFlambdab} of these baryons.

In this work, within the framework of BLFQ~\cite{PhysRevC.81.035205}, we employ an effective light-front Hamiltonian~\cite{Mondal:2019jdg,xusiqi2021b} to solve for the mass eigenstates of the $\Lambda_b$ baryon and its isospin triplet counterparts, the $\Sigma_b$ baryons ($\Sigma_b^+$, $\Sigma_b^0$, and $\Sigma_b^-$). These $\Sigma_b$ baryons, the second lowest-lying bottom baryon states with $J^P = \frac{1}{2}^+$, were first observed in the $\Lambda_b^0 \,\pi$ invariant mass spectrum by the CDF Collaboration~\cite{sigmabfirst}.  
With quarks as the only explicit degrees of freedom, our effective Hamiltonian incorporates a three-dimensional (transverse and longitudinal) confinement potential and a one-gluon exchange (OGE) interaction to account for dynamical spin effects~\cite{LI2016118}. By solving this Hamiltonian in the leading Fock space, using the quark masses, confinement strength, and coupling constant as fitting parameters, we determine the baryon masses as the eigenvalues of the Hamiltonian. Additionally, we obtain the corresponding light-front wave functions (LFWFs) of the baryons as the eigenvectors of the Hamiltonian.
Subsequently, we utilize these resulting LFWFs to investigate the electromagnetic properties and PDFs of the $\Lambda_b$ and its isospin triplet baryons. We apply QCD evolution to the PDFs, evolving them from our model scales to higher scales relevant for the proposed Electron-Ion Colliders (EICs)~\cite{eicc,Accardi:2012qut,eRHIC}. 
While experimental data for these quantities are not yet available, we compare our BLFQ calculations for the magnetic moments with other theoretical results~\cite{EMS,mgntmt37,mgntmt45,mgntmt47,mgntmt50}.

The paper is organized as follows.
We begin with a brief description of the theoretical framework, including a discussion of the corresponding light-front effective Hamitonian in Sec. \ref{sec2}, as well as the resulting mass spectra of $\Lambda_b$ and its isospin triplet baryons. Then we discuss the electromagnetic properties of these bottom baryons in Sec. \ref{sec3}, and present their PDFs in Sec. \ref{sec4}. Finally we conclude in Sec. \ref{sec5}.

\section{\label{sec2}BLFQ AND MASS SPECTRA}

BLFQ is based on the light-front Hamiltonian formalism~\cite{PhysRevC.81.035205}, aiming to solve the light-front eigenvalue equation~\cite{xusiqi2021b},
\begin{equation}
H_\text{LF} \ket{\Psi} = M^2 \ket{\Psi},
\label{eq01}
\end{equation}
where $H_\text{LF} = P^+ P^- - P^2_\perp$ is the effective light-front Hamiltonian, $M^2$ is the squared invariant mass, and $\ket{\Psi}$ is the eigenvector of the bound state, respectively. 

In this paper, we focus on the $\Lambda_b$ and its isospin triplet baryons by solving the corresponding Hamiltonian eigenvalue equation. These bottom baryons can be expanded in the Fock space as:
\begin{equation}
\label{fexpand}
\ket{B} = \psi_{(3q)}\ket{qqq} +\psi_{(3q+g)}\ket{qqqg} + \psi_{(3q+q\bar{q})}\ket{qqqq\bar{q}} + ...
\end{equation}
where $\psi_{(...)}$ denotes the probability amplitudes corresponding to the different parton configurations associated with the Fock components in the baryon. Each Fock component, in turn, consists of an infinite number of basis states within the BLFQ framework.
For the numerical calculations, we employ both a Fock-sector truncation and limitations on the basis states within each Fock component. In this work, we restrict ourselves to the first Fock sector to describe the valence quark contribution to the baryon properties. Under the truncation, the baryon eigenvector $\ket{\Psi}$ with the momentum $P$ and the light-front helicity $\Lambda$ can be expressed in the Fock space representation as~\cite{BRODSKY1998299}
\begin{equation}
\begin{aligned}
 \ket{P,\Lambda}=& \int \prod_{i=1}^3[\frac{d x_i d^2 \vec{k}_{i\perp}}{\sqrt{x_i} 16 \pi^3}] 16 \pi^3 
 \delta(1-\sum_{i=1}^3 x_i) 
 \delta^2 (\sum_{i=1}^3 \vec{k}_{i\perp})\\
 &\Psi_{\{ x_i, \vec{k}_{i\perp}, \lambda_i\}}^\Lambda \ket{\{ x_i P^+, x_i \vec{P}_\perp+\vec{k}_{i\perp},\lambda_i \}}.
\label{iexpand}
\end{aligned}
\end{equation}
Here $x_i=k_i ^+ /P^+$ is the longitudinal momentum fraction, $k_{i\perp}$ represents the relative transverse momentum and $\lambda_i$ labels the light-cone helicity for the $i$-th parton within the Fock sector. The physical transverse momentum is $p_{i\perp}=x_i P_\perp + k_{i\perp}$. $\Psi_{\{ x_i, \vec{k}_{i\perp}, \lambda_i\}}^\Lambda$ are the boost invariant LFWFs. 

With the valence quarks being the only explicit degrees of freedom, the effective light-front Hamiltonian we diagonalize contains an effective Hamitonian $H_\text{eff}$ and a constraint term $H'$~\cite{Mondal:2019jdg,xusiqi2021b,lambdac}:  
\begin{equation}
\begin{aligned}
H_\text{LF}=H_\text{eff}+H'.
\label{hamiltonian}
\end{aligned}
\end{equation}

The effective Hamitonian $H_\text{eff}$ consists of the kinetic energy of quarks, a confining potential, and a one-gluon exchange (OGE) interaction~\cite{Mondal:2019jdg,xusiqi2021b,lambdac}:
\begin{equation}
\begin{aligned}
H_\text{eff}= \sum_{i=1}^3 \frac{\vec{k}_{i\perp}^2+m_{q,i}^2}{x_i}
+\frac{1}{2}\sum_{i\neq j}^3 V_{i,j}^{\text{conf}}
+\frac{1}{2}\sum_{i\neq j}^3 V_{i,j}^{\text{OGE}},
\label{effhami}
\end{aligned}
\end{equation}
where $\sum_i x_i = 1$; $m_q$ is the mass of the constituent quark and $i,j$ denote the index of quarks in the valence Fock sector. For the confining potential, we adopt a form that includes both transverse and longitudinal parts, as employed in Refs.~\cite{Mondal:2019jdg,xusiqi2021b,LI2016118},
\begin{equation}
\begin{aligned}
V_{i,j}^{\text{conf}}=\kappa^4 \Big[\vec{r}_{ij\perp} - \frac{\partial_{x_i}(x_i x_j \partial_{x_j})}{(m_{q,i}+m_{q,j})^2}\Big],
\label{vconf}
\end{aligned}
\end{equation}
where $\kappa$ is the strength of the confinement, $\vec{r}_{ij\perp}=\sqrt{x_i x_j}(\vec{r}_{i\perp}-\vec{r}_{j\perp})$ is the relative coordinate and $\partial_x \equiv (\partial/\partial x )_{r_{ij\perp}}$. Here, the potential is approximately generalized for a many-body system in the single-particle momentum coordinate.

The last term in Eq.~\eqref{effhami} represents the OGE potential 
\begin{equation}
\begin{aligned}
V_{i,j}^{\text{OGE}}=\frac{4\pi C_F \alpha_s}{Q_{ij}^2} \bar{u}_{s'_i}(k'_i)\gamma^\mu u_{s_i}(k_i)\bar{u}_{s'_j}(k'_j)\gamma^\mu u_{s_j}(k_j),
\label{voge}
\end{aligned}
\end{equation}
with fixed coupling constant $\alpha_s$. $C_F = -\frac{2}{3}$ is the color factor which implies that the OGE potential is attractive. $\bar{u}_{s_i}(k_i)$ and $u_{s_i}(k_i)$ are the spinor wave functions. $Q_{ij}^2 = -\frac{1}{2}(p'_i-p_i)^2-\frac{1}{2}(p_j-p'_j)^2$ is the average of 4-momentum square carried by the exchanged gluon which can be expanded in terms of kinematical variables as
\begin{equation}
\begin{aligned}
Q_{ij}^2=\frac{1}{2}&\Big[ \Big( \frac{\vec{p}_{i\perp}^2+m_{q,i}^2}{x_i} - 
\frac{\vec{p}_{i\perp}'^2+m_{q,i}^2}{x'_i}\\
&- \frac{(\vec{p}_{i\perp}^2-\vec{p}_{i\perp}'^2)+\mu_g^2}{x_i-x_i'}\Big) -(i\rightarrow j) \Big],
\label{qij}
\end{aligned}
\end{equation}
here $\mu_g$ represents the gluon mass that regulates the infrared divergence in OGE.

Note that the effective Hamiltonian $H_\text{eff}$, as given in Eq.~\eqref{effhami}, includes the transverse center-of-mass (c.m.) motion, which is mixed with the intrinsic motion. To factor out the transverse c.m. motion from the intrinsic motion, we introduce a constraint term $H'$, defined as
\begin{equation}
\begin{aligned}
H'=\lambda_L (H_\text{c.m.}- 2b^2 I),
\label{hd}
\end{aligned}
\end{equation}
into the effective Hamiltonian. Here, $b$ is the scale of the adopted HO basis (see the discussion of Eq.\eqref{hobasis}), $2b^2$ represents the 2-D HO zero-point energy, $\lambda_L$ is a Lagrange multiplier, and $I$ denotes the identity operator. The c.m. motion is controlled by~\cite{xusiqi2021b,PhysRevD.91.105009}
\begin{equation}
\begin{aligned}
H_\text{c.m.}=\Big( \sum_{i=1}^3 \vec{k}_{i\perp}\Big)^2 + b^4\Big(\sum_{i=1}^3 x_i \vec{r}_{i\perp}\Big)^2,
\label{hcm}
\end{aligned}
\end{equation}
where $\vec{r}_{i\perp}$ is the coordinate of the $i$-th quark. By setting $\lambda_L$ positive and sufficiently large, we can shift the excited states of the c.m. motion to higher energy away from the low-lying states. As a result, the low-lying states have a common c.m. motion configuration which is the lowest state of the HO. They differ among themselves in their orthonormal components of relative motion.

In BLFQ, we adopt the discretized bases. For each quark, its longitudinal motion is described by a plane wave $e^{-\text{i}k^+ x^- /2}$, which is normalized in a one-dimensional box of length $2L$. Thus, we have
\begin{equation}
\begin{aligned}
k_i^+ = \frac{2\pi k_i}{L},
\label{klongi}
\end{aligned}
\end{equation}
with $k_i$ being (half-) integers for (fermions) bosons, and the longitudinal momentum fraction of the $i$-th quark $x_i$ is discretized as $x_i=k_i/K_\text{tot}$, where $K_{\text{tot}}$ is our longitudinal truncation parameter. The transverse motion is represented by a 2-D HO basis function~\cite{PhysRevC.81.035205,ZHAO201465,meng2024}, 
\begin{equation}
\begin{aligned}
\phi_{nm}({\vec{k}_\perp})=&\frac{1}{b}\sqrt{\frac{4\pi n!}{(n+\abs{m})!}}\Big(\frac{\abs{\vec{k}_\perp}}{b}\Big)^{\abs{m}}\\ 
&\times e^{-\frac{1}{2} \vec{k}_\perp ^2/b^2} L^{|m|}_n \Big(\frac{\vec{k}_\perp^2}{b^2} \Big) e^{\imag m\theta},
\end{aligned}
\label{hobasis}
\end{equation}
where $n$ and $m$ are the radial and orbital quantum numbers, respectively. $b$ sets the energy scale of HO, $L_n^{\abs{m}}$ is the associated Laguerre polynomial, and $\theta = \arg (\vec{k}_\perp)$. Thus $\bar \beta_i \equiv \{k_i, n_i, m_i, \lambda_i\}$ denotes the complete single-particle quanta for the $i$-th quark, where we have
\begin{equation}
\sum_{i=1}^3 (\lambda_i + m_i) = \Lambda.
\label{projection}
\end{equation} 
We truncate the infinite 2-D HO basis in transverse directions by introducing the truncation parameter $N_\text{max}$, such that the retained basis states satisfy
\begin{equation}
\sum_{i=1}^3 2n_i +\abs{m_i}+1 \le N_\text{max},
\label{sum}
\end{equation} 
where the summation runs over all the constituents. The basis cutoff $N_\text{max}$ serves as an implicit regulator for the LFWFs in the transverse direction. The infrared (IR) and ultraviolet (UV) cutoffs are given by 
$\Lambda_{\rm IR} \sim \frac{b}{\sqrt{N_\text{max}}}$ and $\Lambda_{\rm UV} \sim b \sqrt{N_\text{max}}$,
respectively~\cite{heavymeson, meng2024}.

With the truncation parameters $K_\text{tot}$ and $N_\text{max}$ set, we solve the Hamiltonian eigenvalue equation, Eq.\eqref{eq01}, within the BLFQ framework to obtain the LFWFs for the valence Fock sector, which can be written as~\cite{huzhi2021,lambdac,meng2024},
\begin{equation}
\begin{aligned}
\Psi_{\{ x_i, \vec{k}_{i\perp}, \lambda_i\}}^\Lambda = \sum_{\{n_i, m_i \}} [\psi^\Lambda _{\{\bar\beta_i \}}
  \prod_{i=1}^3 \phi_{n_i, m_i}({\vec{k}_{i\perp}};b)],
\label{lfwf}
\end{aligned}
\end{equation}
where $\psi^\Lambda _{\{\bar\beta_i \}} = \bra{\{ \bar\beta_i \}}\ket{P,\Lambda}$ are the components of the eigenvectors obtained by diagonalizing the Hamitonian in Eq.~\eqref{eq01}. 
The LFWFs are expected to exhibit parity symmetry ($P$). Although this symmetry is broken due to the Fock space truncation, we employ mirror parity~\cite{Brodsky2006}:  $\hat{P}_x=\hat{R}_x(\pi)P$  as a substitute. Under the mirror parity transformation, our LFWFs satisfy the following relation~\cite{xusiqi2021b},
\begin{equation}
\begin{aligned}
\psi^\downarrow_{\{ x_i,n_i,m_i,\lambda_i\}} = 
(-1)^{\sum_i m_i+1}\psi^\uparrow_{\{ x_i,n_i,-m_i,-\lambda_i\}},
\label{wfdown}
\end{aligned}
\end{equation}
where the arrows indicate the light-cone helicity of the baryons. Note that $\Lambda_b$ and $\Sigma_b$ correspond to the ground state and the first excited state of the Hamiltonian in Eq.~\eqref{eq01}. Consequently, the eigenvectors $\psi^\Lambda _{\{\bar\beta_i \}}$ in Eq.~\eqref{lfwf} differ for $\Lambda_b$ and $\Sigma_b$. The difference between their wave functions can be identified from their different spin-flavor structures dictated by the naive quark model~\cite{lambdac,zhuzhimin2023m},
\begin{equation}
\begin{aligned}
&\ket{\Lambda_b,\uparrow}_{\text{flavor}\otimes \text{spin}}\\
&=\frac{1}{\sqrt{2}}\Big[\frac{1}{2}(bud+ubd-bdu-dbu)\\
&\otimes \frac{1}{\sqrt{6}}(\uparrow \downarrow \uparrow +\downarrow \uparrow \uparrow -2\uparrow \uparrow \downarrow)\\
&+\frac{1}{\sqrt{12}}(dbu-bdu+bud-ubd+2udb-2dub)\\
& \otimes \frac{1}{\sqrt{2}}(\uparrow \downarrow \uparrow -\downarrow \uparrow \uparrow)\Big],
\label{lambdabsf}
\end{aligned}
\end{equation}
and
\begin{equation}
\begin{aligned}
&\ket{\Sigma_b^0,\uparrow}_{\text{flavor}\otimes \text{spin}}\\
&=\frac{1}{\sqrt{2}}\Big[\frac{1}{\sqrt{12}}(bdu+bud+ubd+dbu-2udb-2dub)\\
&\otimes \frac{1}{\sqrt{6}}(\uparrow \downarrow \uparrow +\downarrow \uparrow \uparrow -2\uparrow \uparrow \downarrow)\\
&+\frac{1}{2}(bud-ubd-bdu+dbu)\\
& \otimes \frac{1}{\sqrt{2}}(\uparrow \downarrow \uparrow -\downarrow \uparrow \uparrow)\Big].
\label{sigmab}
\end{aligned}
\end{equation}

\subsection{Mass Spectra}


Our calculation involves the following parameters: the quark mass in the kinetic energy ($m_{q/k}$), the quark mass in the OGE interaction ($m_{q/g}$), the strength of confining potential ($\kappa$), and the coupling constant ($\alpha_s$).
Notably, we adopt a separate quark mass in the OGE interaction ($m_{q/g}$) from that in kinetic energy ($m_{q/k}$).
We outline the reasoning behind the flexibility in choosing the vertex mass. Our approach incorporates an effective OGE interaction that captures the connection between short-distance physics and processes in which valence quarks emit and absorb a gluon, causing the system to fluctuate between the $\ket{qqq}$, $\ket{qqqg}$, and higher Fock sectors. According to mass evolution in renormalization group theory, the dynamical OGE generates contributions to the quark mass from higher momentum scales, which leads to a reduction in the quark mass due to gluon dynamics. Consequently, this suggests that the mass used in the OGE interaction is smaller than the kinetic mass, which is associated with long-range physics in the effective Hamiltonian. This treatment has been noticed and adopted in the literature~\cite{Brisudova:1994it,Burkardt:1998dd,Burkardt:1991tj,Mondal:2019jdg,xusiqi2021b,lambdac}.


We set the HO basis scale parameter $b=0.6\mathrm{GeV}$ in consistent with the previous works on the nucleons~\cite{xusiqi2021b}, the $\Lambda$ and the $\Lambda_c$~\cite{lambdac}, and the truncation parameters to $N_{\rm{max}} = 10$ and $K = 32$, with the model parameters summarized in Table~\ref{para}. The light quark mass ($m_q$) and the strength of the confining potential were determined by fitting the nucleon mass and flavor form factors (FFs)~\cite{Mondal:2019jdg,xusiqi2021b}. Note that these parameters have also been applied to strange ($\Lambda$) and charmed ($\Lambda_c$) baryons~\cite{lambdac}. In this study, one of the light quark masses is replaced with the effective bottom quark mass ($m_b$) for the $\Lambda_b$ baryon. Only the heavy quark masses, $m_{b/k}$ and $m_{b/g}$, are then adjusted to match the experimentally known mass of the $\Lambda_b$, as reported by the Particle Data Group (PDG)~\cite{pdg}. Following the light quark masses, we adopt a $0.1\mathrm{GeV}$ mass difference between $m_{b/k}$ and $m_{b/q}$. The coupling constant $\alpha_s$ is consistent with a previous work~\cite{lambdac}. Additionally, we introduce a $10\%$ uncertainty in $\alpha_s$ to effectively account for the impact of truncations on the system mass ($M$).

In Table~\ref{mass}, we present the resulting mass spectra, comparing them with the experimental data~\cite{pdg}. The uncertainties in our results arise from the error margins we set for the coupling constant $\alpha_s$. We fit only the mass of the $\Lambda_b$, which corresponds to the ground state in our BLFQ approach. For the first excited state, we adopt a mixture of isospin triplet baryons ($\Sigma_b^+$, $\Sigma_b^0$, $\Sigma_b^-$), given the incomplete experimental data available for these states. Notably, the mass of the $\Lambda_b$ aligns well with experimental measurements, whereas the masses of the isospin triplet states are slightly underestimated compared to experimental values. This discrepancy can be attributed to the Fock-space truncation, which prevents us from fully capturing the dynamical role of gluons. Instead, we model the gluon dynamics as effective interactions among the three valence quarks. In future work, we aim to include higher Fock sectors to account for contributions from gluons and sea quarks. This extension will incorporate fundamental QCD interactions and improve our predictions for isospin-dependent mass differences.

Using the model parameters summarized in Table~\ref{para}, we compute the corresponding LFWFs. These LFWFs are then employed to predict the EMFFs and PDFs of the $\Lambda_b$ and its isospin triplet baryons. Additionally, we provide predictions for the electromagnetic radii and magnetic moments of these baryons.

 \begin{table}[ht]
	\caption{\justifying{Model parameters for the basis truncations $N_{\rm{max}}=10$ and $K=32$ for the $\Lambda_b$ and its isospin triplet baryons.}}\label{para}
	\centering 
		\begin{tabular}{|cccc|}
			\hline \hline
			 $\alpha_s$ ~& $m_{q/k}/m_{q/g}$ ~& $m_{b/k}/m_{b/g}$ ~& $\kappa$  \\ 
			\hline 		
			 0.57$\pm0.06$ ~&  0.30/0.20~[GeV] ~& 5.05/4.95~[GeV] ~& 0.337~[GeV] \\ 
			\hline \hline 	
		\end{tabular}
\end{table}


\begin{table}[t]
  \caption{\justifying{The masses of $\Lambda_b$ and its isospin states ($\Sigma_b^+$, $\Sigma_b^0$, and $\Sigma_b^-$). Our results are compared with the experimental data~\cite{pdg}. Currently the mass of $\Sigma_b^0$ is unavailable in experiment. The masses of baryons in units of MeV.}}
    \vspace{0.15cm}
    \label{mass} 
    \centering
\begin{tabular}{ l l l }
    \hline \hline 
    Baryons ~~&~~ $M_{\text {BLFQ }}$ ~~&~~~~~~~ $M_{\text {exp }}$ \\
    \hline
~~~~$\Lambda_b$ ~~&~~ $5624_{-2}^{+2}$ ~~&~~ $5619.60 \pm 0.17 $ \\
~~~~$\Sigma_b$  ~~&~~ $5636_{-2}^{+2}$ ~~&~~ $5810.56 \pm 0.25 $($\Sigma_b^+$) \\
~~&~~ ~~&~~ $\cdots$($\Sigma_b^0$) \\
~~&~~ ~~&~~ $5815.64 \pm 0.27 $($\Sigma_b^-$) \\
\hline \hline
\end{tabular}
\end{table}

\section{\label{sec3}ELECTROMAGNETIC FORM FACTORS}

\begin{figure*}[t!]
    \includegraphics[width=0.49\textwidth]{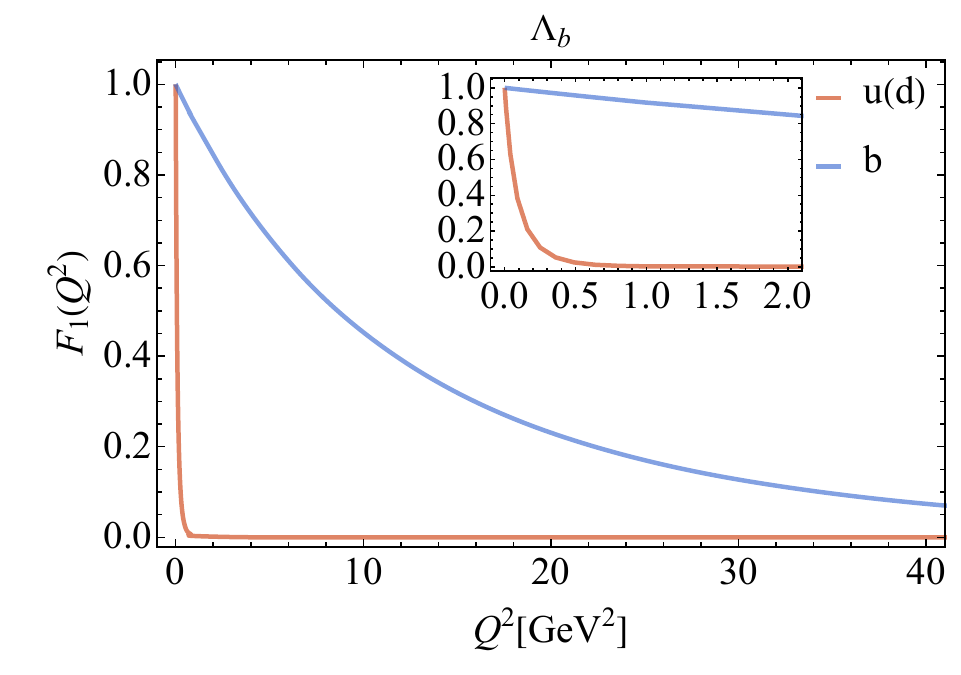}
    \includegraphics[width=0.49\textwidth]{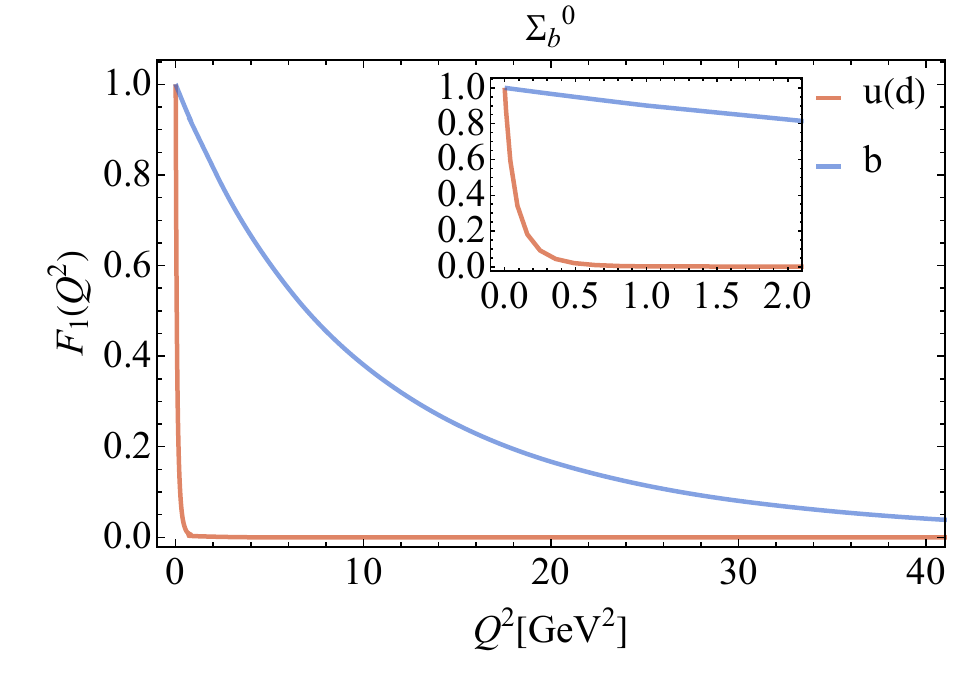}
    \includegraphics[width=0.49\textwidth]{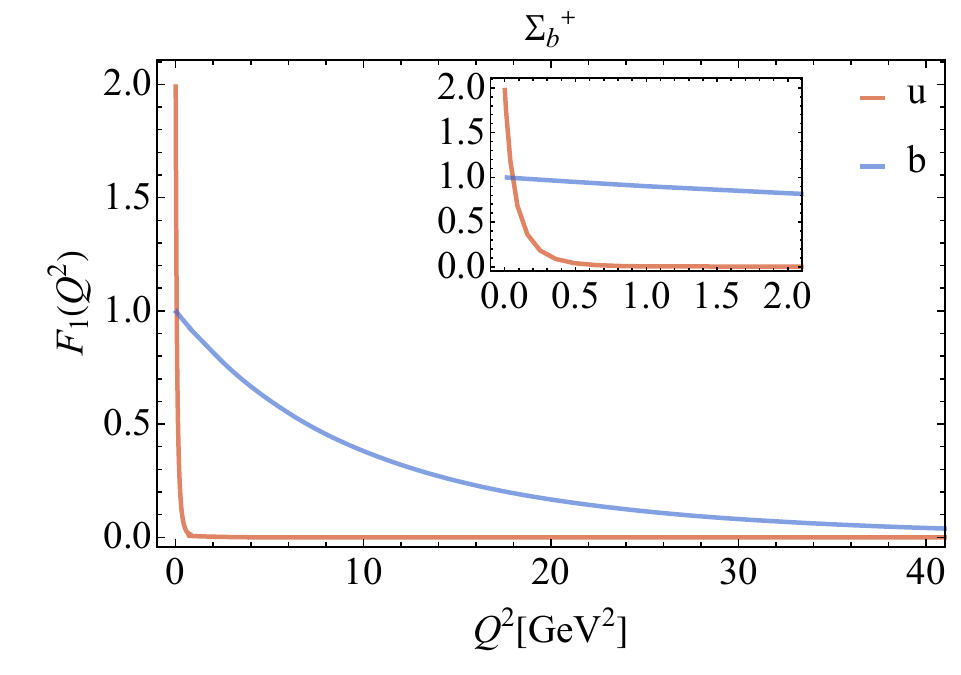}
    \includegraphics[width=0.49\textwidth]{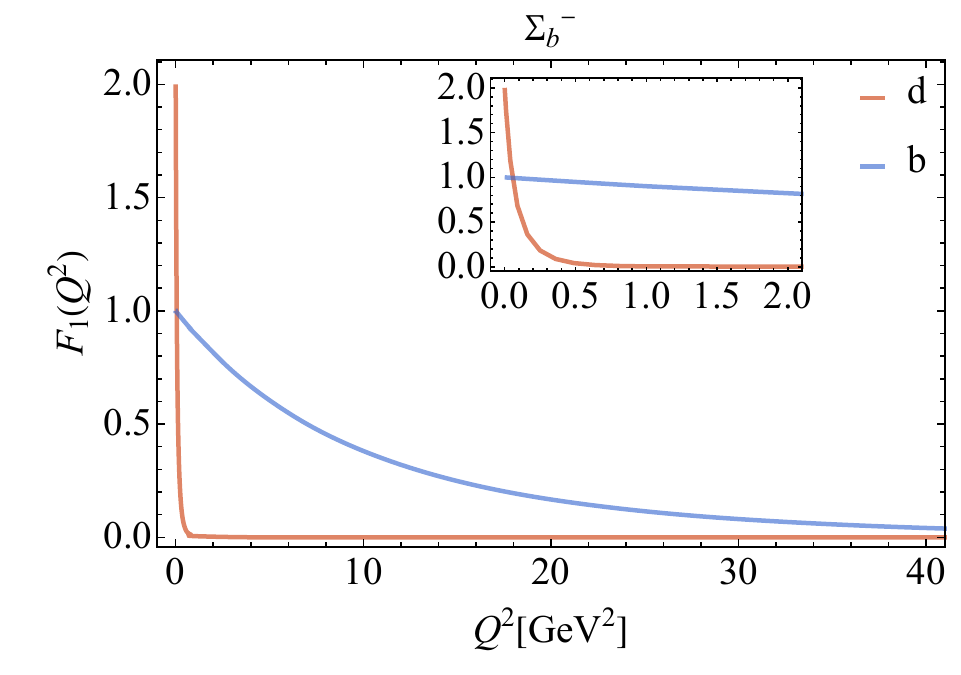}
	\caption{\justifying{Flavor Dirac FFs of $\Lambda_b$ and its isospin states ($\Sigma_b^+$, $\Sigma_b^0$, and $\Sigma_b^-$). The red lines represent the light quark ($u$ and/or $d$) FFs, whereas the blue lines correspond to the bottom quark ($b$) FFs. The insets display the results on an expanded horizontal scale to better display the behaviors at small $Q^2$. The uncertainty band resulting from the variation in the coupling constant $\alpha_s$ is indistinguishable from the lines due to the dominant effect of the very heavy $b$ quark mass.}}
	\label{dirac}
\end{figure*}

\begin{figure*}[t!]
    \includegraphics[width=0.49\textwidth]{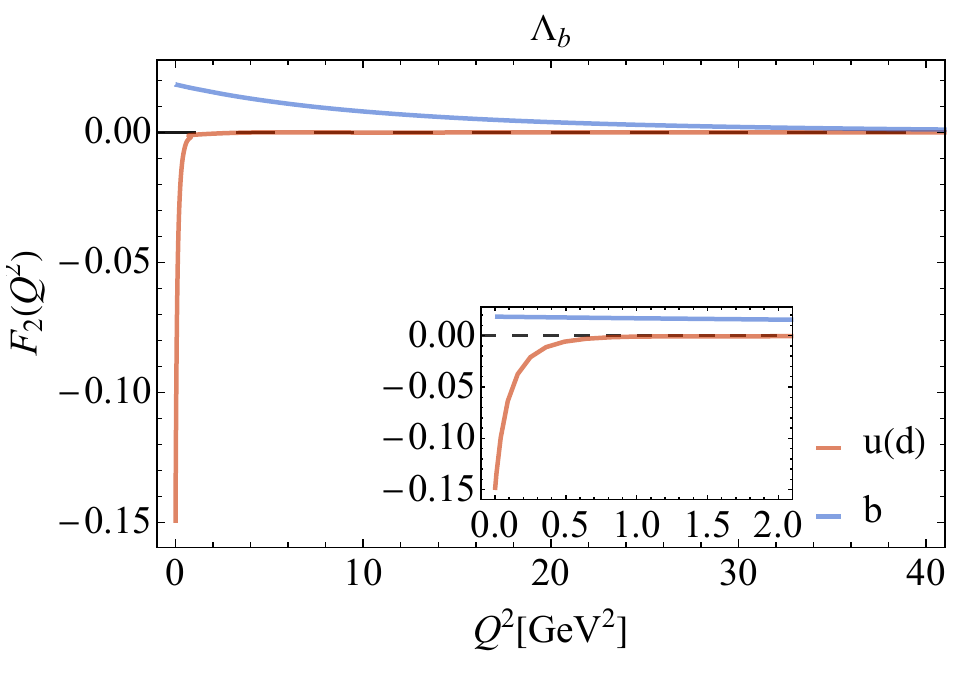}
    \includegraphics[width=0.49\textwidth]{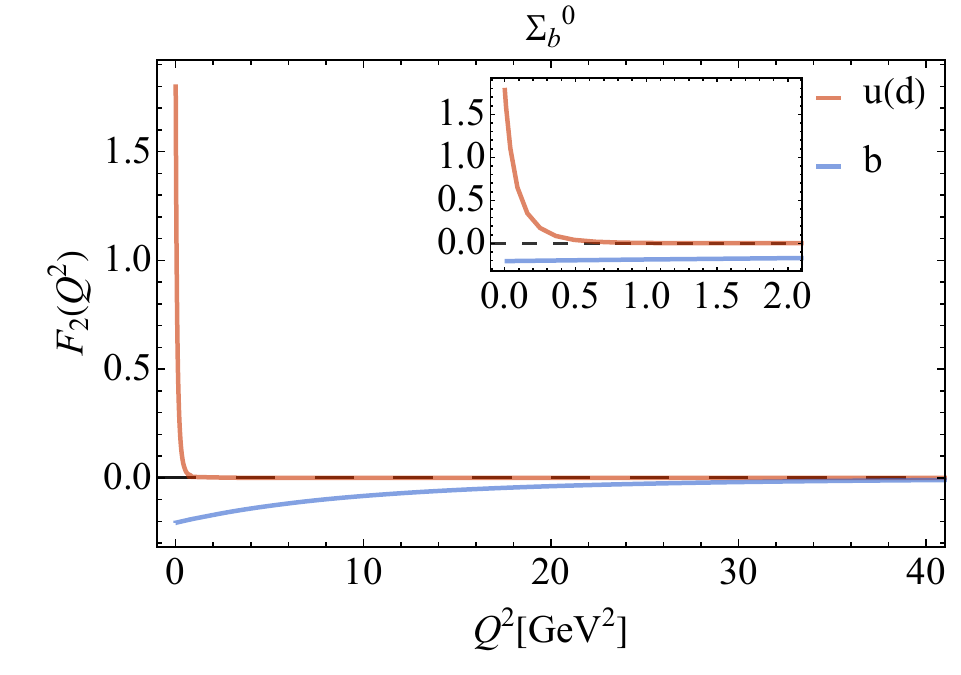}
    \includegraphics[width=0.49\textwidth]{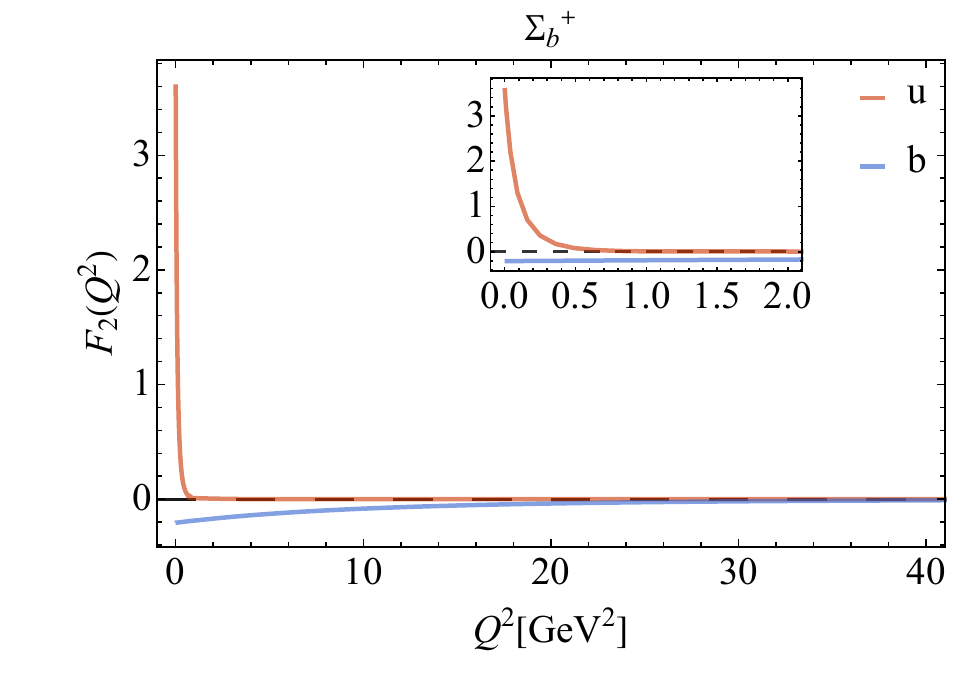}
    \includegraphics[width=0.49\textwidth]{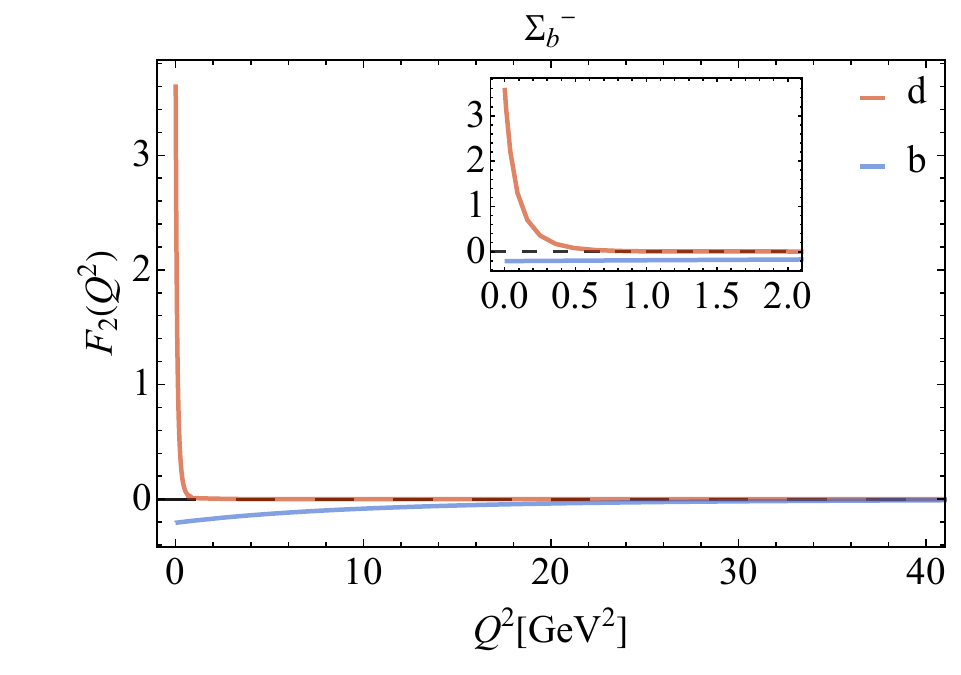}
	\caption{\justifying{Flavor Pauli FFs of $\Lambda_b$ and its isospin states ($\Sigma_b^+$, $\Sigma_b^0$, and $\Sigma_b^-$). The red lines represent the light quark ($u$ and/or $d$) FFs, whereas the blue lines correspond to the bottom quark ($b$) FFs. The insets display the results on an expanded horizontal scale to better display the behaviors at small $Q^2$. The uncertainty band resulting from the variation in the coupling constant $\alpha_s$ is indistinguishable from the lines due to the dominant effect of the very heavy $b$ quark mass. The horizontal dashed gray line indicates the position of zero.}}
	\label{pauli}
\end{figure*}

\begin{figure*}[t!]
    \includegraphics[width=0.49\textwidth]{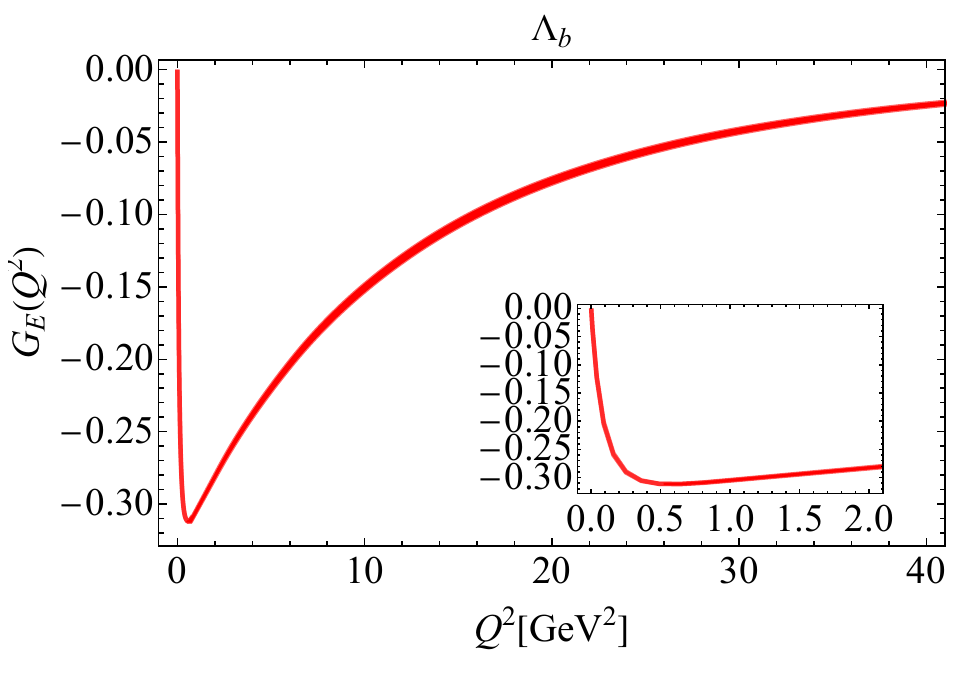}
    \includegraphics[width=0.49\textwidth]{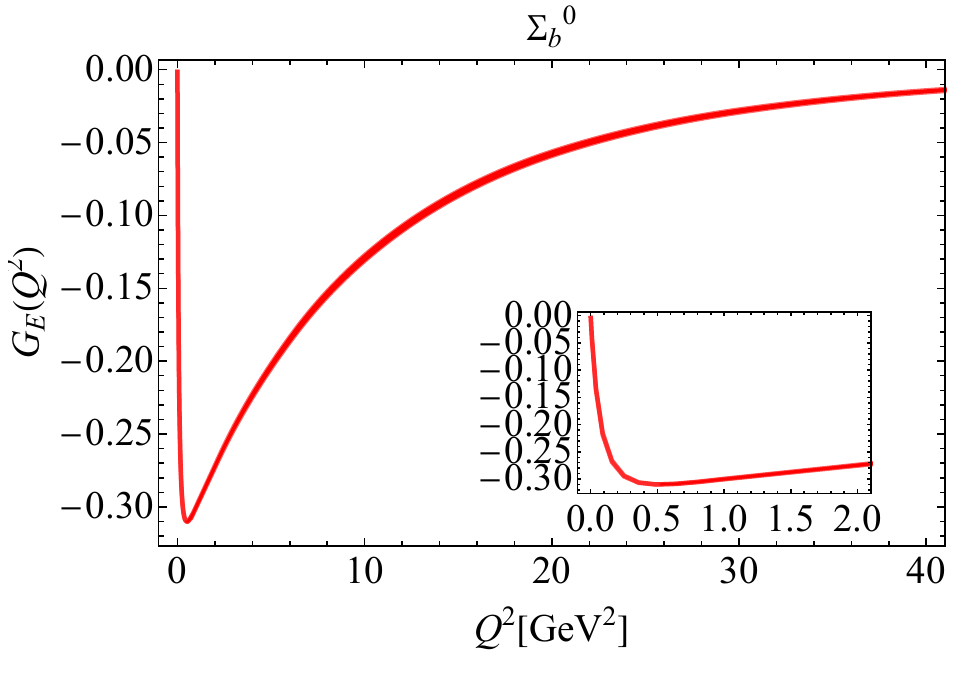}
    \includegraphics[width=0.49\textwidth]{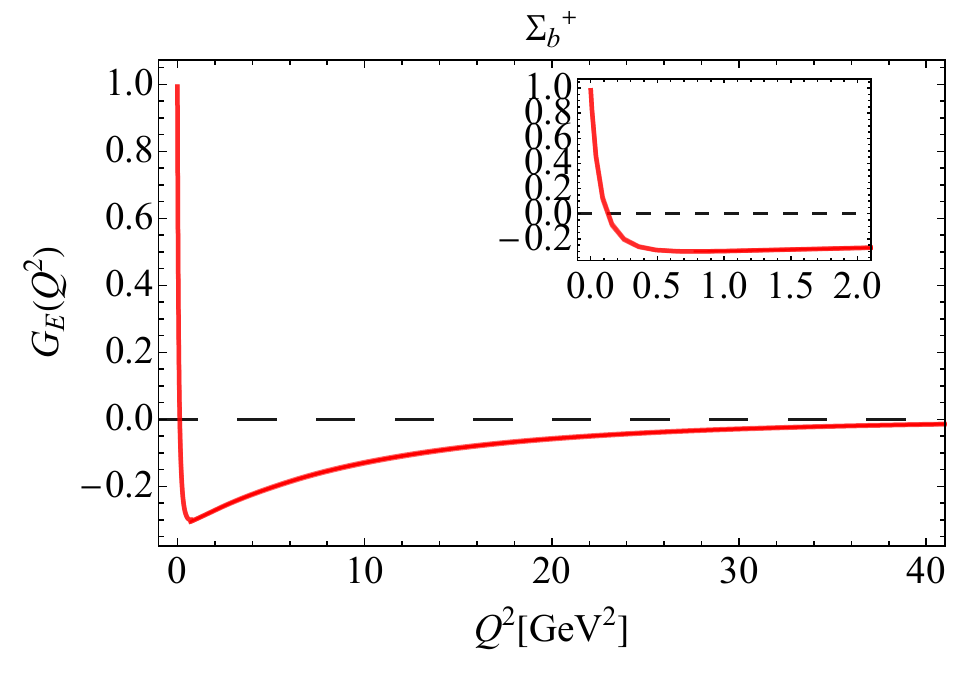}
    \includegraphics[width=0.49\textwidth]{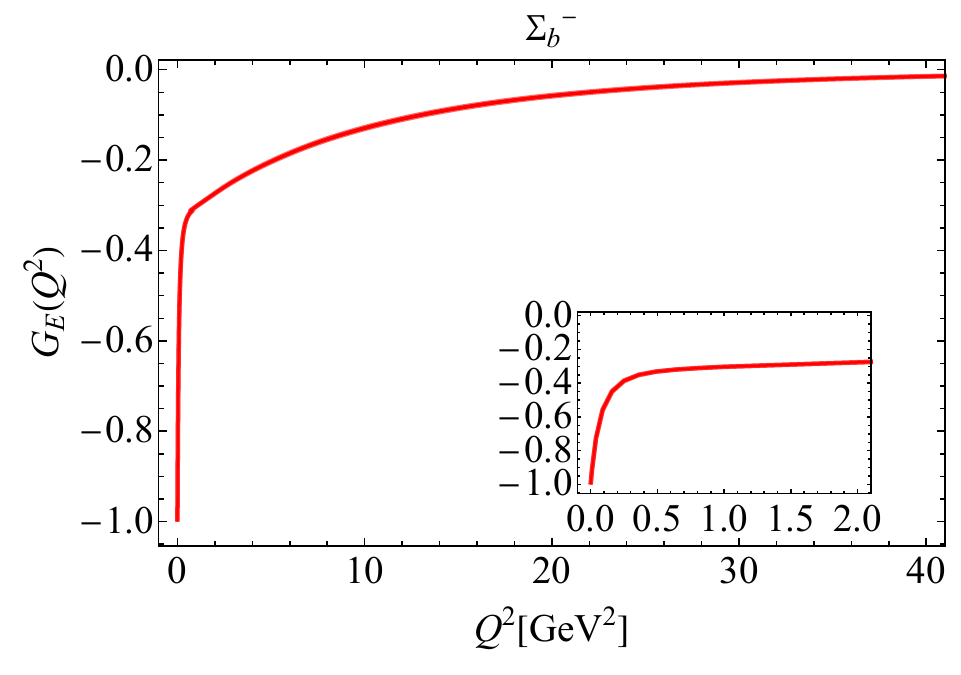}
	\caption{\justifying{Electric Sachs FFs $G_E(Q^2)$ as functions of $Q^2$ for $\Lambda_b$ and its isospin states ($\Sigma_b^+$, $\Sigma_b^0$, and $\Sigma_b^-$). The red bands represent our results obtained within the BLFQ approach, reflecting the $10\%$ uncertainty in the coupling constant $\alpha_s$. The horizontal dashed black line indicates the position of zero. The insets display the results on an expanded horizontal scale to better display the behaviors at small $Q^2$.}}
	\label{ge}
\end{figure*}

\begin{figure*}[t!]
    \includegraphics[width=0.49\textwidth]{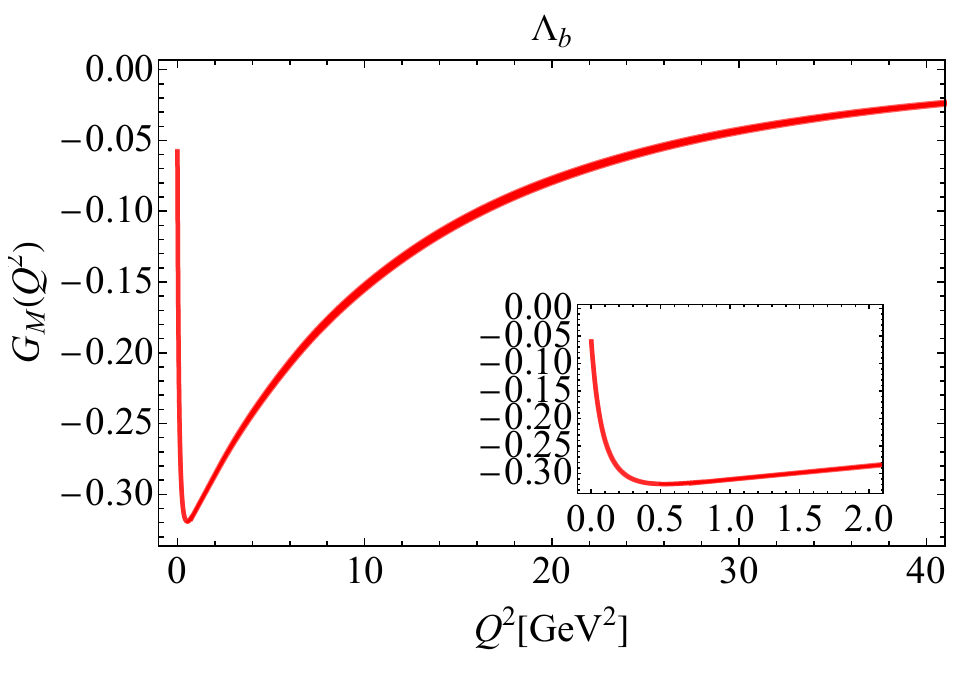}
    \includegraphics[width=0.49\textwidth]{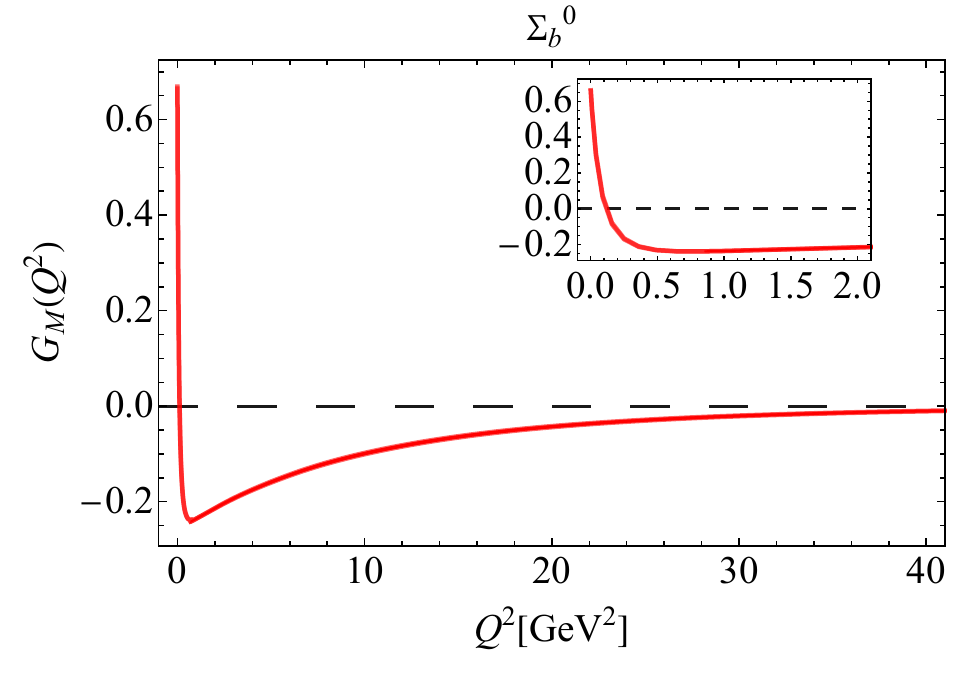}
    \includegraphics[width=0.49\textwidth]{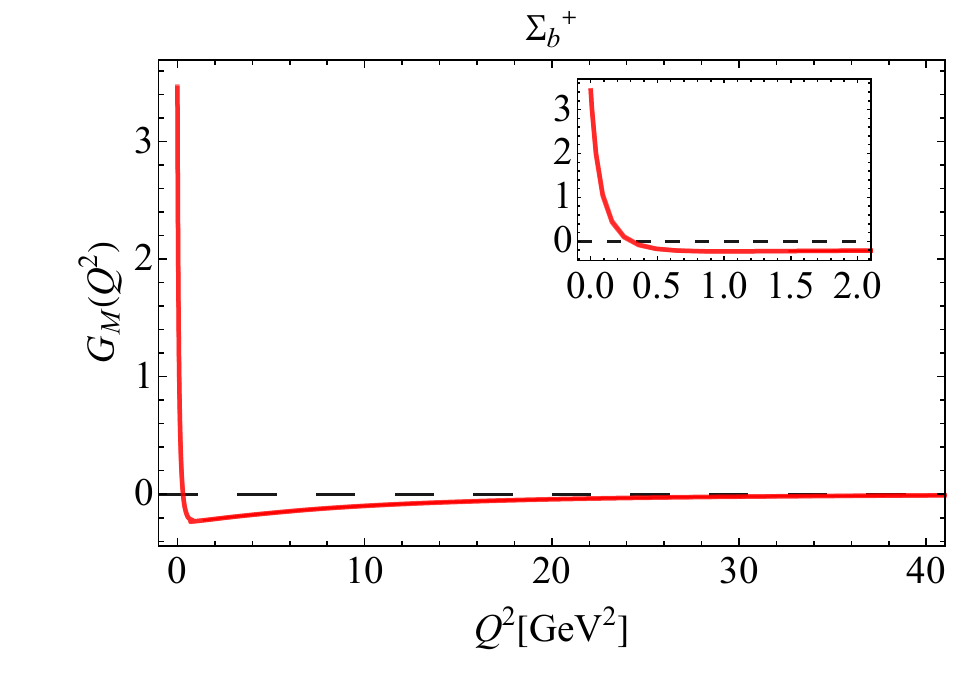}
    \includegraphics[width=0.49\textwidth]{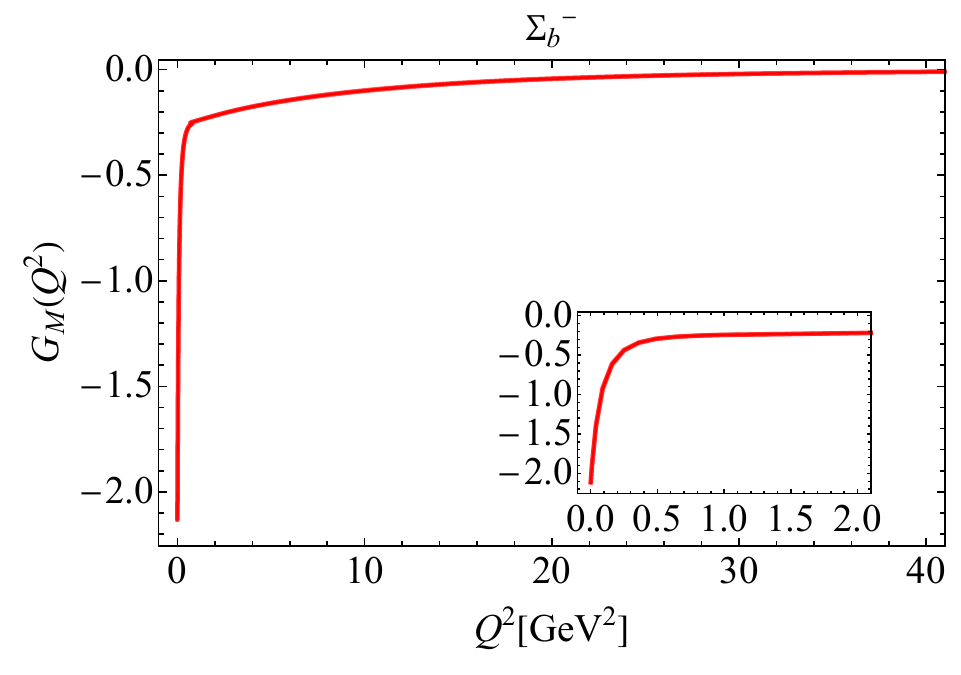}
	\caption{\justifying{Magnetic Sachs FFs $G_M(Q^2)$ as functions of $Q^2$ for $\Lambda_b$ and its isospin states ($\Sigma_b^+$, $\Sigma_b^0$, and $\Sigma_b^-$). The red bands represent our results obtained within the BLFQ approach, reflecting the $10\%$ uncertainty in the coupling constant $\alpha_s$. The horizontal dashed black line indicates the position of zero. The insets display the results on an expanded horizontal scale to better display the behaviors at small $Q^2$.}}
	\label{gm}
\end{figure*}

For spin$-\frac{1}{2}$ composite systems, there are two independent EMFFs $F_1(Q^2)$ and $F_2(Q^2)$, which are known as the Dirac FFs and the Pauli FFs, respectively. In the light-front framework, they are identified with the helicity-conserving and helicity-flip matrix elements of the vector ($J^+ \equiv \sum_q e_q \bar{\psi_q} \gamma^+ \psi_q $) current:
\begin{equation}
\begin{aligned}
\mel{P+q,\uparrow}{\frac{J^+(0)}{2P^+}}{P,\uparrow} & = F_1(Q^2),\\
\mel{P+q,\uparrow}{ \frac{J^+(0)}{2P^+}}{P,\downarrow} & = -\frac{(q_1 -\imag q_2)}{2M} F_2(Q^2).
\end{aligned}
\end{equation}
Here, $Q^2 = -q^2$ represents the square of the momentum transfer. We consider the frame where the longitudinal component of the momentum transfer is zero, i.e., $q = (0, 0, \vec{q}_\perp)$, and thus, $Q^2 = -q^2 = \vec{q}_\perp^2$. $M$ denotes the mass of the system. Within the valence Fock sector, the flavor Dirac and Pauli FFs can be expressed as overlaps of the LFWFs~\cite{BRODSKY200199}:
\begin{equation}
\begin{aligned}
F_1^q (Q^2) = \sum_{\{\lambda_i\}} \int \left[\mathrm{d} \mathcal{X} \mathrm{d} \mathcal{P}_{\perp}\right]
\Psi_{\{ x_i', \vec{k}_{i\perp}', \lambda_i\}}^{\uparrow *} \Psi_{\{ x_i, \vec{k}_{i\perp}, \lambda_i\}}^\uparrow,
\end{aligned}
\end{equation}
\begin{equation}
\begin{aligned}
F_2^q (Q^2) = 
&-\frac{2M}{q^1-iq^2} \sum_{\{\lambda_i\}} \int \left[\mathrm{d} \mathcal{X} \mathrm{d} \mathcal{P}_{\perp}\right] \\
& \times \Psi_{\{ x_i', \vec{k}_{i\perp}', \lambda_i\}}^{\uparrow *} \Psi_{\{ x_i, \vec{k}_{i\perp}, \lambda_i\}}^\downarrow,
\end{aligned}
\end{equation}
where $x_q'=x_q$ and $\vec{k}_{q\perp}'=\vec{k}_{q\perp} + (1-x_q) \vec{q}_{\perp}$ for the struck quark, while $x_i'=x_i$ and $\vec{k}_{i\perp}'=\vec{k}_{i\perp} -x_i \vec{q}_{\perp}$ for the spectators ($i\neq q$), and 
\begin{equation}
\begin{aligned}
\left[\mathrm{d} \mathcal{X} \mathrm{d} \mathcal{P}_{\perp}\right]=
&\prod_{i=1}^3[\frac{d x_i d^2 \vec{k}_{i\perp}}{\sqrt{x_i} 16 \pi^3}] \\
& \times 16 \pi^3 
 \delta(1-\sum_{i=1}^3 x_i) 
 \delta^2 (\sum_{i=1}^3 \vec{k}_{i\perp}).
\end{aligned}
\end{equation}
The Dirac FFs follow the normalizations
\begin{equation}
\begin{aligned}
F_1^q (0) = n_q,
\end{aligned}
\end{equation}
where $n_q$ is the number of valence quarks of flavor $q$ in the baryon. The Pauli FFs  at zero momentum transfer provide the quark anomalous magnetic moments,
\begin{equation}
\begin{aligned}
F_2^q (0) = \kappa_q.
\end{aligned}
\end{equation}

We calculate the Dirac and Pauli FFs for the valence quarks in the $\Lambda_b$ and its isospin states using the LFWFs defined in Eq.\eqref{lfwf}. The results for the flavor Dirac and Pauli FFs are shown in Fig.~\ref{dirac} and Fig.~\ref{pauli}, respectively. The red bands represent the results for the light quarks ($u$ and/or $d$), while the blue bands correspond to the bottom ($b$) quark. The error bands in our results arise from the 10\% uncertainty in the coupling constant. Note that these error bands are narrower compared to those for the $\Lambda$ and $\Lambda_c$ baryons reported in Ref.~\cite{lambdac}.
The slope of the EMFF at $Q^2 \rightarrow 0$ represents the electromagnetic radius of the system arising from the quark motion. In the small $Q^2$ region, the FFs for light quarks decrease more rapidly than those of the bottom quark. This behavior arises from the lighter mass of the up (down) quark compared to the bottom quark, resulting in a larger radius arising from the light quark. Consequently, the $b$ quark is more localized near the center of the baryons, as expected.

Since $\Lambda_b$ and $\Sigma_b^0$ share the same flavor content, their flavor Dirac FFs are similar. Owing to the heavier constituent quark in $b$ baryons, they display greater similarity than those of the $\Lambda$ and $\Sigma^0$ baryons~\cite{lambdac}. Additionally, the flavor FFs for $\Sigma_b^+$ and $\Sigma_b^-$ are identical. Notably, the Pauli FF for the light quark in $\Lambda_b$ is negative, while it is positive for the $b$ quark. In contrast, this pattern is reversed in $\Sigma_b^0$, $\Sigma_b^+$, and $\Sigma_b^-$.

Under charge and isospin symmetry, the baryon FFs can be obtained from the flavor FFs,
\begin{equation}
\label{bff}
\begin{aligned}
F_{1(2)}^\text{B} (Q^2) = \sum_q e_q F_{1(2)}^q (Q^2),
\end{aligned}
\end{equation}
where the charges of the quarks are $e_u = \frac{2}{3}$, $e_d = -\frac{1}{3}$ and $e_b = -\frac{1}{3}$, respectively. With Eq.~\eqref{bff}, the Sachs FFs can be expressed in terms of the Dirac and the Pauli FFs:
\begin{equation}
\begin{aligned}
& G_E^\text{B} (Q^2) = F_{1}^\text{B} (Q^2) - \frac{Q^2}{4M^2}F_2^\text{B} (Q^2),\\
& G_M^\text{B} (Q^2) = F_{1}^\text{B} (Q^2) + F_2^\text{B} (Q^2).
\end{aligned}
\end{equation}

\begin{table*}
  \caption{Our predictions for the magnetic moments of $\Lambda_b$ and its isospin states ($\Sigma_b^+$, $\Sigma_b^0$, and $\Sigma_b^-$) in units of the nuclear magneton $\mu_\text{N}$. We compare our predications with other theoretical calculations in Refs.~\cite{EMS,mgntmt37,mgntmt45,mgntmt47,mgntmt50}. In Ref.~\cite{mgntmt45}, M-I and M-II represent the results obtained using two different models.}
    \vspace{0.15cm}
    \label{magnetic moments} 
    \centering
\begin{tabular}{ l r r r r r r r }
    \hline \hline 
    Baryons & {~~~~~~~~~$\mu_{\text {BLFQ}}$~~~~~~} & {~~~~~~~~~~\cite{EMS}~~~} & {~~~~~~~~~~\cite{mgntmt37}~~~}
    & {~~~~~~~~~~M-I\cite{mgntmt45}~~~}
    & {~~~~~~~~~~M-II\cite{mgntmt45}~~~} & {~~~~~~~~~~\cite{mgntmt47}~~~} & {~~~~~~~~~~\cite{mgntmt50}~~~}\\
    \hline
{$\Lambda_b$} & {$-0.0562_{-0.0002}^{+0.0002}$} & $-0.0620$ & {$\cdots$} & $-0.060$ &$-0.066$ & $-0.060$ & {$\cdots$}\\
$\Sigma^0_b$  & $0.6719_{-0.0023}^{+0.0020}$ & 0.5653 & 0.659 & 0.640 & 0.422 & 0.603 & 0.390\\
$\Sigma^+_b$  & $3.4809_{-0.0085}^{+0.0076}$ & 2.1989 & 2.575 & 2.500 & 1.622 & 2.250 & 1.590\\
$\Sigma^-_b$  & $-2.1372_{-0.0036}^{+0.0040}$ & $-1.0684$ & $-1.256$ & $-1.220$ & $-0.778$ & $-1.150$ & $-0.810$\\
\hline \hline
\end{tabular}
\end{table*}

\begin{table}
  \caption{\justifying{Our predictions for the mean square charge and magnetic radii $\expval{r_\text{E}^2}$ and $\expval{r_\text{M}^2}$ of $\Lambda_b$ and its isospin states ($\Sigma_b^+$, $\Sigma_b^0$, and $\Sigma_b^-$) in the units of $\text{fm}^2$.}}
    \vspace{0.15cm}
    \label{chargeradiu} 
    \centering
\begin{tabular}{ l r r }
    \hline \hline 
    Baryons & ~~~~~~~~~$\expval{r_\text{E}^2}_{\text {BLFQ}}$~~~~& ~~~~~~~~~$\expval{r_\text{M}^2}_{\text {BLFQ}}$~~~~\\
    \hline
$\Lambda_b$ & $0.894_{-0.014}^{+0.015}$ & $-13.710_{-0.174}^{+0.159}$ \\
$\Sigma^0_b$  & $1.001_{-0.016}^{+0.017}$ & $4.119_{-0.057}^{+0.059}$ \\
$\Sigma^+_b$  & $4.051_{-0.065}^{+0.066}$ & $3.187_{-0.041}^{+0.043}$ \\
$\Sigma^-_b$  & $2.037_{-0.033}^{+0.033}$ & $2.599_{-0.033}^{+0.033}$ \\
\hline \hline
\end{tabular}
\end{table}


In Fig.~\ref{ge} and Fig.~\ref{gm}, we present the electric and magnetic Sachs FFs of the $\Lambda_b$ and its isospin triplet states, respectively. At $Q^2=0$, the electric Sachs FFs of $\Lambda_b$ and $\Sigma_b^0$ are zero, whereas those of $\Sigma_b^+$ and $\Sigma_b^-$ are 1 and -1, respectively. A strong peak is observed at $Q^2 \approx 0.5$ for both $\Lambda_b$ and $\Sigma_b^0$. Similar peaks are also found for $\Lambda$ and $\Sigma^0$~\cite{lambdac}, likely due to comparable constituent charges. However, the slopes of the electric Sachs FFs for $\Lambda_b$ and $\Sigma_b^0$ are significantly steeper, reflecting the impact of the heavy quark mass.

The magnetic Sachs FFs at $Q^2=0$ correspond to the magnetic moments of the baryons. Table~\ref{magnetic moments} summarizes our calculated magnetic moments and compares them with other theoretical predictions~\cite{EMS,mgntmt37,mgntmt45,mgntmt47,mgntmt50}. Our results align with previous studies in terms of both signs and approximate magnitudes. While the magnetic moments of $\Lambda_b$ and $\Sigma_b^0$ show strong agreement with prior calculations, more significant differences are observed for $\Sigma_b^+$ and $\Sigma_b^-$.

From the Sachs FFs, we can also compute the charge and magnetic radii of the baryons which are defined by
\begin{equation}
\begin{aligned}
\left\langle r_{\mathrm{E}}^2\right\rangle^{\mathrm{B}}=-\left.\frac{6}{G_{\mathrm{E}}^{\mathrm{B}}(0)} \frac{\mathrm{d} G_{\mathrm{E}}^{\mathrm{B}}\left(Q^2\right)}{\mathrm{d} Q^2}\right|_{Q^2=0},
\end{aligned}
\end{equation}
and
\begin{equation}
\begin{aligned}
\left\langle r_{\mathrm{M}}^2\right\rangle^{\mathrm{B}}=-\left.\frac{6}{G_{\mathrm{M}}^{\mathrm{B}}(0)} \frac{\mathrm{d} G_{\mathrm{M}}^{\mathrm{B}}\left(Q^2\right)}{\mathrm{d} Q^2}\right|_{Q^2=0},
\end{aligned}
\end{equation}
respectively. Note that we need to replace the $G_{\mathrm{E}}^{\mathrm{B}}(0)$ with unity when the charge of the baryon is zero. We present our predictions of the charge and the magnetic radii in Table~\ref{chargeradiu}. The magnetic radius of the $\Lambda_b$ is much larger than those of the $\Lambda_c$ and the $\Lambda$~\cite{lambdac}, as expected, owing to the large increase of the heavy quark mass.

\section{\label{sec4}PARTON DISTRIBUTION FUNCTIONS}


\begin{figure*}[t!]
    \includegraphics[width=0.49\textwidth]{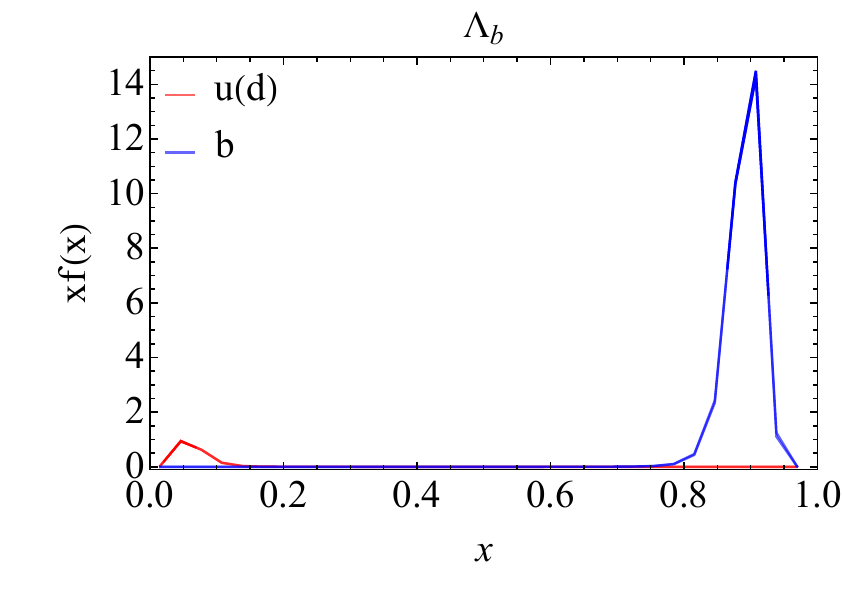}
    \includegraphics[width=0.49\textwidth]{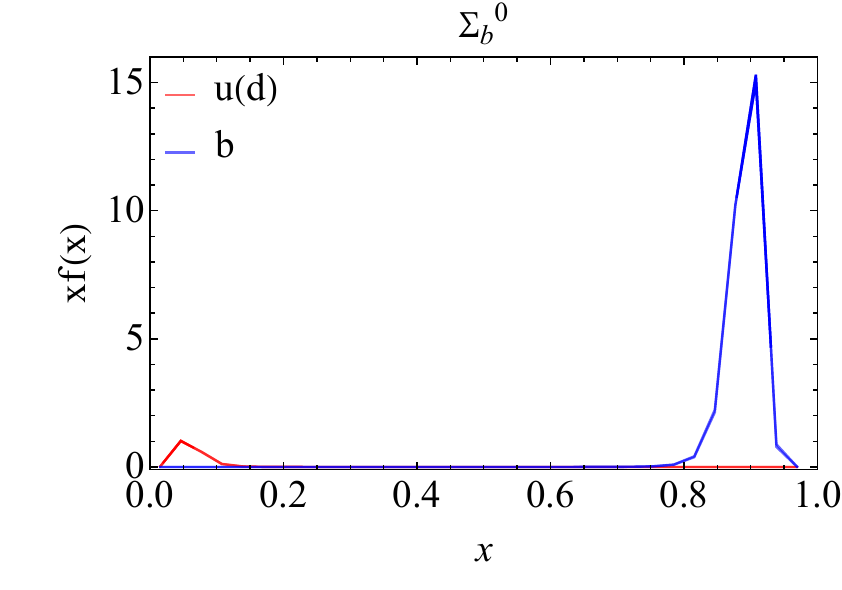}
    \includegraphics[width=0.49\textwidth]{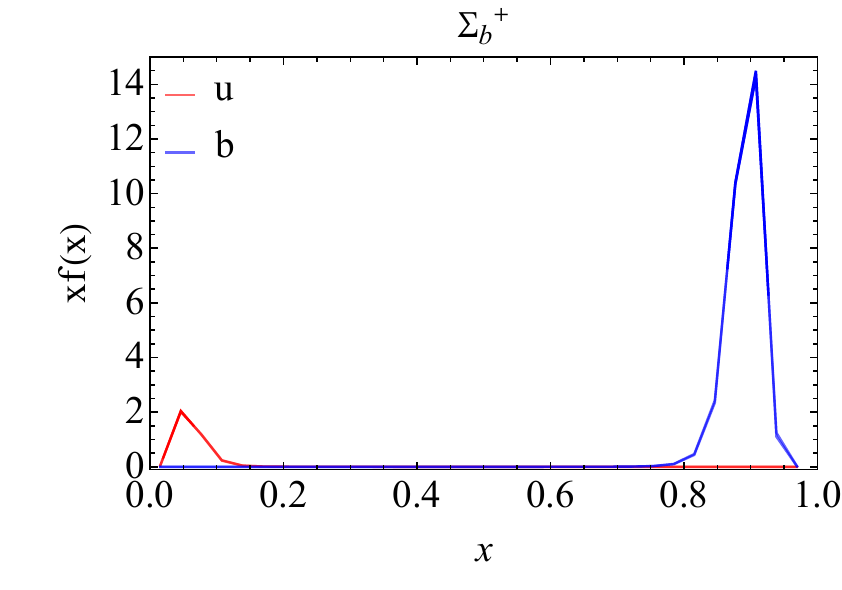}
    \includegraphics[width=0.49\textwidth]{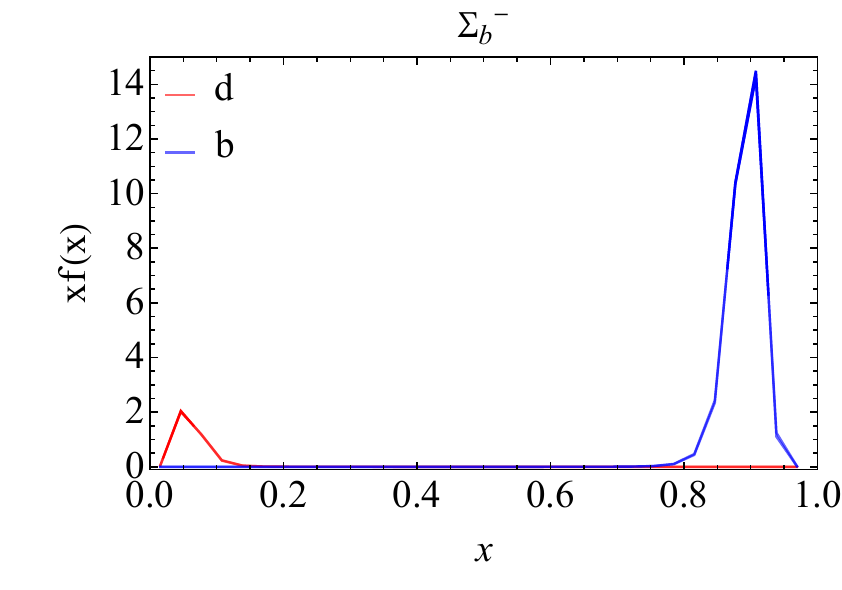}
	\caption{\justifying{Valence quarks’ unpolarized PDFs of $\Lambda_b$ and its isospin states ($\Sigma_b^+$, $\Sigma_b^0$, and $\Sigma_b^-$) at the model scale multiplied by $x$ as functions of $x$. The blue and red bands represent distributions for the bottom quark and the light quark ($u$ and/or $d$), respectively. The bands, which are only slightly larger than the thickness of the primary line, reflect the $10\%$ uncertainty in the coupling constant $\alpha_s$.}}
	\label{fxini}
\end{figure*}

\begin{figure*}[t!]
    \includegraphics[width=0.49\textwidth]{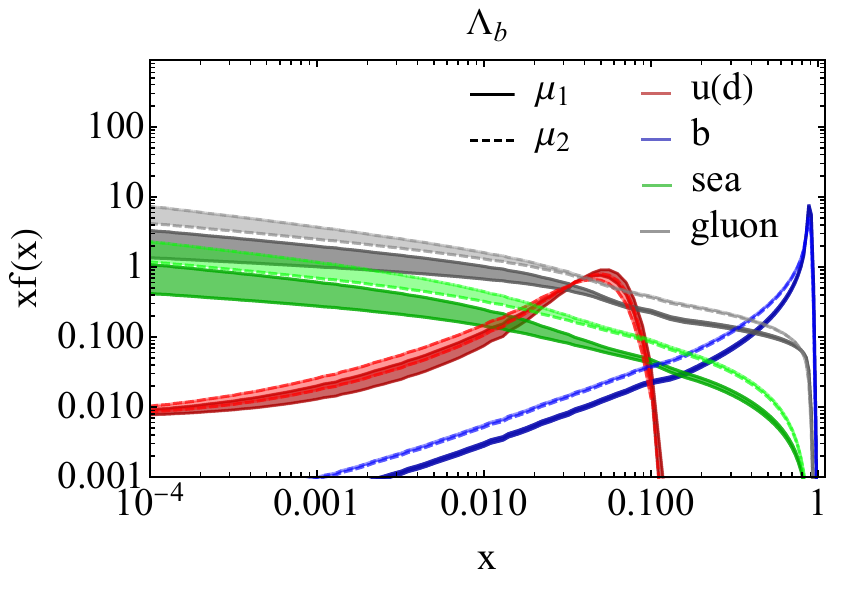}
    \includegraphics[width=0.49\textwidth]{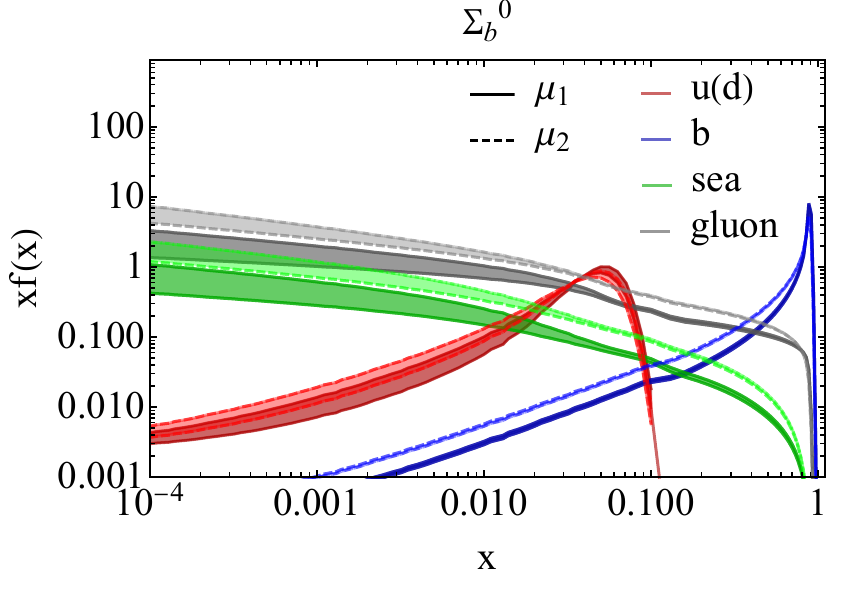}
    \includegraphics[width=0.49\textwidth]{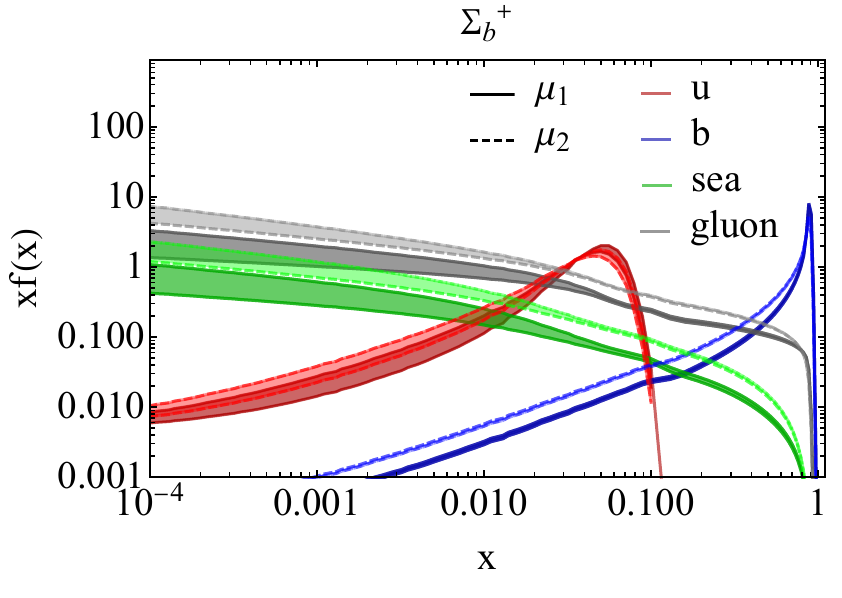}
    \includegraphics[width=0.49\textwidth]{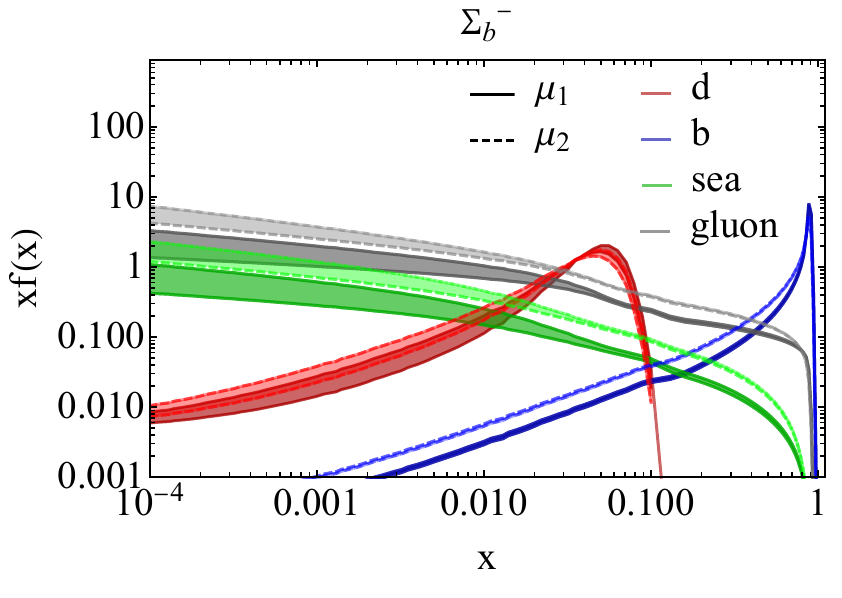}
	\caption{\justifying{Unpolarized PDFs at different final scales ($\mu_1=20\,\text{GeV}, \mu_2=80\,\text{GeV}$) multiplied by $x$ as functions of $x$. The red, blue, green and gray bands represent distributions for the valence light quark ($u$ and/or $d$), valence bottom quark, sea quarks, and gluon, respectively. The bands reflect the uncertainty in the initial scale $\mu_0$ from $1.90\,\text{GeV}$ to $5.05\,\text{GeV}$.}}
	\label{fxevo}
\end{figure*}

\begin{figure*}[t!]
    \includegraphics[width=0.49\textwidth]{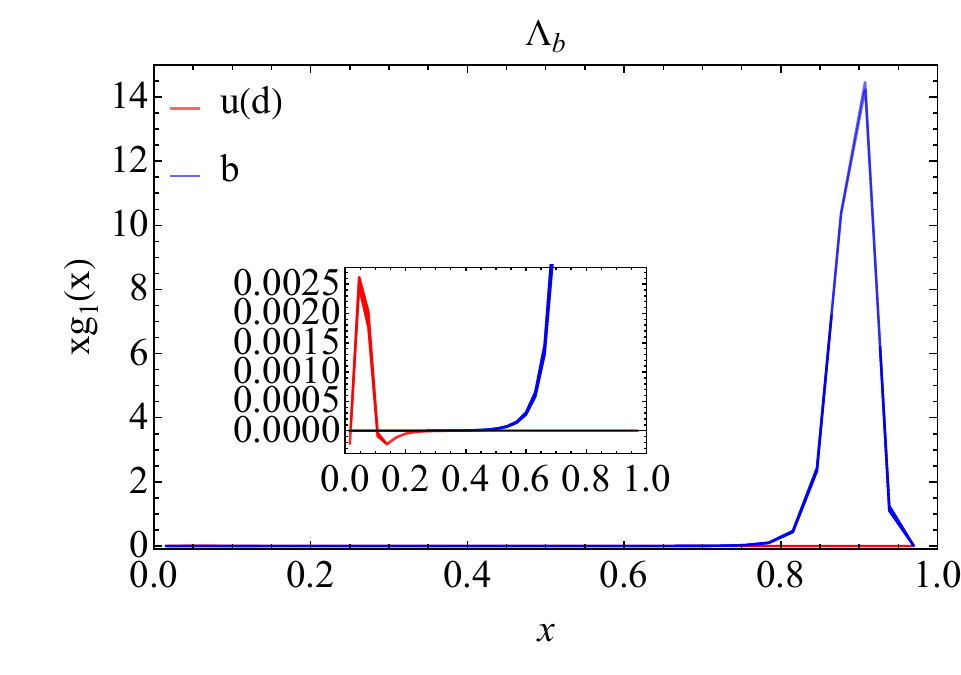}
    \includegraphics[width=0.49\textwidth]{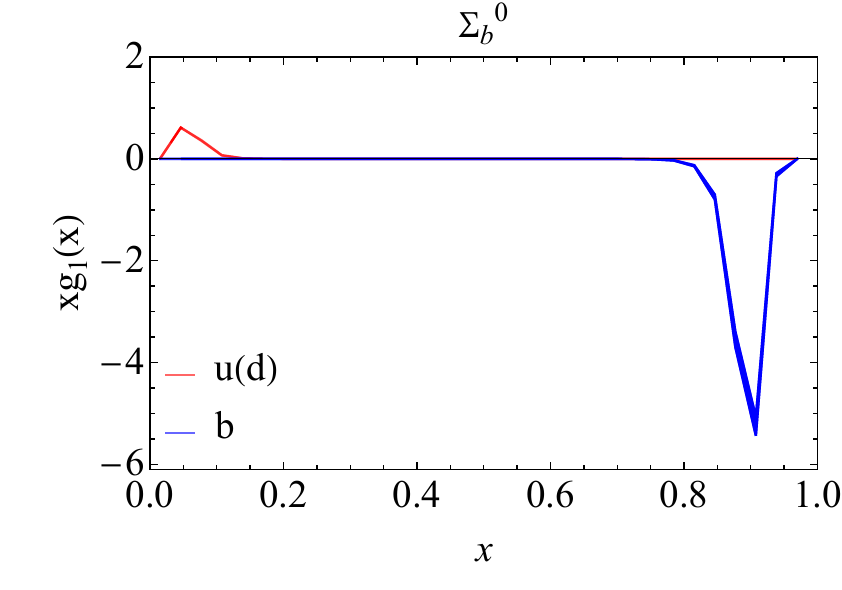}
    \includegraphics[width=0.49\textwidth]{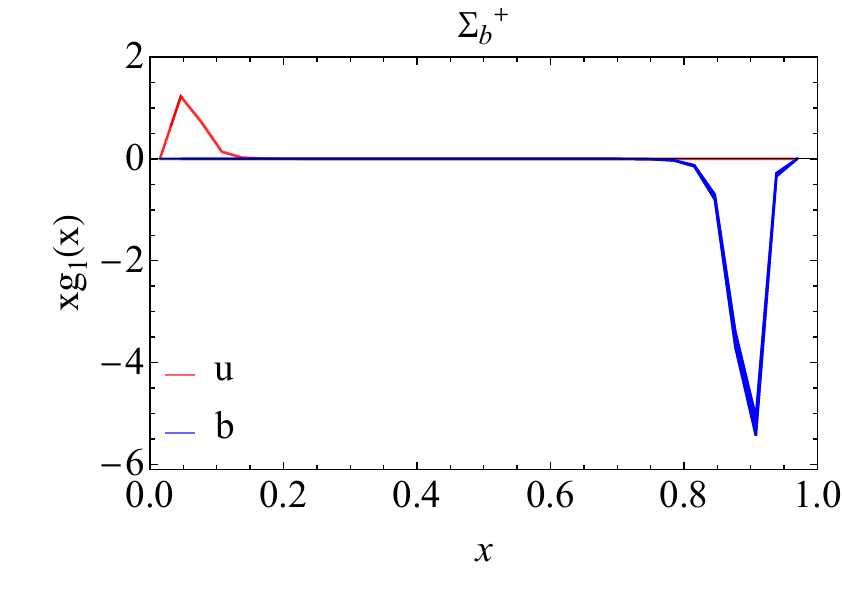}
    \includegraphics[width=0.49\textwidth]{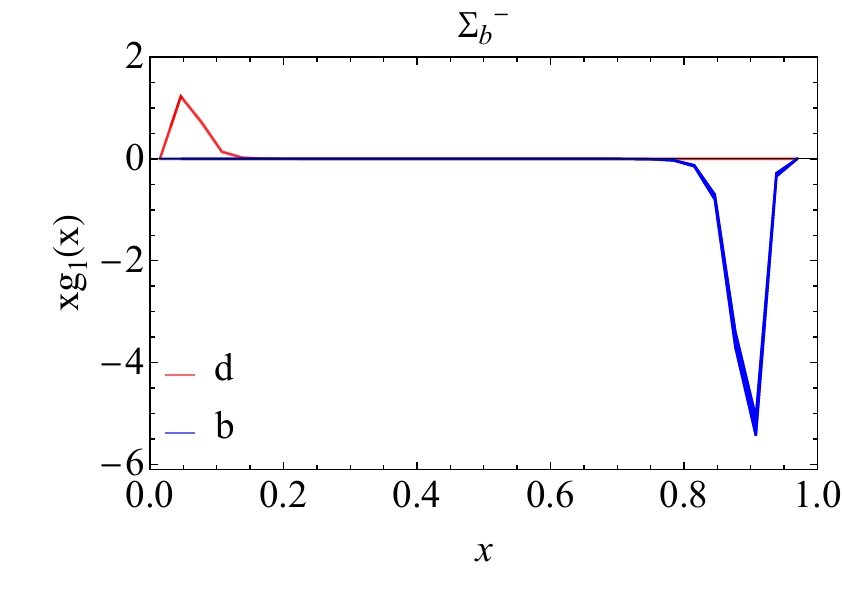}
	\caption{\justifying{Valence quarks' helicity PDFs of $\Lambda_b$ and its isospin states ($\Sigma_b^+$, $\Sigma_b^0$, and $\Sigma_b^-$) at the model scale multiplied by $x$ as functions of $x$. The blue and red bands represent distributions for the bottom quark and the light quark ($u$ and/or $d$), respectively. The inset displays the results on an expanded vertical scale to better display the behavior at small $x$. The bands, which are barely visible, reflect the $10\%$ uncertainty in the coupling constant $\alpha_s$.}}
	\label{gxini}
\end{figure*}

\begin{figure*}[t!]
    \includegraphics[width=0.49\textwidth]{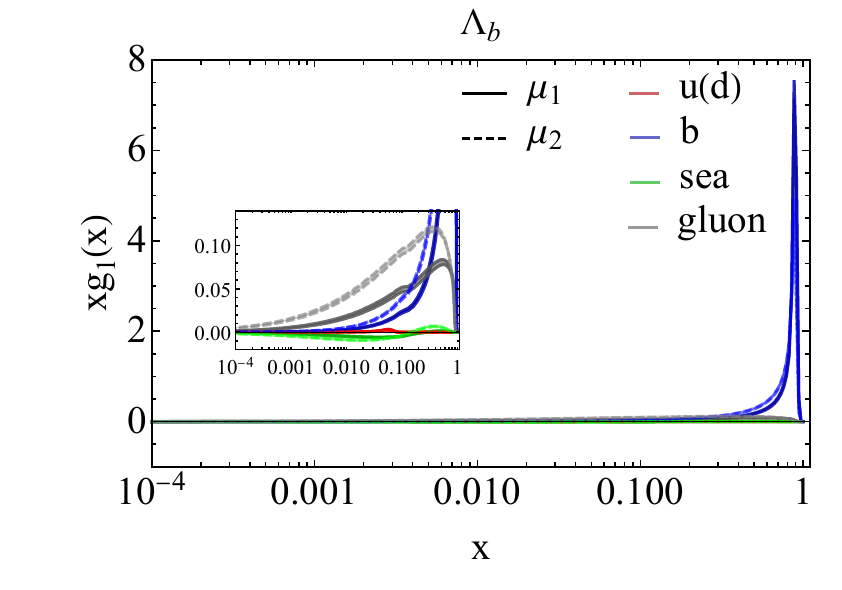}
    \includegraphics[width=0.49\textwidth]{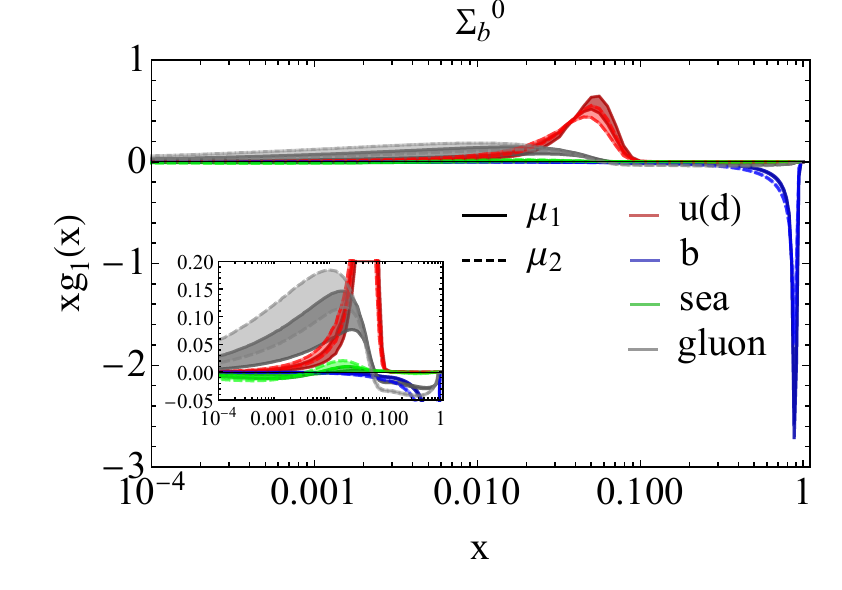}
    \includegraphics[width=0.49\textwidth]{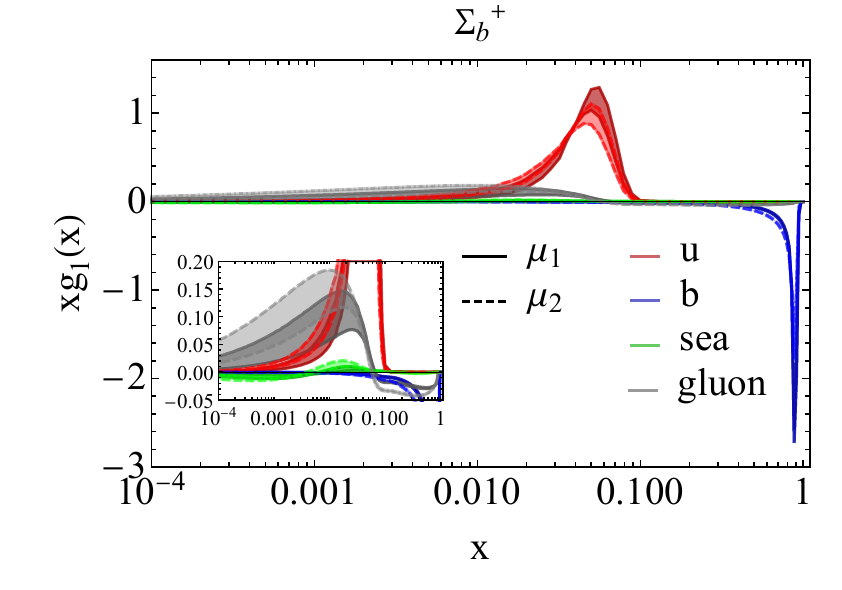}
    \includegraphics[width=0.49\textwidth]{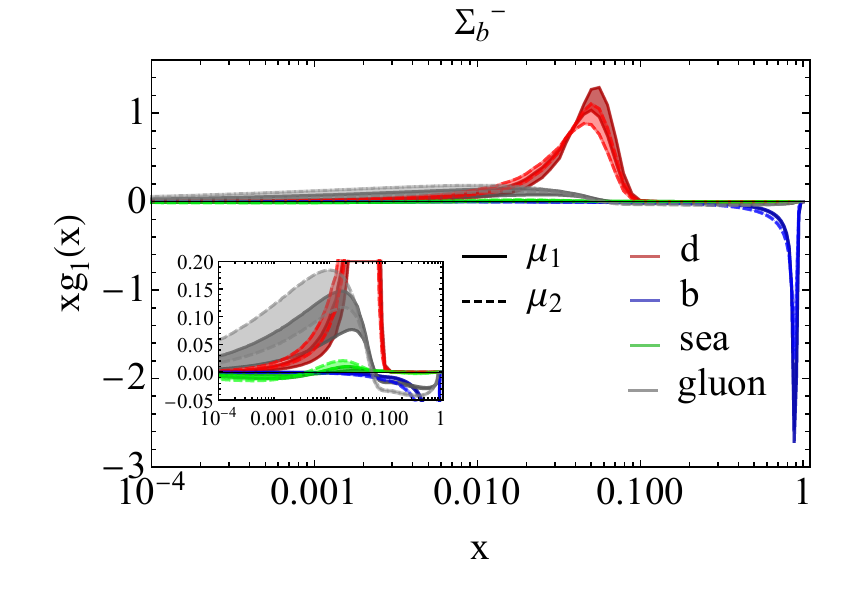}
	\caption{\justifying{Helicity PDFs at different final scales ($\mu_1=20\,\text{GeV}, \mu_2=80\,\text{GeV}$) multiplied by $x$ as functions of $x$. The red, blue, green and gray bands represent distributions for the valence light quark ($u$ and/or $d$), valence bottom quark, sea quarks, and gluon, respectively. The insets display the results on an expanded vertical scale to better display the behaviors at small $x$. The bands reflect the uncertainty in the initial scale $\mu_0$ from $1.90\,\text{GeV}$ to $5.05\,\text{GeV}$.}}
	\label{gxevo}
\end{figure*}

\begin{figure*}[t!]
    \includegraphics[width=0.49\textwidth]{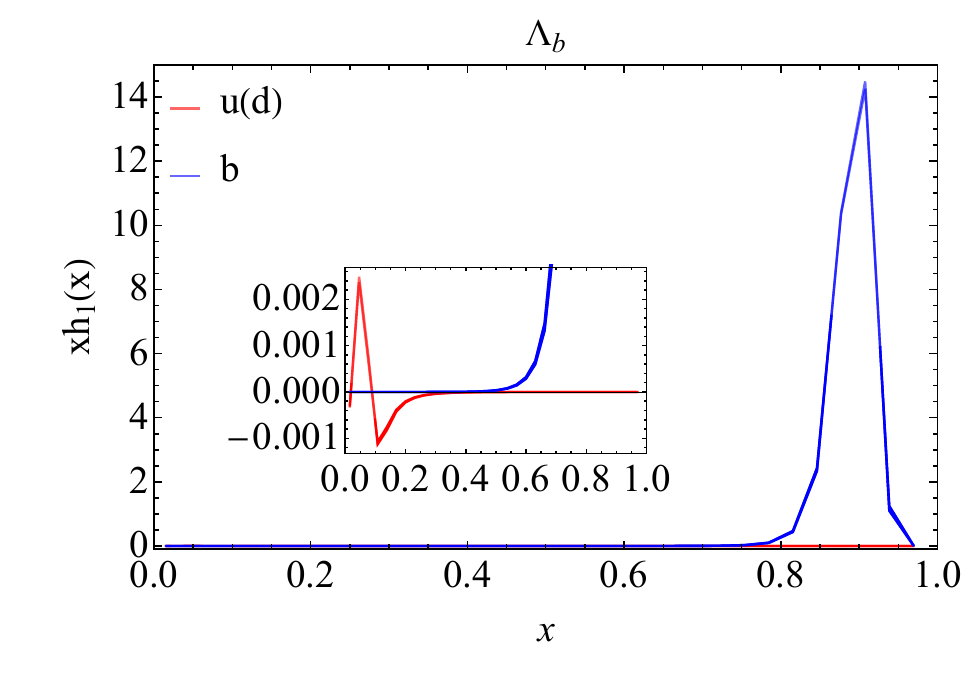}
    \includegraphics[width=0.49\textwidth]{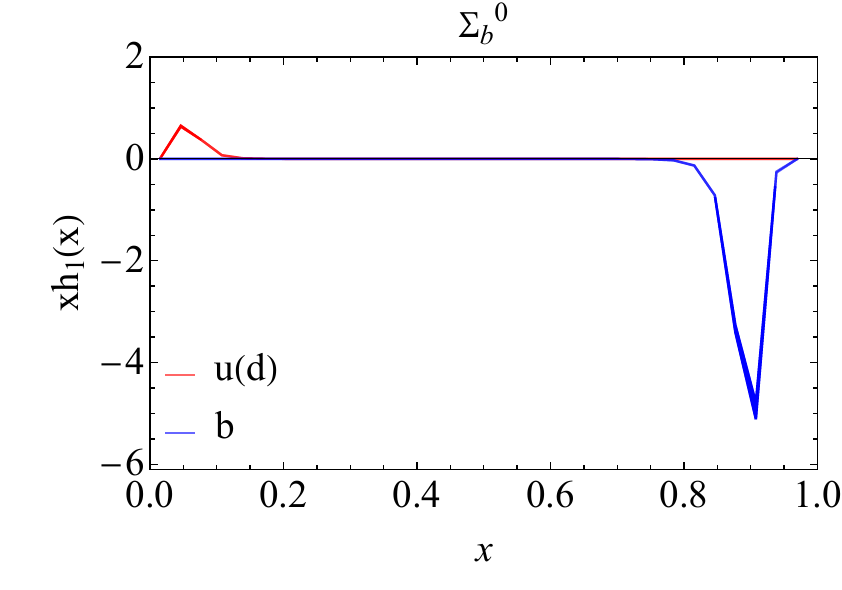}
    \includegraphics[width=0.49\textwidth]{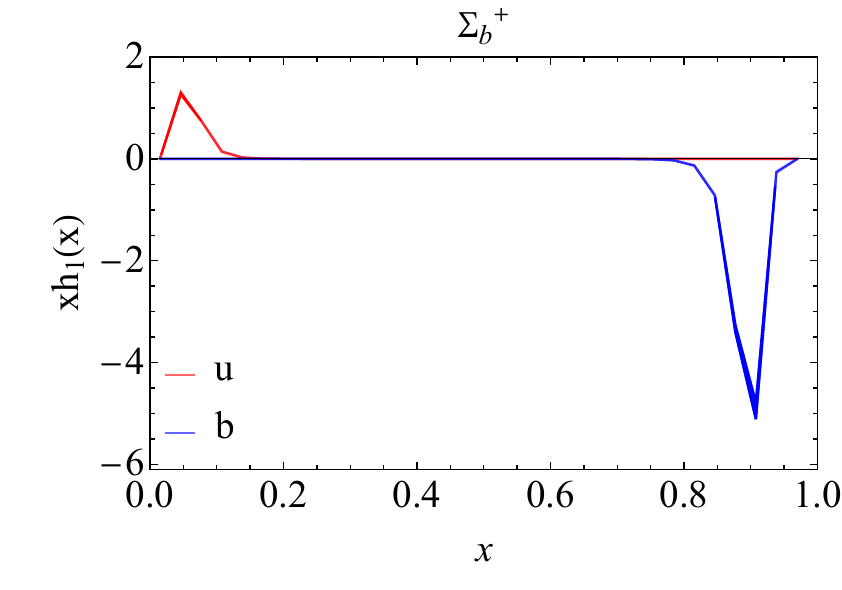}
    \includegraphics[width=0.49\textwidth]{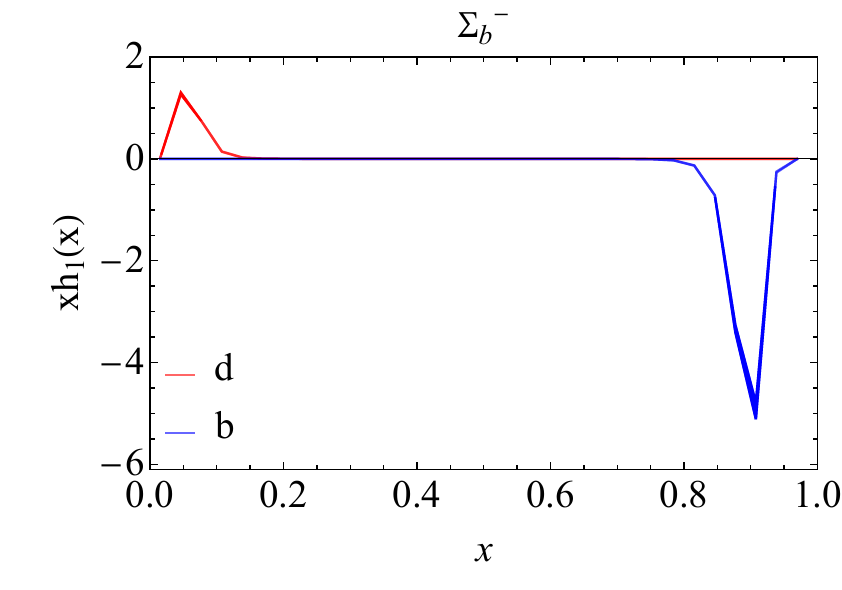}
	\caption{\justifying{Valence quarks' transversity PDFs of $\Lambda_b$ and its isospin states ($\Sigma_b^+$, $\Sigma_b^0$, and $\Sigma_b^-$) at the model scale multiplied by $x$ as functions of $x$. The blue and red bands represent distributions for the bottom quark and the light quark ($u$ and/or $d$), respectively. The inset displays the results on an expanded vertical scale to better display the behavior at small $x$. The bands, which are barely visible, reflect the $10\%$ uncertainty in the coupling constant $\alpha_s$.}}
	\label{hxini}
\end{figure*}

\begin{figure*}[t!]
    \includegraphics[width=0.49\textwidth]{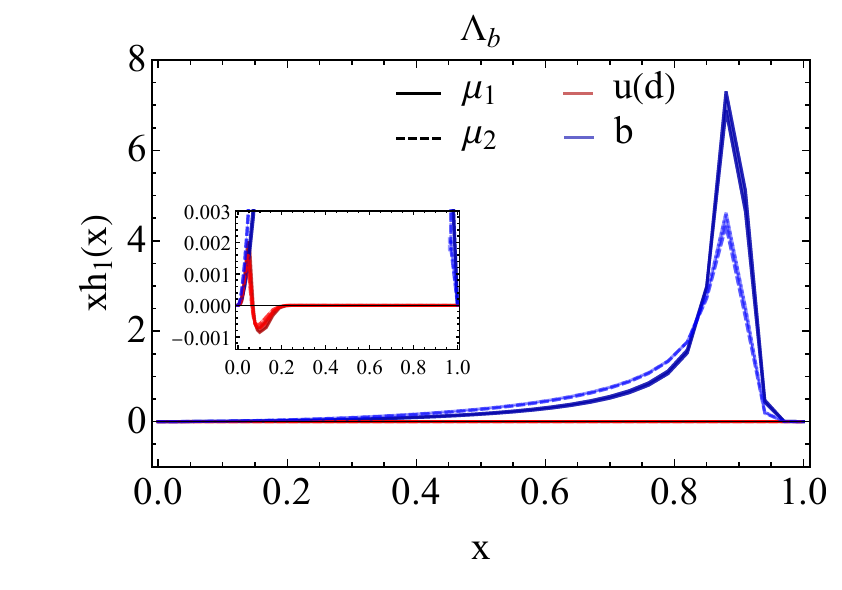}
    \includegraphics[width=0.49\textwidth]{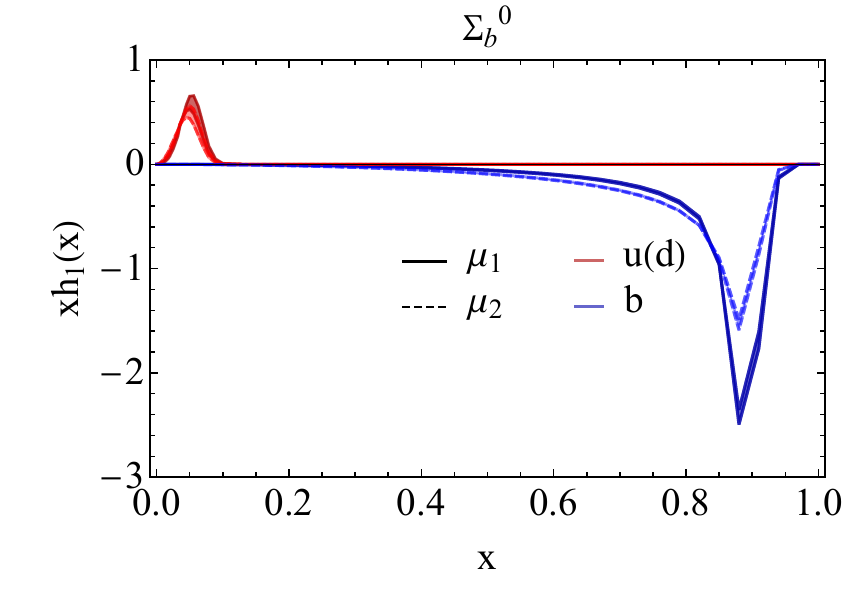}
    \includegraphics[width=0.49\textwidth]{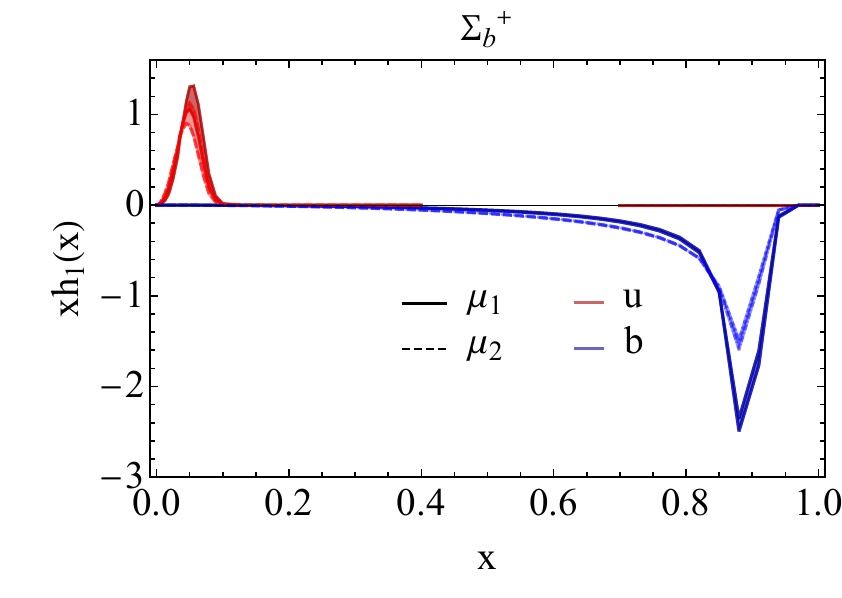}
    \includegraphics[width=0.49\textwidth]{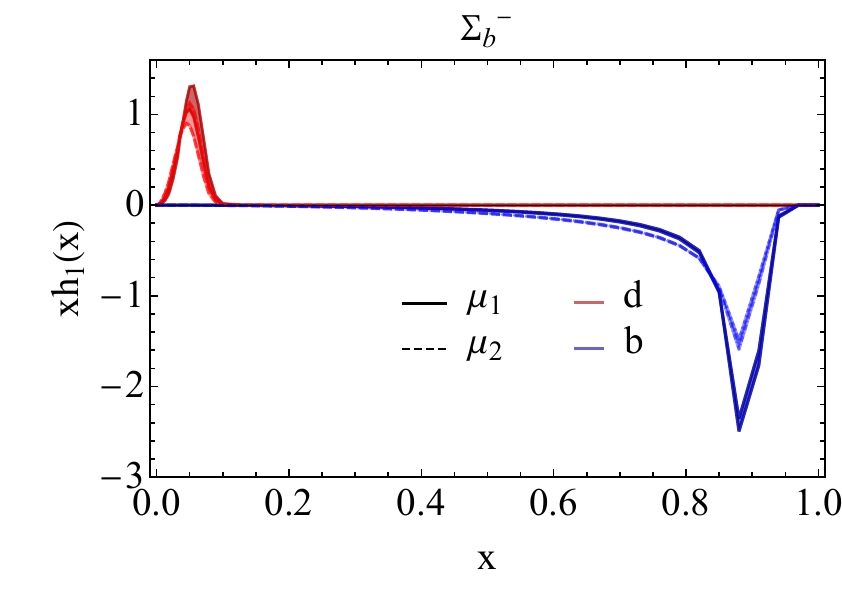}
	\caption{\justifying{Transversity PDFs at different final scales ($\mu_1=20\,\text{GeV}, \mu_2=80\,\text{GeV}$) multiplied by $x$ as functions of $x$. The red and blue bands represent distributions for the valence light quark ($u$ and/or $d$) and valence bottom quark, respectively. The inset displays the results on an expanded vertical scale to better display the behavior at small $x$. The bands reflect the uncertainty in the initial scale $\mu_0$ from $1.90\,\text{GeV}$ to $5.05\,\text{GeV}$.}}
	\label{hxevo}
\end{figure*}

\begin{figure*}[t!]
    \includegraphics[width=0.49\textwidth]{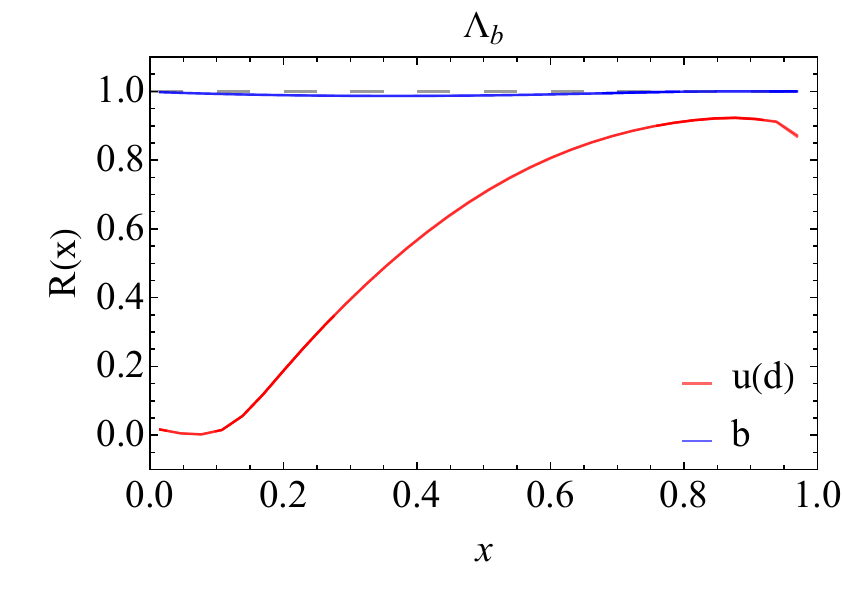}
    \includegraphics[width=0.49\textwidth]{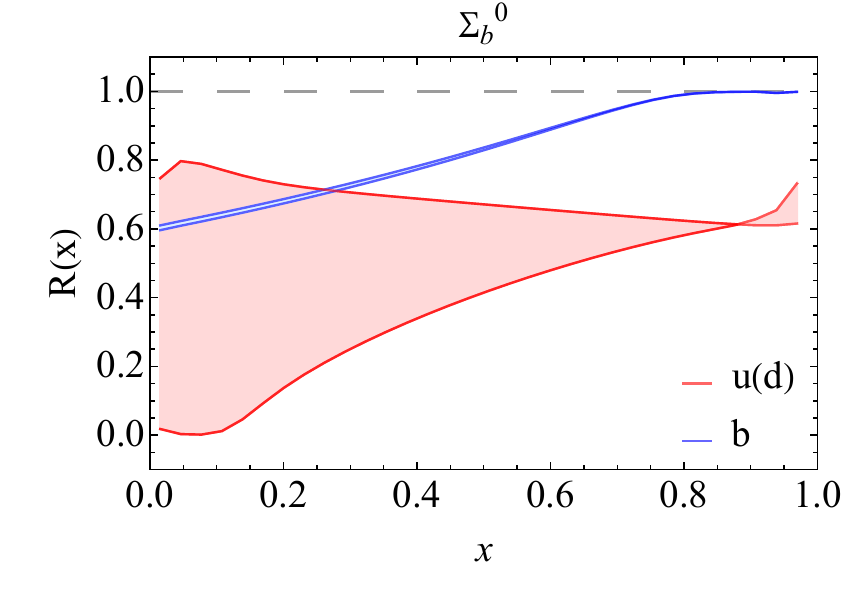}
    \includegraphics[width=0.49\textwidth]{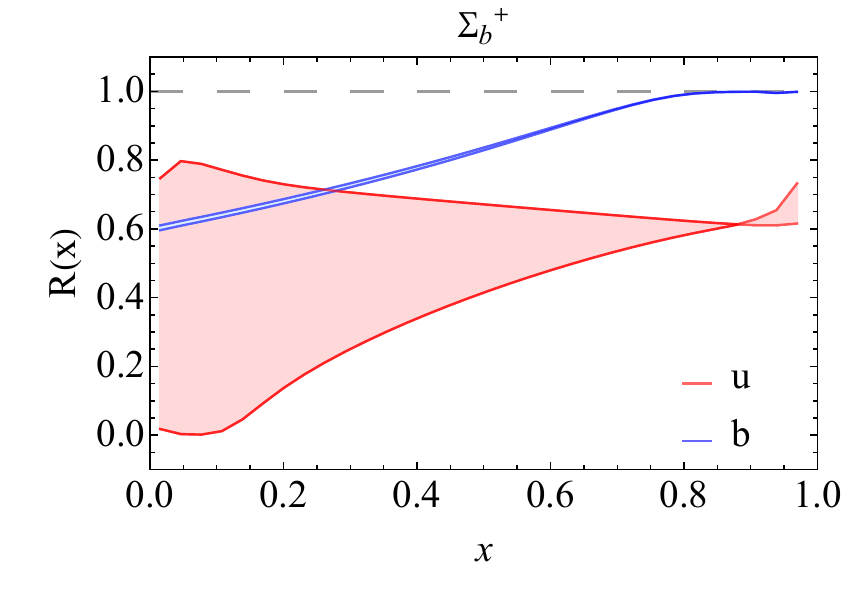}
    \includegraphics[width=0.49\textwidth]{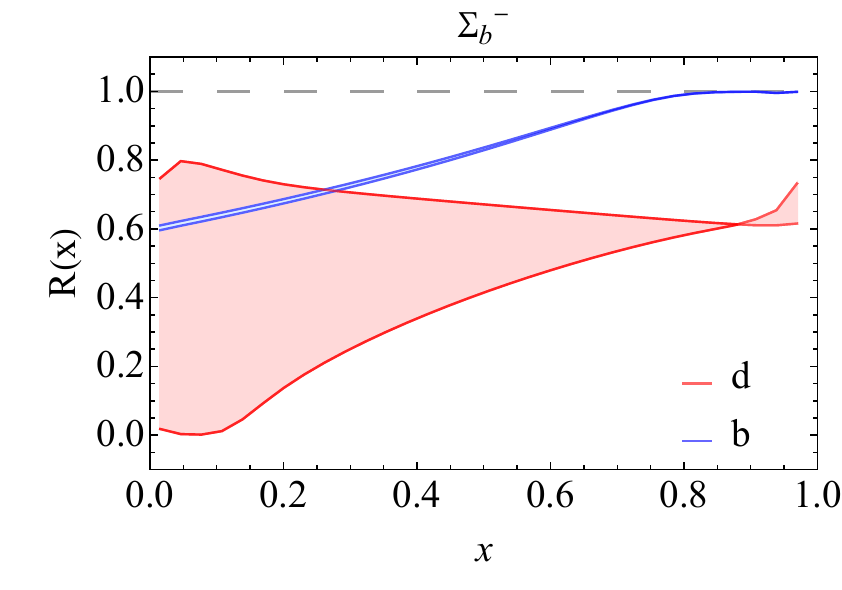}
	\caption{\justifying{The Soffer bound for $\Lambda_b$ and its isospin states ($\Sigma_b^+$, $\Sigma_b^0$, and $\Sigma_b^-$). The blue and red bands represent ratio $R^q(x)$ for the bottom quark and the light quark ($u$ and/or $d$), respectively. The bands reflect the $10\%$ uncertainty in the coupling constant $\alpha_s$. The horizontal dashed gray line indicates the position of unity.}}
	\label{pdfratio}
\end{figure*}

The PDF represents the probability that a parton carries a specific fraction of the total light-front longitudinal momentum of a hadron, offering insights into the nonperturbative structure of hadrons. The quark PDF of the baryon, which encodes the distribution of longitudinal momentum and polarization carried by the quark in the baryon, is defined as:
\begin{equation}
\begin{aligned}
\Phi^{\Gamma(q)}(x) = &\frac{1}{2}\int \frac{dz^-}{4\pi} e^{ip^+z^-/2}\\
&\times\mel{P,\Lambda}{\bar{\psi}_q (0) \Gamma \psi_q (z^-)}{P,\Lambda}|_{z^+=\vec{z}_\perp=0}, 
\label{pdf}
\end{aligned}
\end{equation}
where $x$ is the longitudinal momentum fraction carried by the quark. By selecting the Dirac structure $\Gamma$ in different forms, such as $\gamma^+$, $\gamma^+\gamma^5$, and $i\sigma^{j+}\gamma^5$, one obtains the unpolarized PDF $f(x)$, the helicity distribution $g_1(x)$, and the transversity distribution $h_1(x)$, respectively. We work in the light-front gauge ($A^+ = 0$), where the gauge link appearing between the quark fields in Eq.~\eqref{pdf} becomes unity. The quark PDFs are computed from the eigenstates of our light-front effective Hamiltonian, Eq.~\eqref{hamiltonian}, within the constituent valence quark representation, suitable for low-momentum scale applications.

By performing QCD evolution, the valence quark PDFs at higher scales can be obtained from the input PDFs at the initial scale. The QCD evolution is governed by the well-known Dokshitzer-Gribov-Lipatov-Altarelli-Parisi (DGLAP) equations~\cite{dglap1,dglap2,dglap3}. To numerically solve the DGLAP equations at different orders, we use the Higher-Order Perturbative Parton Evolution toolkit~\cite{hoppet}.
Since the mass of the constituent heavy quark is exceptionally large, the influence of the uncertainty in the coupling constant $\alpha_s$ is negligible. Therefore, we select the initial scale $\mu_0$ within a large range to account for the overall uncertainty. Following previous studies on heavy mesons~\cite{heavymeson}, we perform the scale evolution of the PDFs by varying the initial scale $\mu_0$ from $1.90 \, \mathrm{GeV}$ to $5.05 \, \mathrm{GeV}$, which correspond to the UV cutoff ($b\sqrt{N_{\rm max}}$) of the transverse momentum and the heavy quark mass $m_{b/k}$ in the baryon, respectively.
The difference in the results serves as an indicator of the sensitivity to the choice of the initial scale. The PDFs are evolved to final scales of 20 GeV and 80 GeV, which are the relevant scales for the proposed EICs~\cite{eicc,Accardi:2012qut,eRHIC}.
\subsection{Unpolarized PDFs and QCD evolution}
Following the two-point quark correlation function defined in Eq.~\eqref{pdf}, the unpolarized PDFs $f^q(x)$ within the valence Fock sector at the initial scale ($\mu_0$) are expressed in the LFWF overlap representation as:
\begin{equation}
\begin{aligned}
f^q (x) = &\sum_{\{\lambda_i\}} \int \left[\mathrm{d} \mathcal{X} \mathrm{d} \mathcal{P}_{\perp}\right]\\
&\times\Psi_{\{ x_i', \vec{k}_{i\perp}', \lambda_i\}}^{\uparrow *} \Psi_{\{ x_i, \vec{k}_{i\perp}, \lambda_i\}}^\uparrow \delta(x-x_q).
\end{aligned}
\end{equation}
Using the LFWFs within the BLFQ approach given in Eq.~\eqref{lfwf}, we evaluate the unpolarized PDFs for the valence quarks in the baryon, which follow the normalization:
\begin{equation}
\begin{aligned}
\int^1_0dxf^q(x)=F^q_1(0)=n_q,
\end{aligned}
\end{equation}
where $n_q$ represents the number of quarks of flavor $q$ in the
baryon. Additionally, at the model scale, our PDFs satisfy the following momentum sum rule:
\begin{equation}
\begin{aligned}
\sum_q\int^1_0 dxxf^q(x)=1.
\end{aligned}
\end{equation}

Figure~\ref{fxini} shows our results for the unpolarized valence quark PDFs of the $\Lambda_b$ and its isospin triplet states at the model scale, computed using the LFWFs given in Eq.\eqref{lfwf}. The red bands correspond to the light quarks (u and/or d), while the blue bands represent the bottom quark. The bands in our results reflect a $10\%$ uncertainty in the coupling constant $\alpha_s$ and are not easily visible in Fig.~\ref{fxini}. The heavy quark contributes significantly to the longitudinal momentum of the baryons as may be expected. 
It is worth noting that the bottom quark distribution is narrower in the larger $x$ region compared to the $s$ quark in $\Lambda$ and the $c$ quark in $\Lambda_c$\cite{lambdac} which is consistent with the expected trend towards the classical limit with increasing mass differences between the system's quarks. Due to isospin symmetry, the valence quark distributions in $\Sigma_b^+$ and $\Sigma_b^-$ are nearly identical. Meanwhile, the magnitude of the single light quark distribution is higher compared to that in $\Lambda_b$ and $\Sigma_b^0$, since there are two up (down) quarks in $\Sigma_b^+$ ($\Sigma_b^-$).

We present our results for the evolved unpolarized PDFs of the $\Lambda_b$ and its isospin triplet states in Fig.~\ref{fxevo}. Here, we use the next-to-next-to-leading order (NNLO) DGLAP equations\cite{hoppet} of QCD to evolve our valence quark PDFs from the model scale $\mu_0$ to higher scales. 
At the final scales, as $x$ increases, the distributions of the valence quarks grow slowly at low $x$ and fall rapidly at higher $x$.
Due to the very different masses of the bottom and light quarks, the peaks of their distributions follow the expectations of classical limits and appear at correspondingly different $x$ values. The gluon and sea quark PDFs at low $x$ increase much faster than the valence quark PDFs. Effectively, in the low-$x$ region, the distributions are dominated by the gluon and sea quark PDFs, while at large $x$, the valence quarks dominate the distributions. We observe that the qualitative behaviors of the gluon and sea quark PDFs obtained dynamically through the scale evolution in $\Lambda_b$ and its isospin states are very similar. As the final scale increases from $\mu_1$ to $\mu_2$, all distributions grow at low $x$ and decrease at higher $x$.

\subsection{Helicity PDFs and QCD evolution}

The helicity PDFs, $g_1^q (x)$, provide insight into the spin structure of baryons, capturing contributions from their quark and gluon constituents. Polarized PDFs describe the difference in probability densities between quarks with helicities aligned parallel and antiparallel to the spin of the baryon. In terms of the overlap of the LFWFs, the helicity distribution can be expressed as
\begin{equation}
\begin{aligned}
g_1^q (x) = &\sum_{\{\lambda_i\}} \int \left[\mathrm{d} \mathcal{X} \mathrm{d} \mathcal{P}_{\perp}\right]\\
&\times \lambda _1 \Psi_{\{ x_i', \vec{k}_{i\perp}', \lambda_i\}}^{\uparrow *} \Psi_{\{ x_i, \vec{k}_{i\perp}, \lambda_i\}}^\downarrow \delta(x-x_q),
\end{aligned}
\end{equation}
where $\lambda_1 = 1 (-1)$ is the helicity of the struck quark.

In Fig.~\ref{gxini}, we present the valence quark helicity PDFs, $g_1(x)$, for the $\Lambda_b$ and its isospin triplet states at the model scale. The red and blue bands correspond to the results for the light quarks ($u$ and/or $d$) and the bottom quark, respectively. As expected, the bottom quark distributions peak at higher $x$, while the light quark distributions peak at lower $x$. This behavior is consistent across all these baryons.
Notably, the $b$-quark distribution remains positive and dominates the helicity PDFs in the $\Lambda_b$. In contrast, for its isospin triplet states ($\Sigma_b^+$, $\Sigma_b^0$, and $\Sigma_b^-$), the $b$-quark distributions are negative, while the light quarks contribute significantly. This distinction can be attributed to the fact that $\Lambda_b$ represents the ground state, whereas the triplet baryons are excited states in our model. Consequently, in $\Lambda_b$, the total spin of the bound system is predominantly determined by the sum of the quark spins, with the bottom quark providing the dominant contribution. For the triplet states, which exhibit larger orbital angular momentum (OAM), the OAM contributes to the total spin, thereby reducing the contribution from the sum of the quark spins. As a result, the quark spins can become negative to achieve a smaller overall summation.
Our results for the $\Lambda_b$ lie in the vicinity of the analysis in Ref.~\cite{EMFFlambdab}, which suggests that the spin of the light quarks should be zero.

We evolve the helicity PDFs and present our results in Fig.~\ref{gxevo}. Here we use the next-to-leading order (NLO) DGLAP equations of QCD~\cite{hoppet} to evolve our valence quark PDFs. In the $\Lambda_b$, the bottom quark remains dominant even after QCD evolution. The gluon distribution becomes significant at low $x$, while the light and sea quarks contribute minimally. In the triplet baryons, the gluon distributions grow at lower $x$ and diminish at higher $x$. The light and bottom quark distributions contribute primarily at large $x$, with the peaks of the bottom quark distributions occurring at higher $x$ than those of the light quarks. Being consistent with the evolved unpolarized PDFs, all the distributions increase at low $x$ and decrease at high $x$ as the final scale evolves from $\mu_1$ to $\mu_2$. This trend is a characteristic feature observed consistently across all these baryons.

\subsection{Transversity PDFs and QCD evolution}

The transversity PDFs $h_1(x)$ describe the correlation between the transverse polarization of the baryon and that of its constituent quarks. Recently, these distributions have gained increasing attention due to the importance of a precise determination of its integral, the tensor charge ($g_T$) of the nucleon. With the LFWFs, the transversity PDFs read
\begin{equation}
\begin{aligned}
h_1^q (x) = &\sum_{\{\lambda_i\}} \int \left[\mathrm{d} \mathcal{X} \mathrm{d} \mathcal{P}_{\perp}\right]\\
&\times (\Psi_{\{ x_i', \vec{k}_{i\perp}', \lambda_i\}}^{\uparrow *} \Psi_{\{ x_i, \vec{k}_{i\perp}, \lambda_i\}}^\uparrow +(\uparrow \leftrightarrow \downarrow)) \delta(x-x_q),
\end{aligned}
\end{equation}
where the quark helicities $\lambda_1' = -\lambda_1$ for the struck quark and $\lambda_{2,3}'=\lambda_{2,3}$ for the spectators.

We compute $h_1^q(x)$ using the LFWFs given in Eq.~\eqref{lfwf} and Eq.~\eqref{wfdown}. The results of the transversity PDFs are shown in Fig.~\ref{hxini}. 
The transversity PDFs similarly reflect the internal polarization structure of the baryons, and therefore exhibit trends analogous to those observed in the helicity PDFs. Specifically, the peaks of the bottom quark distributions appear at higher $x$, while those of the light quarks occur at lower $x$. This trend is consistently seen in all these baryons.
In the $\Lambda_b$, the $b$-quark distribution is positive and dominates the transversity PDFs. Conversely, in the isospin triplet states ($\Sigma_b^+, \Sigma_b^0$ and $\Sigma_b^-$), the $b$-quark distributions are negative, with the light quarks making significant contributions. This difference can be explained using similar reasoning as that applied above to the helicity PDFs.

Figure~\ref{hxevo} presents the evolved transversity PDFs. We utilize the leading-order (LO) DGLAP equations of QCD~\cite{hoppet} to evolve the valence quark PDFs. Since higher order evolution of the transversity distributions is not well established, the evolution is restricted to the first order, providing only the valence quark distributions. Being consistent with their behavior at the model scale, the bottom quark distributions peak at higher $x$, while the light quark distributions peak at lower $x$, a trend observed across all these baryons. As the final scale evolves from $\mu_1$ to $\mu_2$, all distributions increase in the low-$x$ region and decrease in the high-$x$ region, maintaining this pattern across all baryons.

The results for the PDFs are consistent with the Soffer bound~\cite{Sofferbound}, which is defined at an arbitrary scale $\mu$ as:
\begin{equation}
\begin{aligned}
\abs{h_1^q (x,\mu)} \le \frac{1}{2}\abs{f_1^q (x,\mu)+g_1^q (x,\mu)}.
\label{Sofferboundeq}
\end{aligned}
\end{equation}
Since the PDFs are evolved at various orders, we verify the Soffer bound using the PDFs at the model scale, i.e., $\mu = \mu_0$. In Fig.~\ref{pdfratio}, we present the ratio $R^q(x)$, defined as:
\begin{equation}
\begin{aligned}
R^q(x)=\abs{\frac{h_1^q (x,\mu)}{\frac{1}{2}f_1^q (x,\mu)+g_1^q (x,\mu)}}.
\end{aligned}
\end{equation}
As illustrated in Fig.\ref{pdfratio}, the ratio $R^q(x)$ for both the light and bottom valence quarks is found to be less than or equal to unity, confirming that the condition in Eq.~\eqref{Sofferboundeq} is satisfied.

\section{Summary}
\label{sec5}
In this work, we investigated the $\Lambda_b$ and its isospin triplet baryons ($\Sigma_b^+$, $\Sigma_b^0$, and $\Sigma_b^-$) using the BLFQ approach. By truncating the Fock state expansion of the baryons to their valence states and diagonalizing the corresponding light-front effective Hamiltonian, we obtained the masses and LFWFs of these bottom baryons. These LFWFs were then used to predict their electromagnetic properties and PDFs. The Hamiltonian incorporates confinement effects in both transverse and longitudinal directions, along with a one-gluon exchange interaction for the constituent valence quarks, making it well-suited for describing low-resolution properties. The mass spectra were determined as eigenvalues of the Hamiltonian, with the corresponding LFWFs as eigenvectors. For the electromagnetic properties, we studied the electromagnetic FFs of the baryons and their flavor decompositions, as well as their magnetic moments, electric and magnetic radii. The masses obtained in our results align reasonably well with experimental data~\cite{pdg}, and our predictions for the electromagnetic properties are consistent with other theoretical calculations~\cite{EMS,mgntmt37,mgntmt45,mgntmt47,mgntmt50}.

We also predicted the unpolarized, helicity, and transversity PDFs at a low-resolution scale using our LFWFs. We found that the bottom quark remains predominant in all the distributions inside the $\Lambda_b$, especially for the polarized PDFs, while in its isospin triplet baryons, the light quarks make considerable contributions. The PDFs at higher scales, relevant to the proposed EICs, were computed based on the DGLAP equations. The QCD evolution of the PDFs provides insights into the gluon and sea quark distributions. For the unpolarized and helicity PDFs, we observed that, although the valence quarks dominate the region where $x > 0.1$, the distributions at small-$x$ are mainly controlled by the gluon distribution. As the scale $\mu$ increases, all these PDFs increase at low $x$ and decrease at higher $x$. We also observed a sense of universality in the gluon PDFs across different baryon states. Overall, the QCD evolution of the valence quark PDFs offers valuable predictions about the gluon and sea quarks, which originate from higher Fock components. For further improvement, future developments should aim to incorporate higher Fock sectors directly into the Hamiltonian eigenvalue problem, enabling an explicit representation of gluon and sea quark degrees of freedom at suitable initial scales. Our predictions for the PDFs of baryons containing one bottom quark provide guidance for future experimental investigations, as well as providing a baseline for theoretical studies involving higher Fock sectors.

Since our LFWFs account for the spin, flavor, and three-dimensional spatial distribution of all three active quarks simultaneously, we also plan to explore other parton distribution functions, such as generalized parton distributions, transverse momentum dependent parton distribution functions, Wigner distributions, and double parton distribution functions, as well as the mechanical properties of the baryons. Moreover, by utilizing the established LFWFs for other baryons within BLFQ, such as the proton~\cite{xusiqi2021b} and $\Lambda_c$~\cite{lambdac}, we can investigate baryonic decay processes involving the $\Lambda_b$ baryon.

\section{Acknowledgments}			
We thank Jiatong Wu, Satvir Kaur and Ziyang Lin for fruitful discussions, advice and help.
T. P. is supported by the Gansu Province Postgraduate Innovation Star Program No. 2025CXZX-043. 
J. L. is supported by Special Research Assistant Funding Project, Chinese Academy of Sciences, by the National Natural Science Foundation of China under Grant No. 12305095, and the Natural Science Foundation of Gansu Province, China, Grant No. 23JRRA631.
C. M. is supported by new faculty start-up funding by the Institute of Modern Physics, Chinese Academy of Sciences, Grant No. E129952YR0. 
G. W. is supported by the National Natural Science Founda-tion of China (NSFC) under the Grant No.12075073.
X. Z. is supported by new faculty startup funding by the Institute of Modern Physics, Chinese Academy of Sciences, by Key Re- search Program of Frontier Sciences, Chinese Academy of Sciences, Grant No. ZDBS-LY-7020, by the Foundation for Key Talents of Gansu Province, by the Central Funds Guiding the Local Science and Technology Development of Gansu Province, Grant No. 22ZY1QA006, by international partnership program of the Chinese Academy of Sciences, Grant No. 016GJHZ2022103FN, by National Key R\&D Program of China, Grant No. 2023YFA1606903, by the Strategic Priority Research Program of the Chinese Academy of Sciences, Grant No. XDB34000000, Grant No. 25RCKA008, by the Senior Scientist Program funded by Gansu Province, and by the National Natural Science Foundation of China under Grant No.12375143. 
J. P. V. is supported by the Department of Energy under Grant No. DE-SC0023692. 
This research is supported by Gansu International Collaboration and Talents Recruitment Base of Particle Physics (2023–2027). A major portion of the computational resources were also provided by Sugon Advanced Computing Center.

\bibliography{lambdab.bib}

\begin{thebibliography}{66}%
\makeatletter
\providecommand \@ifxundefined [1]{%
 \@ifx{#1\undefined}
}%
\providecommand \@ifnum [1]{%
 \ifnum #1\expandafter \@firstoftwo
 \else \expandafter \@secondoftwo
 \fi
}%
\providecommand \@ifx [1]{%
 \ifx #1\expandafter \@firstoftwo
 \else \expandafter \@secondoftwo
 \fi
}%
\providecommand \natexlab [1]{#1}%
\providecommand \enquote  [1]{``#1''}%
\providecommand \bibnamefont  [1]{#1}%
\providecommand \bibfnamefont [1]{#1}%
\providecommand \citenamefont [1]{#1}%
\providecommand \href@noop [0]{\@secondoftwo}%
\providecommand \href [0]{\begingroup \@sanitize@url \@href}%
\providecommand \@href[1]{\@@startlink{#1}\@@href}%
\providecommand \@@href[1]{\endgroup#1\@@endlink}%
\providecommand \@sanitize@url [0]{\catcode `\\12\catcode `\$12\catcode `\&12\catcode `\#12\catcode `\^12\catcode `\_12\catcode `\%12\relax}%
\providecommand \@@startlink[1]{}%
\providecommand \@@endlink[0]{}%
\providecommand \url  [0]{\begingroup\@sanitize@url \@url }%
\providecommand \@url [1]{\endgroup\@href {#1}{\urlprefix }}%
\providecommand \urlprefix  [0]{URL }%
\providecommand \Eprint [0]{\href }%
\providecommand \doibase [0]{https://doi.org/}%
\providecommand \selectlanguage [0]{\@gobble}%
\providecommand \bibinfo  [0]{\@secondoftwo}%
\providecommand \bibfield  [0]{\@secondoftwo}%
\providecommand \translation [1]{[#1]}%
\providecommand \BibitemOpen [0]{}%
\providecommand \bibitemStop [0]{}%
\providecommand \bibitemNoStop [0]{.\EOS\space}%
\providecommand \EOS [0]{\spacefactor3000\relax}%
\providecommand \BibitemShut  [1]{\csname bibitem#1\endcsname}%
\let\auto@bib@innerbib\@empty
\bibitem [{\citenamefont {De~R\'ujula}\ \emph {et~al.}(1975)\citenamefont {De~R\'ujula}, \citenamefont {Georgi},\ and\ \citenamefont {Glashow}}]{qcd1}%
  \BibitemOpen
  \bibfield  {author} {\bibinfo {author} {\bibfnamefont {A.}~\bibnamefont {De~R\'ujula}}, \bibinfo {author} {\bibfnamefont {H.}~\bibnamefont {Georgi}}\ and\ \bibinfo {author} {\bibfnamefont {S.~L.}\ \bibnamefont {Glashow}},\ }\bibfield  {title} {\bibinfo {title} {Hadron masses in a gauge theory},\ }\href {https://doi.org/10.1103/PhysRevD.12.147} {\bibfield  {journal} {\bibinfo  {journal} {Phys. Rev. D}\ }\textbf {\bibinfo {volume} {12}},\ \bibinfo {pages} {147} (\bibinfo {year} {1975})}\BibitemShut {NoStop}%
\bibitem [{\citenamefont {Navas}\ \emph {et~al.}(2024)\citenamefont {Navas} \emph {et~al.}}]{pdg}%
  \BibitemOpen
  \bibfield  {author} {\bibinfo {author} {\bibfnamefont {S.}~\bibnamefont {Navas}} \emph {et~al.} (\bibinfo {collaboration} {Particle Data Group}),\ }\bibfield  {title} {\bibinfo {title} {{Review of particle physics}},\ }\href {https://doi.org/10.1103/PhysRevD.110.030001} {\bibfield  {journal} {\bibinfo  {journal} {Phys. Rev. D}\ }\textbf {\bibinfo {volume} {110}},\ \bibinfo {pages} {030001} (\bibinfo {year} {2024})}\BibitemShut {NoStop}%
\bibitem [{\citenamefont {Abe}\ \emph {et~al.}(2001)\citenamefont {Abe} \emph {et~al.}}]{Belle:2001zzw}%
  \BibitemOpen
  \bibfield  {author} {\bibinfo {author} {\bibfnamefont {K.}~\bibnamefont {Abe}} \emph {et~al.} (\bibinfo {collaboration} {Belle}),\ }\bibfield  {title} {\bibinfo {title} {{Observation of large CP violation in the neutral $B$ meson system}},\ }\href {https://doi.org/10.1103/PhysRevLett.87.091802} {\bibfield  {journal} {\bibinfo  {journal} {Phys. Rev. Lett.}\ }\textbf {\bibinfo {volume} {87}},\ \bibinfo {pages} {091802} (\bibinfo {year} {2001})},\ \Eprint {https://arxiv.org/abs/hep-ex/0107061} {arXiv:hep-ex/0107061} \BibitemShut {NoStop}%
\bibitem [{\citenamefont {Brambilla}\ \emph {et~al.}(2014)\citenamefont {Brambilla} \emph {et~al.}}]{qcd2}%
  \BibitemOpen
  \bibfield  {author} {\bibinfo {author} {\bibfnamefont {N.}~\bibnamefont {Brambilla}} \emph {et~al.},\ }\bibfield  {title} {\bibinfo {title} {{QCD and Strongly Coupled Gauge Theories: Challenges and Perspectives}},\ }\href {https://doi.org/10.1140/epjc/s10052-014-2981-5} {\bibfield  {journal} {\bibinfo  {journal} {Eur. Phys. J. C}\ }\textbf {\bibinfo {volume} {74}},\ \bibinfo {pages} {2981} (\bibinfo {year} {2014})},\ \Eprint {https://arxiv.org/abs/1404.3723} {arXiv:1404.3723 [hep-ph]} \BibitemShut {NoStop}%
\bibitem [{\citenamefont {Hsiao}\ and\ \citenamefont {Geng}(2015)}]{QCDF1}%
  \BibitemOpen
  \bibfield  {author} {\bibinfo {author} {\bibfnamefont {Y.~K.}\ \bibnamefont {Hsiao}}\ and\ \bibinfo {author} {\bibfnamefont {C.~Q.}\ \bibnamefont {Geng}},\ }\bibfield  {title} {\bibinfo {title} {Direct $cp$ violation in ${\mathrm{\ensuremath{\Lambda}}}_{b}$ decays},\ }\href {https://doi.org/10.1103/PhysRevD.91.116007} {\bibfield  {journal} {\bibinfo  {journal} {Phys. Rev. D}\ }\textbf {\bibinfo {volume} {91}},\ \bibinfo {pages} {116007} (\bibinfo {year} {2015})}\BibitemShut {NoStop}%
\bibitem [{\citenamefont {Hsiao}\ \emph {et~al.}(2017)\citenamefont {Hsiao}, \citenamefont {Yao},\ and\ \citenamefont {Geng}}]{QCDF2}%
  \BibitemOpen
  \bibfield  {author} {\bibinfo {author} {\bibfnamefont {Y.~K.}\ \bibnamefont {Hsiao}}, \bibinfo {author} {\bibfnamefont {Y.}~\bibnamefont {Yao}}\ and\ \bibinfo {author} {\bibfnamefont {C.~Q.}\ \bibnamefont {Geng}},\ }\bibfield  {title} {\bibinfo {title} {Charmless two-body antitriplet $b$-baryon decays},\ }\href {https://doi.org/10.1103/PhysRevD.95.093001} {\bibfield  {journal} {\bibinfo  {journal} {Phys. Rev. D}\ }\textbf {\bibinfo {volume} {95}},\ \bibinfo {pages} {093001} (\bibinfo {year} {2017})}\BibitemShut {NoStop}%
\bibitem [{\citenamefont {Geng}\ \emph {et~al.}(2021)\citenamefont {Geng}, \citenamefont {Liu},\ and\ \citenamefont {Tsai}}]{QCDF3}%
  \BibitemOpen
  \bibfield  {author} {\bibinfo {author} {\bibfnamefont {C.}~\bibnamefont {Geng}}, \bibinfo {author} {\bibfnamefont {C.-W.}\ \bibnamefont {Liu}}\ and\ \bibinfo {author} {\bibfnamefont {T.-H.}\ \bibnamefont {Tsai}},\ }\bibfield  {title} {\bibinfo {title} {Non-leptonic two-body decays of ${\mathrm{\ensuremath{\Lambda}}}_{b}^0$ in light-front quark model},\ }\href {https://doi.org/https://doi.org/10.1016/j.physletb.2021.136125} {\bibfield  {journal} {\bibinfo  {journal} {Physics Letters B}\ }\textbf {\bibinfo {volume} {815}},\ \bibinfo {pages} {136125} (\bibinfo {year} {2021})}\BibitemShut {NoStop}%
\bibitem [{\citenamefont {L\"u}\ \emph {et~al.}(2009)\citenamefont {L\"u}, \citenamefont {Wang}, \citenamefont {Zou}, \citenamefont {Ali},\ and\ \citenamefont {Kramer}}]{PQCD1}%
  \BibitemOpen
  \bibfield  {author} {\bibinfo {author} {\bibfnamefont {C.-D.}\ \bibnamefont {L\"u}}, \bibinfo {author} {\bibfnamefont {Y.-M.}\ \bibnamefont {Wang}}, \bibinfo {author} {\bibfnamefont {H.}~\bibnamefont {Zou}}, \bibinfo {author} {\bibfnamefont {A.}~\bibnamefont {Ali}},\ and\ \bibinfo {author} {\bibfnamefont {G.}~\bibnamefont {Kramer}},\ }\bibfield  {title} {\bibinfo {title} {Anatomy of the perturbative qcd approach to the baryonic decays ${\ensuremath{\Lambda}}_{b}\ensuremath{\rightarrow}p\ensuremath{\pi}$, $pk$},\ }\href {https://doi.org/10.1103/PhysRevD.80.034011} {\bibfield  {journal} {\bibinfo  {journal} {Phys. Rev. D}\ }\textbf {\bibinfo {volume} {80}},\ \bibinfo {pages} {034011} (\bibinfo {year} {2009})}\BibitemShut {NoStop}%
\bibitem [{\citenamefont {Han}\ \emph {et~al.}(2022)\citenamefont {Han}, \citenamefont {Li}, \citenamefont {Li}, \citenamefont {Shen}, \citenamefont {Xiao},\ and\ \citenamefont {Yu}}]{PQCD2}%
  \BibitemOpen
  \bibfield  {author} {\bibinfo {author} {\bibfnamefont {J.-J.}\ \bibnamefont {Han}}, \bibinfo {author} {\bibfnamefont {Y.}~\bibnamefont {Li}}, \bibinfo {author} {\bibfnamefont {H.-n.}\ \bibnamefont {Li}}, \bibinfo {author} {\bibfnamefont {Y.-L.}\ \bibnamefont {Shen}}, \bibinfo {author} {\bibfnamefont {Z.-J.}\ \bibnamefont {Xiao}},\ and\ \bibinfo {author} {\bibfnamefont {F.-S.}\ \bibnamefont {Yu}},\ }\bibfield  {title} {\bibinfo {title} {${\ensuremath{\Lambda}}_{b}\ensuremath{\rightarrow}p$ transition form factors in perturbative qcd},\ }\href {https://doi.org/10.1140/epjc/s10052-022-10642-0} {\bibfield  {journal} {\bibinfo  {journal} {The European Physical Journal C}\ }\textbf {\bibinfo {volume} {82}},\ \bibinfo {pages} {686} (\bibinfo {year} {2022})}\BibitemShut {NoStop}%
\bibitem [{\citenamefont {Han}\ \emph {et~al.}(2025)\citenamefont {Han}, \citenamefont {Yu}, \citenamefont {Li}, \citenamefont {nan Li}, \citenamefont {Wang}, \citenamefont {Xiao},\ and\ \citenamefont {Yu}}]{cpv:fsyu}%
  \BibitemOpen
  \bibfield  {author} {\bibinfo {author} {\bibfnamefont {J.-J.}\ \bibnamefont {Han}}, \bibinfo {author} {\bibfnamefont {J.-X.}\ \bibnamefont {Yu}}, \bibinfo {author} {\bibfnamefont {Y.}~\bibnamefont {Li}}, \bibinfo {author} {\bibfnamefont {H.}~\bibnamefont {nan Li}}, \bibinfo {author} {\bibfnamefont {J.-P.}\ \bibnamefont {Wang}}, \bibinfo {author} {\bibfnamefont {Z.-J.}\ \bibnamefont {Xiao}},\ and\ \bibinfo {author} {\bibfnamefont {F.-S.}\ \bibnamefont {Yu}},\ }\href {https://arxiv.org/abs/2409.02821} {\bibinfo {title} {Establishing cp violation in $b$-baryon decays}} (\bibinfo {year} {2025}),\ \Eprint {https://arxiv.org/abs/2409.02821} {arXiv:2409.02821 [hep-ph]} \BibitemShut {NoStop}%
\bibitem [{\citenamefont {Hagler}(2010)}]{Hagler:2009ni}%
  \BibitemOpen
  \bibfield  {author} {\bibinfo {author} {\bibfnamefont {P.}~\bibnamefont {Hagler}},\ }\bibfield  {title} {\bibinfo {title} {{Hadron structure from lattice quantum chromodynamics}},\ }\href {https://doi.org/10.1016/j.physrep.2009.12.008} {\bibfield  {journal} {\bibinfo  {journal} {Phys. Rept.}\ }\textbf {\bibinfo {volume} {490}},\ \bibinfo {pages} {49} (\bibinfo {year} {2010})},\ \Eprint {https://arxiv.org/abs/0912.5483} {arXiv:0912.5483 [hep-lat]} \BibitemShut {NoStop}%
\bibitem [{\citenamefont {Bashir}\ \emph {et~al.}(2012)\citenamefont {Bashir}, \citenamefont {Chang}, \citenamefont {Cloet}, \citenamefont {El-Bennich}, \citenamefont {Liu}, \citenamefont {Roberts},\ and\ \citenamefont {Tandy}}]{Bashir:2012fs}%
  \BibitemOpen
  \bibfield  {author} {\bibinfo {author} {\bibfnamefont {A.}~\bibnamefont {Bashir}}, \bibinfo {author} {\bibfnamefont {L.}~\bibnamefont {Chang}}, \bibinfo {author} {\bibfnamefont {I.~C.}\ \bibnamefont {Cloet}}, \bibinfo {author} {\bibfnamefont {B.}~\bibnamefont {El-Bennich}}, \bibinfo {author} {\bibfnamefont {Y.-X.}\ \bibnamefont {Liu}}, \bibinfo {author} {\bibfnamefont {C.~D.}\ \bibnamefont {Roberts}},\ and\ \bibinfo {author} {\bibfnamefont {P.~C.}\ \bibnamefont {Tandy}},\ }\bibfield  {title} {\bibinfo {title} {{Collective perspective on advances in Dyson-Schwinger Equation QCD}},\ }\href {https://doi.org/10.1088/0253-6102/58/1/16} {\bibfield  {journal} {\bibinfo  {journal} {Commun. Theor. Phys.}\ }\textbf {\bibinfo {volume} {58}},\ \bibinfo {pages} {79} (\bibinfo {year} {2012})},\ \Eprint {https://arxiv.org/abs/1201.3366} {arXiv:1201.3366 [nucl-th]} \BibitemShut {NoStop}%
\bibitem [{\citenamefont {Brodsky}\ \emph {et~al.}(2015)\citenamefont {Brodsky}, \citenamefont {de~Teramond}, \citenamefont {Dosch},\ and\ \citenamefont {Erlich}}]{Brodsky:2014yha}%
  \BibitemOpen
  \bibfield  {author} {\bibinfo {author} {\bibfnamefont {S.~J.}\ \bibnamefont {Brodsky}}, \bibinfo {author} {\bibfnamefont {G.~F.}\ \bibnamefont {de~Teramond}}, \bibinfo {author} {\bibfnamefont {H.~G.}\ \bibnamefont {Dosch}},\ and\ \bibinfo {author} {\bibfnamefont {J.}~\bibnamefont {Erlich}},\ }\bibfield  {title} {\bibinfo {title} {{Light-Front Holographic QCD and Emerging Confinement}},\ }\href {https://doi.org/10.1016/j.physrep.2015.05.001} {\bibfield  {journal} {\bibinfo  {journal} {Phys. Rept.}\ }\textbf {\bibinfo {volume} {584}},\ \bibinfo {pages} {1} (\bibinfo {year} {2015})},\ \Eprint {https://arxiv.org/abs/1407.8131} {arXiv:1407.8131 [hep-ph]} \BibitemShut {NoStop}%
\bibitem [{\citenamefont {Vary}\ \emph {et~al.}(2010)\citenamefont {Vary}, \citenamefont {Honkanen}, \citenamefont {Li}, \citenamefont {Maris}, \citenamefont {Brodsky}, \citenamefont {Harindranath}, \citenamefont {de~Teramond}, \citenamefont {Sternberg}, \citenamefont {Ng},\ and\ \citenamefont {Yang}}]{PhysRevC.81.035205}%
  \BibitemOpen
  \bibfield  {author} {\bibinfo {author} {\bibfnamefont {J.~P.}\ \bibnamefont {Vary}}, \bibinfo {author} {\bibfnamefont {H.}~\bibnamefont {Honkanen}}, \bibinfo {author} {\bibfnamefont {J.}~\bibnamefont {Li}}, \bibinfo {author} {\bibfnamefont {P.}~\bibnamefont {Maris}}, \bibinfo {author} {\bibfnamefont {S.~J.}\ \bibnamefont {Brodsky}}, \bibinfo {author} {\bibfnamefont {A.}~\bibnamefont {Harindranath}}, \bibinfo {author} {\bibfnamefont {G.~F.}\ \bibnamefont {de~Teramond}}, \bibinfo {author} {\bibfnamefont {P.}~\bibnamefont {Sternberg}}, \bibinfo {author} {\bibfnamefont {E.~G.}\ \bibnamefont {Ng}},\ and\ \bibinfo {author} {\bibfnamefont {C.}~\bibnamefont {Yang}},\ }\bibfield  {title} {\bibinfo {title} {Hamiltonian light-front field theory in a basis function approach},\ }\href {https://doi.org/10.1103/PhysRevC.81.035205} {\bibfield  {journal} {\bibinfo  {journal} {Phys. Rev. C}\ }\textbf {\bibinfo {volume} {81}},\ \bibinfo {pages} {035205} (\bibinfo {year} {2010})}\BibitemShut {NoStop}%
\bibitem [{\citenamefont {Honkanen}\ \emph {et~al.}(2011)\citenamefont {Honkanen}, \citenamefont {Maris}, \citenamefont {Vary},\ and\ \citenamefont {Brodsky}}]{honkanen2011}%
  \BibitemOpen
  \bibfield  {author} {\bibinfo {author} {\bibfnamefont {H.}~\bibnamefont {Honkanen}}, \bibinfo {author} {\bibfnamefont {P.}~\bibnamefont {Maris}}, \bibinfo {author} {\bibfnamefont {J.~P.}\ \bibnamefont {Vary}},\ and\ \bibinfo {author} {\bibfnamefont {S.~J.}\ \bibnamefont {Brodsky}},\ }\bibfield  {title} {\bibinfo {title} {Electron in a transverse harmonic cavity},\ }\href {https://doi.org/10.1103/PhysRevLett.106.061603} {\bibfield  {journal} {\bibinfo  {journal} {Phys. Rev. Lett.}\ }\textbf {\bibinfo {volume} {106}},\ \bibinfo {pages} {061603} (\bibinfo {year} {2011})}\BibitemShut {NoStop}%
\bibitem [{\citenamefont {Chakrabarti}\ \emph {et~al.}(2014)\citenamefont {Chakrabarti}, \citenamefont {Zhao}, \citenamefont {Honkanen}, \citenamefont {Manohar}, \citenamefont {Maris},\ and\ \citenamefont {Vary}}]{PhysRevD.89.116004}%
  \BibitemOpen
  \bibfield  {author} {\bibinfo {author} {\bibfnamefont {D.}~\bibnamefont {Chakrabarti}}, \bibinfo {author} {\bibfnamefont {X.}~\bibnamefont {Zhao}}, \bibinfo {author} {\bibfnamefont {H.}~\bibnamefont {Honkanen}}, \bibinfo {author} {\bibfnamefont {R.}~\bibnamefont {Manohar}}, \bibinfo {author} {\bibfnamefont {P.}~\bibnamefont {Maris}},\ and\ \bibinfo {author} {\bibfnamefont {J.~P.}\ \bibnamefont {Vary}},\ }\bibfield  {title} {\bibinfo {title} {Generalized parton distributions in a light-front nonperturbative approach},\ }\href {https://doi.org/10.1103/PhysRevD.89.116004} {\bibfield  {journal} {\bibinfo  {journal} {Phys. Rev. D}\ }\textbf {\bibinfo {volume} {89}},\ \bibinfo {pages} {116004} (\bibinfo {year} {2014})}\BibitemShut {NoStop}%
\bibitem [{\citenamefont {Zhao}\ \emph {et~al.}(2014)\citenamefont {Zhao}, \citenamefont {Honkanen}, \citenamefont {Maris}, \citenamefont {Vary},\ and\ \citenamefont {Brodsky}}]{ZHAO201465}%
  \BibitemOpen
  \bibfield  {author} {\bibinfo {author} {\bibfnamefont {X.}~\bibnamefont {Zhao}}, \bibinfo {author} {\bibfnamefont {H.}~\bibnamefont {Honkanen}}, \bibinfo {author} {\bibfnamefont {P.}~\bibnamefont {Maris}}, \bibinfo {author} {\bibfnamefont {J.~P.}\ \bibnamefont {Vary}},\ and\ \bibinfo {author} {\bibfnamefont {S.~J.}\ \bibnamefont {Brodsky}},\ }\bibfield  {title} {\bibinfo {title} {Electron g-2 in light-front quantization},\ }\href {https://doi.org/https://doi.org/10.1016/j.physletb.2014.08.020} {\bibfield  {journal} {\bibinfo  {journal} {Physics Letters B}\ }\textbf {\bibinfo {volume} {737}},\ \bibinfo {pages} {65 } (\bibinfo {year} {2014})}\BibitemShut {NoStop}%
\bibitem [{\citenamefont {Wiecki}\ \emph {et~al.}(2015)\citenamefont {Wiecki}, \citenamefont {Li}, \citenamefont {Zhao}, \citenamefont {Maris},\ and\ \citenamefont {Vary}}]{PhysRevD.91.105009}%
  \BibitemOpen
  \bibfield  {author} {\bibinfo {author} {\bibfnamefont {P.}~\bibnamefont {Wiecki}}, \bibinfo {author} {\bibfnamefont {Y.}~\bibnamefont {Li}}, \bibinfo {author} {\bibfnamefont {X.}~\bibnamefont {Zhao}}, \bibinfo {author} {\bibfnamefont {P.}~\bibnamefont {Maris}},\ and\ \bibinfo {author} {\bibfnamefont {J.~P.}\ \bibnamefont {Vary}},\ }\bibfield  {title} {\bibinfo {title} {Basis light-front quantization approach to positronium},\ }\href {https://doi.org/10.1103/PhysRevD.91.105009} {\bibfield  {journal} {\bibinfo  {journal} {Phys. Rev. D}\ }\textbf {\bibinfo {volume} {91}},\ \bibinfo {pages} {105009} (\bibinfo {year} {2015})}\BibitemShut {NoStop}%
\bibitem [{\citenamefont {Lan}\ \emph {et~al.}(2020{\natexlab{a}})\citenamefont {Lan}, \citenamefont {Mondal}, \citenamefont {Li}, \citenamefont {Li}, \citenamefont {Tang}, \citenamefont {Zhao},\ and\ \citenamefont {Vary}}]{heavymeson}%
  \BibitemOpen
  \bibfield  {author} {\bibinfo {author} {\bibfnamefont {J.}~\bibnamefont {Lan}}, \bibinfo {author} {\bibfnamefont {C.}~\bibnamefont {Mondal}}, \bibinfo {author} {\bibfnamefont {M.}~\bibnamefont {Li}}, \bibinfo {author} {\bibfnamefont {Y.}~\bibnamefont {Li}}, \bibinfo {author} {\bibfnamefont {S.}~\bibnamefont {Tang}}, \bibinfo {author} {\bibfnamefont {X.}~\bibnamefont {Zhao}},\ and\ \bibinfo {author} {\bibfnamefont {J.~P.}\ \bibnamefont {Vary}} (\bibinfo {collaboration} {BLFQ Collaboration}),\ }\bibfield  {title} {\bibinfo {title} {Parton distribution functions of heavy mesons on the light front},\ }\href {https://doi.org/10.1103/PhysRevD.102.014020} {\bibfield  {journal} {\bibinfo  {journal} {Phys. Rev. D}\ }\textbf {\bibinfo {volume} {102}},\ \bibinfo {pages} {014020} (\bibinfo {year} {2020}{\natexlab{a}})}\BibitemShut {NoStop}%
\bibitem [{\citenamefont {Lan}\ \emph {et~al.}(2020{\natexlab{b}})\citenamefont {Lan}, \citenamefont {Mondal}, \citenamefont {Jia}, \citenamefont {Zhao},\ and\ \citenamefont {Vary}}]{Lan:2019rba}%
  \BibitemOpen
  \bibfield  {author} {\bibinfo {author} {\bibfnamefont {J.}~\bibnamefont {Lan}}, \bibinfo {author} {\bibfnamefont {C.}~\bibnamefont {Mondal}}, \bibinfo {author} {\bibfnamefont {S.}~\bibnamefont {Jia}}, \bibinfo {author} {\bibfnamefont {X.}~\bibnamefont {Zhao}},\ and\ \bibinfo {author} {\bibfnamefont {J.~P.}\ \bibnamefont {Vary}},\ }\bibfield  {title} {\bibinfo {title} {{Pion and kaon parton distribution functions from basis light front quantization and QCD evolution}},\ }\href {https://doi.org/10.1103/PhysRevD.101.034024} {\bibfield  {journal} {\bibinfo  {journal} {Phys. Rev. D}\ }\textbf {\bibinfo {volume} {101}},\ \bibinfo {pages} {034024} (\bibinfo {year} {2020}{\natexlab{b}})},\ \Eprint {https://arxiv.org/abs/1907.01509} {arXiv:1907.01509 [nucl-th]} \BibitemShut {NoStop}%
\bibitem [{\citenamefont {Nair}\ \emph {et~al.}(2022)\citenamefont {Nair}, \citenamefont {Mondal}, \citenamefont {Zhao}, \citenamefont {Mukherjee},\ and\ \citenamefont {Vary}}]{sreeraj2022}%
  \BibitemOpen
  \bibfield  {author} {\bibinfo {author} {\bibfnamefont {S.}~\bibnamefont {Nair}}, \bibinfo {author} {\bibfnamefont {C.}~\bibnamefont {Mondal}}, \bibinfo {author} {\bibfnamefont {X.}~\bibnamefont {Zhao}}, \bibinfo {author} {\bibfnamefont {A.}~\bibnamefont {Mukherjee}},\ and\ \bibinfo {author} {\bibfnamefont {J.~P.}\ \bibnamefont {Vary}},\ }\bibfield  {title} {\bibinfo {title} {Basis light-front quantization approach to photon},\ }\href {https://doi.org/https://doi.org/10.1016/j.physletb.2022.137005} {\bibfield  {journal} {\bibinfo  {journal} {Physics Letters B}\ }\textbf {\bibinfo {volume} {827}},\ \bibinfo {pages} {137005} (\bibinfo {year} {2022})}\BibitemShut {NoStop}%
\bibitem [{\citenamefont {Nair}\ \emph {et~al.}(2023)\citenamefont {Nair}, \citenamefont {Mondal}, \citenamefont {Zhao}, \citenamefont {Mukherjee},\ and\ \citenamefont {Vary}}]{sreeraj2023}%
  \BibitemOpen
  \bibfield  {author} {\bibinfo {author} {\bibfnamefont {S.}~\bibnamefont {Nair}}, \bibinfo {author} {\bibfnamefont {C.}~\bibnamefont {Mondal}}, \bibinfo {author} {\bibfnamefont {X.}~\bibnamefont {Zhao}}, \bibinfo {author} {\bibfnamefont {A.}~\bibnamefont {Mukherjee}},\ and\ \bibinfo {author} {\bibfnamefont {J.~P.}\ \bibnamefont {Vary}} (\bibinfo {collaboration} {BLFQ Collaboration}),\ }\bibfield  {title} {\bibinfo {title} {Massless and massive photons within basis light-front quantization},\ }\href {https://doi.org/10.1103/PhysRevD.108.116005} {\bibfield  {journal} {\bibinfo  {journal} {Phys. Rev. D}\ }\textbf {\bibinfo {volume} {108}},\ \bibinfo {pages} {116005} (\bibinfo {year} {2023})}\BibitemShut {NoStop}%
\bibitem [{\citenamefont {Li}\ \emph {et~al.}(2013)\citenamefont {Li}, \citenamefont {Wiecki}, \citenamefont {Zhao}, \citenamefont {Maris},\ and\ \citenamefont {Vary}}]{Li:2013cga}%
  \BibitemOpen
  \bibfield  {author} {\bibinfo {author} {\bibfnamefont {Y.}~\bibnamefont {Li}}, \bibinfo {author} {\bibfnamefont {P.~W.}\ \bibnamefont {Wiecki}}, \bibinfo {author} {\bibfnamefont {X.}~\bibnamefont {Zhao}}, \bibinfo {author} {\bibfnamefont {P.}~\bibnamefont {Maris}},\ and\ \bibinfo {author} {\bibfnamefont {J.~P.}\ \bibnamefont {Vary}},\ }\bibfield  {title} {\bibinfo {title} {{Introduction to Basis Light-Front Quantization Approach to QCD Bound State Problems}},\ }in\ \href {http://www.ntse-2013.khb.ru/Proc/YLi.pdf} {\emph {\bibinfo {booktitle} {{Proceedings, International Conference on Nuclear Theory in the Supercomputing Era (NTSE-2013): Ames, Iowa, USA, May 13-17, 2013}}}}\ (\bibinfo {year} {2013})\ p.\ \bibinfo {pages} {136},\ \Eprint {https://arxiv.org/abs/1311.2980} {arXiv:1311.2980 [nucl-th]} \BibitemShut {NoStop}%
\bibitem [{\citenamefont {Li}\ \emph {et~al.}(2017)\citenamefont {Li}, \citenamefont {Maris},\ and\ \citenamefont {Vary}}]{PhysRevD.96.016022}%
  \BibitemOpen
  \bibfield  {author} {\bibinfo {author} {\bibfnamefont {Y.}~\bibnamefont {Li}}, \bibinfo {author} {\bibfnamefont {P.}~\bibnamefont {Maris}}\ and\ \bibinfo {author} {\bibfnamefont {J.~P.}\ \bibnamefont {Vary}},\ }\bibfield  {title} {\bibinfo {title} {Quarkonium as a relativistic bound state on the light front},\ }\href {https://doi.org/10.1103/PhysRevD.96.016022} {\bibfield  {journal} {\bibinfo  {journal} {Phys. Rev. D}\ }\textbf {\bibinfo {volume} {96}},\ \bibinfo {pages} {016022} (\bibinfo {year} {2017})}\BibitemShut {NoStop}%
\bibitem [{\citenamefont {Li}\ \emph {et~al.}(2016)\citenamefont {Li}, \citenamefont {Maris}, \citenamefont {Zhao},\ and\ \citenamefont {Vary}}]{LI2016118}%
  \BibitemOpen
  \bibfield  {author} {\bibinfo {author} {\bibfnamefont {Y.}~\bibnamefont {Li}}, \bibinfo {author} {\bibfnamefont {P.}~\bibnamefont {Maris}}, \bibinfo {author} {\bibfnamefont {X.}~\bibnamefont {Zhao}},\ and\ \bibinfo {author} {\bibfnamefont {J.~P.}\ \bibnamefont {Vary}},\ }\bibfield  {title} {\bibinfo {title} {Heavy quarkonium in a holographic basis},\ }\href {https://doi.org/https://doi.org/10.1016/j.physletb.2016.04.065} {\bibfield  {journal} {\bibinfo  {journal} {Physics Letters B}\ }\textbf {\bibinfo {volume} {758}},\ \bibinfo {pages} {118 } (\bibinfo {year} {2016})}\BibitemShut {NoStop}%
\bibitem [{\citenamefont {Tang}\ \emph {et~al.}(2018)\citenamefont {Tang}, \citenamefont {Li}, \citenamefont {Maris},\ and\ \citenamefont {Vary}}]{PhysRevD.98.114038}%
  \BibitemOpen
  \bibfield  {author} {\bibinfo {author} {\bibfnamefont {S.}~\bibnamefont {Tang}}, \bibinfo {author} {\bibfnamefont {Y.}~\bibnamefont {Li}}, \bibinfo {author} {\bibfnamefont {P.}~\bibnamefont {Maris}},\ and\ \bibinfo {author} {\bibfnamefont {J.~P.}\ \bibnamefont {Vary}},\ }\bibfield  {title} {\bibinfo {title} {${B}_{c}$ mesons and their properties on the light front},\ }\href {https://doi.org/10.1103/PhysRevD.98.114038} {\bibfield  {journal} {\bibinfo  {journal} {Phys. Rev. D}\ }\textbf {\bibinfo {volume} {98}},\ \bibinfo {pages} {114038} (\bibinfo {year} {2018})}\BibitemShut {NoStop}%
\bibitem [{\citenamefont {Jia}\ and\ \citenamefont {Vary}(2019)}]{PhysRevC.99.035206}%
  \BibitemOpen
  \bibfield  {author} {\bibinfo {author} {\bibfnamefont {S.}~\bibnamefont {Jia}}\ and\ \bibinfo {author} {\bibfnamefont {J.~P.}\ \bibnamefont {Vary}},\ }\bibfield  {title} {\bibinfo {title} {Basis light front quantization for the charged light mesons with color singlet nambu--jona-lasinio interactions},\ }\href {https://doi.org/10.1103/PhysRevC.99.035206} {\bibfield  {journal} {\bibinfo  {journal} {Phys. Rev. C}\ }\textbf {\bibinfo {volume} {99}},\ \bibinfo {pages} {035206} (\bibinfo {year} {2019})}\BibitemShut {NoStop}%
\bibitem [{\citenamefont {Lan}\ \emph {et~al.}(2019)\citenamefont {Lan}, \citenamefont {Mondal}, \citenamefont {Jia}, \citenamefont {Zhao},\ and\ \citenamefont {Vary}}]{lanjiangshan2019m}%
  \BibitemOpen
  \bibfield  {author} {\bibinfo {author} {\bibfnamefont {J.}~\bibnamefont {Lan}}, \bibinfo {author} {\bibfnamefont {C.}~\bibnamefont {Mondal}}, \bibinfo {author} {\bibfnamefont {S.}~\bibnamefont {Jia}}, \bibinfo {author} {\bibfnamefont {X.}~\bibnamefont {Zhao}},\ and\ \bibinfo {author} {\bibfnamefont {J.~P.}\ \bibnamefont {Vary}} (\bibinfo {collaboration} {BLFQ Collaboration}),\ }\bibfield  {title} {\bibinfo {title} {Parton distribution functions from a light front hamiltonian and qcd evolution for light mesons},\ }\href {https://doi.org/10.1103/PhysRevLett.122.172001} {\bibfield  {journal} {\bibinfo  {journal} {Phys. Rev. Lett.}\ }\textbf {\bibinfo {volume} {122}},\ \bibinfo {pages} {172001} (\bibinfo {year} {2019})}\BibitemShut {NoStop}%
\bibitem [{\citenamefont {Mondal}\ \emph {et~al.}(2020)\citenamefont {Mondal}, \citenamefont {Xu}, \citenamefont {Lan}, \citenamefont {Zhao}, \citenamefont {Li}, \citenamefont {Chakrabarti},\ and\ \citenamefont {Vary}}]{Mondal:2019jdg}%
  \BibitemOpen
  \bibfield  {author} {\bibinfo {author} {\bibfnamefont {C.}~\bibnamefont {Mondal}}, \bibinfo {author} {\bibfnamefont {S.}~\bibnamefont {Xu}}, \bibinfo {author} {\bibfnamefont {J.}~\bibnamefont {Lan}}, \bibinfo {author} {\bibfnamefont {X.}~\bibnamefont {Zhao}}, \bibinfo {author} {\bibfnamefont {Y.}~\bibnamefont {Li}}, \bibinfo {author} {\bibfnamefont {D.}~\bibnamefont {Chakrabarti}},\ and\ \bibinfo {author} {\bibfnamefont {J.~P.}\ \bibnamefont {Vary}},\ }\bibfield  {title} {\bibinfo {title} {{Proton structure from a light-front Hamiltonian}},\ }\href {https://doi.org/10.1103/PhysRevD.102.016008} {\bibfield  {journal} {\bibinfo  {journal} {Phys. Rev. D}\ }\textbf {\bibinfo {volume} {102}},\ \bibinfo {pages} {016008} (\bibinfo {year} {2020})},\ \Eprint {https://arxiv.org/abs/1911.10913} {arXiv:1911.10913 [hep-ph]} \BibitemShut {NoStop}%
\bibitem [{\citenamefont {Xu}\ \emph {et~al.}(2021)\citenamefont {Xu}, \citenamefont {Mondal}, \citenamefont {Lan}, \citenamefont {Zhao}, \citenamefont {Li},\ and\ \citenamefont {Vary}}]{xusiqi2021b}%
  \BibitemOpen
  \bibfield  {author} {\bibinfo {author} {\bibfnamefont {S.}~\bibnamefont {Xu}}, \bibinfo {author} {\bibfnamefont {C.}~\bibnamefont {Mondal}}, \bibinfo {author} {\bibfnamefont {J.}~\bibnamefont {Lan}}, \bibinfo {author} {\bibfnamefont {X.}~\bibnamefont {Zhao}}, \bibinfo {author} {\bibfnamefont {Y.}~\bibnamefont {Li}},\ and\ \bibinfo {author} {\bibfnamefont {J.~P.}\ \bibnamefont {Vary}} (\bibinfo {collaboration} {BLFQ Collaboration}),\ }\bibfield  {title} {\bibinfo {title} {Nucleon structure from basis light-front quantization},\ }\href {https://doi.org/10.1103/PhysRevD.104.094036} {\bibfield  {journal} {\bibinfo  {journal} {Phys. Rev. D}\ }\textbf {\bibinfo {volume} {104}},\ \bibinfo {pages} {094036} (\bibinfo {year} {2021})}\BibitemShut {NoStop}%
\bibitem [{\citenamefont {Adhikari}\ \emph {et~al.}(2021)\citenamefont {Adhikari}, \citenamefont {Mondal}, \citenamefont {Nair}, \citenamefont {Xu}, \citenamefont {Jia}, \citenamefont {Zhao},\ and\ \citenamefont {Vary}}]{adhikari2021m}%
  \BibitemOpen
  \bibfield  {author} {\bibinfo {author} {\bibfnamefont {L.}~\bibnamefont {Adhikari}}, \bibinfo {author} {\bibfnamefont {C.}~\bibnamefont {Mondal}}, \bibinfo {author} {\bibfnamefont {S.}~\bibnamefont {Nair}}, \bibinfo {author} {\bibfnamefont {S.}~\bibnamefont {Xu}}, \bibinfo {author} {\bibfnamefont {S.}~\bibnamefont {Jia}}, \bibinfo {author} {\bibfnamefont {X.}~\bibnamefont {Zhao}},\ and\ \bibinfo {author} {\bibfnamefont {J.~P.}\ \bibnamefont {Vary}} (\bibinfo {collaboration} {BLFQ Collaboration}),\ }\bibfield  {title} {\bibinfo {title} {Generalized parton distributions and spin structures of light mesons from a light-front hamiltonian approach},\ }\href {https://doi.org/10.1103/PhysRevD.104.114019} {\bibfield  {journal} {\bibinfo  {journal} {Phys. Rev. D}\ }\textbf {\bibinfo {volume} {104}},\ \bibinfo {pages} {114019} (\bibinfo {year} {2021})}\BibitemShut {NoStop}%
\bibitem [{\citenamefont {Lan}\ \emph {et~al.}(2022)\citenamefont {Lan}, \citenamefont {Fu}, \citenamefont {Mondal}, \citenamefont {Zhao},\ and\ \citenamefont {Vary}}]{lanjiangshan2022m}%
  \BibitemOpen
  \bibfield  {author} {\bibinfo {author} {\bibfnamefont {J.}~\bibnamefont {Lan}}, \bibinfo {author} {\bibfnamefont {K.}~\bibnamefont {Fu}}, \bibinfo {author} {\bibfnamefont {C.}~\bibnamefont {Mondal}}, \bibinfo {author} {\bibfnamefont {X.}~\bibnamefont {Zhao}},\ and\ \bibinfo {author} {\bibfnamefont {J.~P.}\ \bibnamefont {Vary}},\ }\bibfield  {title} {\bibinfo {title} {Light mesons with one dynamical gluon on the light front},\ }\href {https://doi.org/https://doi.org/10.1016/j.physletb.2022.136890} {\bibfield  {journal} {\bibinfo  {journal} {Physics Letters B}\ }\textbf {\bibinfo {volume} {825}},\ \bibinfo {pages} {136890} (\bibinfo {year} {2022})}\BibitemShut {NoStop}%
\bibitem [{\citenamefont {Liu}\ \emph {et~al.}(2022)\citenamefont {Liu}, \citenamefont {Xu}, \citenamefont {Mondal}, \citenamefont {Zhao},\ and\ \citenamefont {Vary}}]{liuyiping2022b}%
  \BibitemOpen
  \bibfield  {author} {\bibinfo {author} {\bibfnamefont {Y.}~\bibnamefont {Liu}}, \bibinfo {author} {\bibfnamefont {S.}~\bibnamefont {Xu}}, \bibinfo {author} {\bibfnamefont {C.}~\bibnamefont {Mondal}}, \bibinfo {author} {\bibfnamefont {X.}~\bibnamefont {Zhao}},\ and\ \bibinfo {author} {\bibfnamefont {J.~P.}\ \bibnamefont {Vary}} (\bibinfo {collaboration} {BLFQ Collaboration}),\ }\bibfield  {title} {\bibinfo {title} {Angular momentum and generalized parton distributions for the proton with basis light-front quantization},\ }\href {https://doi.org/10.1103/PhysRevD.105.094018} {\bibfield  {journal} {\bibinfo  {journal} {Phys. Rev. D}\ }\textbf {\bibinfo {volume} {105}},\ \bibinfo {pages} {094018} (\bibinfo {year} {2022})}\BibitemShut {NoStop}%
\bibitem [{\citenamefont {Kuang}\ \emph {et~al.}(2022)\citenamefont {Kuang}, \citenamefont {Serafin}, \citenamefont {Zhao},\ and\ \citenamefont {Vary}}]{kuangzhongkui2022t}%
  \BibitemOpen
  \bibfield  {author} {\bibinfo {author} {\bibfnamefont {Z.}~\bibnamefont {Kuang}}, \bibinfo {author} {\bibfnamefont {K.}~\bibnamefont {Serafin}}, \bibinfo {author} {\bibfnamefont {X.}~\bibnamefont {Zhao}},\ and\ \bibinfo {author} {\bibfnamefont {J.~P.}\ \bibnamefont {Vary}} (\bibinfo {collaboration} {BLFQ Collaboration}),\ }\bibfield  {title} {\bibinfo {title} {All-charm tetraquark in front form dynamics},\ }\href {https://doi.org/10.1103/PhysRevD.105.094028} {\bibfield  {journal} {\bibinfo  {journal} {Phys. Rev. D}\ }\textbf {\bibinfo {volume} {105}},\ \bibinfo {pages} {094028} (\bibinfo {year} {2022})}\BibitemShut {NoStop}%
\bibitem [{\citenamefont {Peng}\ \emph {et~al.}(2022{\natexlab{a}})\citenamefont {Peng}, \citenamefont {Zhu}, \citenamefont {Xu}, \citenamefont {Liu}, \citenamefont {Mondal}, \citenamefont {Zhao},\ and\ \citenamefont {Vary}}]{pengtiancai2022b}%
  \BibitemOpen
  \bibfield  {author} {\bibinfo {author} {\bibfnamefont {T.}~\bibnamefont {Peng}}, \bibinfo {author} {\bibfnamefont {Z.}~\bibnamefont {Zhu}}, \bibinfo {author} {\bibfnamefont {S.}~\bibnamefont {Xu}}, \bibinfo {author} {\bibfnamefont {X.}~\bibnamefont {Liu}}, \bibinfo {author} {\bibfnamefont {C.}~\bibnamefont {Mondal}}, \bibinfo {author} {\bibfnamefont {X.}~\bibnamefont {Zhao}},\ and\ \bibinfo {author} {\bibfnamefont {J.~P.}\ \bibnamefont {Vary}} (\bibinfo {collaboration} {BLFQ Collaboration}),\ }\bibfield  {title} {\bibinfo {title} {Basis light-front quantization approach to $\mathrm{\ensuremath{\Lambda}}$ and ${\mathrm{\ensuremath{\Lambda}}}_{c}$ and their isospin triplet baryons},\ }\href {https://doi.org/10.1103/PhysRevD.106.114040} {\bibfield  {journal} {\bibinfo  {journal} {Phys. Rev. D}\ }\textbf {\bibinfo {volume} {106}},\ \bibinfo {pages} {114040} (\bibinfo {year} {2022}{\natexlab{a}})}\BibitemShut {NoStop}%
\bibitem [{\citenamefont {Hu}\ \emph {et~al.}(2022)\citenamefont {Hu}, \citenamefont {Xu}, \citenamefont {Mondal}, \citenamefont {Zhao},\ and\ \citenamefont {Vary}}]{huzhi2022b}%
  \BibitemOpen
  \bibfield  {author} {\bibinfo {author} {\bibfnamefont {Z.}~\bibnamefont {Hu}}, \bibinfo {author} {\bibfnamefont {S.}~\bibnamefont {Xu}}, \bibinfo {author} {\bibfnamefont {C.}~\bibnamefont {Mondal}}, \bibinfo {author} {\bibfnamefont {X.}~\bibnamefont {Zhao}},\ and\ \bibinfo {author} {\bibfnamefont {J.~P.}\ \bibnamefont {Vary}},\ }\bibfield  {title} {\bibinfo {title} {Transverse momentum structure of proton within the basis light-front quantization framework},\ }\href {https://doi.org/https://doi.org/10.1016/j.physletb.2022.137360} {\bibfield  {journal} {\bibinfo  {journal} {Physics Letters B}\ }\textbf {\bibinfo {volume} {833}},\ \bibinfo {pages} {137360} (\bibinfo {year} {2022})}\BibitemShut {NoStop}%
\bibitem [{\citenamefont {Zhu}\ \emph {et~al.}(2023{\natexlab{a}})\citenamefont {Zhu}, \citenamefont {Peng}, \citenamefont {Hu}, \citenamefont {Xu}, \citenamefont {Mondal}, \citenamefont {Zhao},\ and\ \citenamefont {Vary}}]{zhuzhimin2023m}%
  \BibitemOpen
  \bibfield  {author} {\bibinfo {author} {\bibfnamefont {Z.}~\bibnamefont {Zhu}}, \bibinfo {author} {\bibfnamefont {T.}~\bibnamefont {Peng}}, \bibinfo {author} {\bibfnamefont {Z.}~\bibnamefont {Hu}}, \bibinfo {author} {\bibfnamefont {S.}~\bibnamefont {Xu}}, \bibinfo {author} {\bibfnamefont {C.}~\bibnamefont {Mondal}}, \bibinfo {author} {\bibfnamefont {X.}~\bibnamefont {Zhao}},\ and\ \bibinfo {author} {\bibfnamefont {J.~P.}\ \bibnamefont {Vary}} (\bibinfo {collaboration} {BLFQ Collaboration}),\ }\bibfield  {title} {\bibinfo {title} {Transverse momentum structure of strange and charmed baryons: A light-front hamiltonian approach},\ }\href {https://doi.org/10.1103/PhysRevD.108.036009} {\bibfield  {journal} {\bibinfo  {journal} {Phys. Rev. D}\ }\textbf {\bibinfo {volume} {108}},\ \bibinfo {pages} {036009} (\bibinfo {year} {2023}{\natexlab{a}})}\BibitemShut {NoStop}%
\bibitem [{\citenamefont {Zhu}\ \emph {et~al.}(2023{\natexlab{b}})\citenamefont {Zhu}, \citenamefont {Hu}, \citenamefont {Lan}, \citenamefont {Mondal}, \citenamefont {Zhao},\ and\ \citenamefont {Vary}}]{zhuzhimin2023m1}%
  \BibitemOpen
  \bibfield  {author} {\bibinfo {author} {\bibfnamefont {Z.}~\bibnamefont {Zhu}}, \bibinfo {author} {\bibfnamefont {Z.}~\bibnamefont {Hu}}, \bibinfo {author} {\bibfnamefont {J.}~\bibnamefont {Lan}}, \bibinfo {author} {\bibfnamefont {C.}~\bibnamefont {Mondal}}, \bibinfo {author} {\bibfnamefont {X.}~\bibnamefont {Zhao}},\ and\ \bibinfo {author} {\bibfnamefont {J.~P.}\ \bibnamefont {Vary}},\ }\bibfield  {title} {\bibinfo {title} {Transverse structure of the pion beyond leading twist with basis light-front quantization},\ }\href {https://doi.org/https://doi.org/10.1016/j.physletb.2023.137808} {\bibfield  {journal} {\bibinfo  {journal} {Physics Letters B}\ }\textbf {\bibinfo {volume} {839}},\ \bibinfo {pages} {137808} (\bibinfo {year} {2023}{\natexlab{b}})}\BibitemShut {NoStop}%
\bibitem [{\citenamefont {Xu}\ \emph {et~al.}(2023)\citenamefont {Xu}, \citenamefont {Mondal}, \citenamefont {Zhao}, \citenamefont {Li},\ and\ \citenamefont {Vary}}]{xusiqi2023b}%
  \BibitemOpen
  \bibfield  {author} {\bibinfo {author} {\bibfnamefont {S.}~\bibnamefont {Xu}}, \bibinfo {author} {\bibfnamefont {C.}~\bibnamefont {Mondal}}, \bibinfo {author} {\bibfnamefont {X.}~\bibnamefont {Zhao}}, \bibinfo {author} {\bibfnamefont {Y.}~\bibnamefont {Li}},\ and\ \bibinfo {author} {\bibfnamefont {J.~P.}\ \bibnamefont {Vary}} (\bibinfo {collaboration} {BLFQ Collaboration}),\ }\bibfield  {title} {\bibinfo {title} {Quark and gluon spin and orbital angular momentum in the proton},\ }\href {https://doi.org/10.1103/PhysRevD.108.094002} {\bibfield  {journal} {\bibinfo  {journal} {Phys. Rev. D}\ }\textbf {\bibinfo {volume} {108}},\ \bibinfo {pages} {094002} (\bibinfo {year} {2023})}\BibitemShut {NoStop}%
\bibitem [{\citenamefont {Lin}\ \emph {et~al.}(2023)\citenamefont {Lin}, \citenamefont {Nair}, \citenamefont {Xu}, \citenamefont {Hu}, \citenamefont {Mondal}, \citenamefont {Zhao},\ and\ \citenamefont {Vary}}]{linbolang2023b}%
  \BibitemOpen
  \bibfield  {author} {\bibinfo {author} {\bibfnamefont {B.}~\bibnamefont {Lin}}, \bibinfo {author} {\bibfnamefont {S.}~\bibnamefont {Nair}}, \bibinfo {author} {\bibfnamefont {S.}~\bibnamefont {Xu}}, \bibinfo {author} {\bibfnamefont {Z.}~\bibnamefont {Hu}}, \bibinfo {author} {\bibfnamefont {C.}~\bibnamefont {Mondal}}, \bibinfo {author} {\bibfnamefont {X.}~\bibnamefont {Zhao}},\ and\ \bibinfo {author} {\bibfnamefont {J.~P.}\ \bibnamefont {Vary}},\ }\bibfield  {title} {\bibinfo {title} {Generalized parton distributions of gluon in proton: A light-front quantization approach},\ }\href {https://doi.org/https://doi.org/10.1016/j.physletb.2023.138305} {\bibfield  {journal} {\bibinfo  {journal} {Physics Letters B}\ }\textbf {\bibinfo {volume} {847}},\ \bibinfo {pages} {138305} (\bibinfo {year} {2023})}\BibitemShut {NoStop}%
\bibitem [{\citenamefont {Gross}\ \emph {et~al.}(2023)\citenamefont {Gross} \emph {et~al.}}]{gross2023}%
  \BibitemOpen
  \bibfield  {author} {\bibinfo {author} {\bibfnamefont {F.}~\bibnamefont {Gross}} \emph {et~al.},\ }\bibfield  {title} {\bibinfo {title} {{50 Years of Quantum Chromodynamics}},\ }\href {https://doi.org/10.1140/epjc/s10052-023-11949-2} {\bibfield  {journal} {\bibinfo  {journal} {Eur. Phys. J. C}\ }\textbf {\bibinfo {volume} {83}},\ \bibinfo {pages} {1125} (\bibinfo {year} {2023})},\ \Eprint {https://arxiv.org/abs/2212.11107} {arXiv:2212.11107 [hep-ph]} \BibitemShut {NoStop}%
\bibitem [{\citenamefont {Meng}\ \emph {et~al.}(2024)\citenamefont {Meng}, \citenamefont {Tang}, \citenamefont {Hu}, \citenamefont {Wang}, \citenamefont {Li}, \citenamefont {Zhao},\ and\ \citenamefont {Vary}}]{meng2024}%
  \BibitemOpen
  \bibfield  {author} {\bibinfo {author} {\bibfnamefont {L.}~\bibnamefont {Meng}}, \bibinfo {author} {\bibfnamefont {S.}~\bibnamefont {Tang}}, \bibinfo {author} {\bibfnamefont {Z.}~\bibnamefont {Hu}}, \bibinfo {author} {\bibfnamefont {G.-L.}\ \bibnamefont {Wang}}, \bibinfo {author} {\bibfnamefont {Y.}~\bibnamefont {Li}}, \bibinfo {author} {\bibfnamefont {X.}~\bibnamefont {Zhao}},\ and\ \bibinfo {author} {\bibfnamefont {J.~P.}\ \bibnamefont {Vary}},\ }\href {https://arxiv.org/abs/2405.16995} {\bibinfo {title} {Electron form factors in basis light-front quantization}} (\bibinfo {year} {2024}),\ \Eprint {https://arxiv.org/abs/2405.16995} {arXiv:2405.16995 [hep-ph]} \BibitemShut {NoStop}%
\bibitem [{\citenamefont {Peng}\ \emph {et~al.}(2022{\natexlab{b}})\citenamefont {Peng}, \citenamefont {Zhu}, \citenamefont {Xu}, \citenamefont {Liu}, \citenamefont {Mondal}, \citenamefont {Zhao},\ and\ \citenamefont {Vary}}]{lambdac}%
  \BibitemOpen
  \bibfield  {author} {\bibinfo {author} {\bibfnamefont {T.}~\bibnamefont {Peng}}, \bibinfo {author} {\bibfnamefont {Z.}~\bibnamefont {Zhu}}, \bibinfo {author} {\bibfnamefont {S.}~\bibnamefont {Xu}}, \bibinfo {author} {\bibfnamefont {X.}~\bibnamefont {Liu}}, \bibinfo {author} {\bibfnamefont {C.}~\bibnamefont {Mondal}}, \bibinfo {author} {\bibfnamefont {X.}~\bibnamefont {Zhao}},\ and\ \bibinfo {author} {\bibfnamefont {J.~P.}\ \bibnamefont {Vary}} (\bibinfo {collaboration} {BLFQ Collaboration}),\ }\bibfield  {title} {\bibinfo {title} {Basis light-front quantization approach to $\mathrm{\ensuremath{\Lambda}}$ and ${\mathrm{\ensuremath{\Lambda}}}_{c}$ and their isospin triplet baryons},\ }\href {https://doi.org/10.1103/PhysRevD.106.114040} {\bibfield  {journal} {\bibinfo  {journal} {Phys. Rev. D}\ }\textbf {\bibinfo {volume} {106}},\ \bibinfo {pages} {114040} (\bibinfo {year} {2022}{\natexlab{b}})}\BibitemShut {NoStop}%
\bibitem [{\citenamefont {Franklin}\ \emph {et~al.}(1981)\citenamefont {Franklin}, \citenamefont {Lichtenberg}, \citenamefont {Namgung},\ and\ \citenamefont {Carydas}}]{mgntmt36}%
  \BibitemOpen
  \bibfield  {author} {\bibinfo {author} {\bibfnamefont {J.}~\bibnamefont {Franklin}}, \bibinfo {author} {\bibfnamefont {D.~B.}\ \bibnamefont {Lichtenberg}}, \bibinfo {author} {\bibfnamefont {W.}~\bibnamefont {Namgung}},\ and\ \bibinfo {author} {\bibfnamefont {D.}~\bibnamefont {Carydas}},\ }\bibfield  {title} {\bibinfo {title} {Wave-function mixing of flavor-degenerate baryons},\ }\href {https://doi.org/10.1103/PhysRevD.24.2910} {\bibfield  {journal} {\bibinfo  {journal} {Phys. Rev. D}\ }\textbf {\bibinfo {volume} {24}},\ \bibinfo {pages} {2910} (\bibinfo {year} {1981})}\BibitemShut {NoStop}%
\bibitem [{\citenamefont {Hazra}\ \emph {et~al.}(2021)\citenamefont {Hazra}, \citenamefont {Rakshit},\ and\ \citenamefont {Dhir}}]{EMS}%
  \BibitemOpen
  \bibfield  {author} {\bibinfo {author} {\bibfnamefont {A.}~\bibnamefont {Hazra}}, \bibinfo {author} {\bibfnamefont {S.}~\bibnamefont {Rakshit}}\ and\ \bibinfo {author} {\bibfnamefont {R.}~\bibnamefont {Dhir}},\ }\bibfield  {title} {\bibinfo {title} {Radiative $m1$ transitions of heavy baryons: Effective quark mass scheme},\ }\href {https://doi.org/10.1103/PhysRevD.104.053002} {\bibfield  {journal} {\bibinfo  {journal} {Phys. Rev. D}\ }\textbf {\bibinfo {volume} {104}},\ \bibinfo {pages} {053002} (\bibinfo {year} {2021})}\BibitemShut {NoStop}%
\bibitem [{\citenamefont {Barik}\ and\ \citenamefont {Das}(1983)}]{mgntmt37}%
  \BibitemOpen
  \bibfield  {author} {\bibinfo {author} {\bibfnamefont {N.}~\bibnamefont {Barik}}\ and\ \bibinfo {author} {\bibfnamefont {M.}~\bibnamefont {Das}},\ }\bibfield  {title} {\bibinfo {title} {Magnetic moments of confined quarks and baryons in an independent-quark model based on dirac equation with power-law potential},\ }\href {https://doi.org/10.1103/PhysRevD.28.2823} {\bibfield  {journal} {\bibinfo  {journal} {Phys. Rev. D}\ }\textbf {\bibinfo {volume} {28}},\ \bibinfo {pages} {2823} (\bibinfo {year} {1983})}\BibitemShut {NoStop}%
\bibitem [{\citenamefont {Bernotas}\ and\ \citenamefont {Šimonis}(2012)}]{mgntmt45}%
  \BibitemOpen
  \bibfield  {author} {\bibinfo {author} {\bibfnamefont {A.}~\bibnamefont {Bernotas}}\ and\ \bibinfo {author} {\bibfnamefont {V.}~\bibnamefont {Šimonis}},\ }\href {https://arxiv.org/abs/1209.2900} {\bibinfo {title} {Magnetic moments of heavy baryons in the bag model reexamined}} (\bibinfo {year} {2012}),\ \Eprint {https://arxiv.org/abs/1209.2900} {arXiv:1209.2900 [hep-ph]} \BibitemShut {NoStop}%
\bibitem [{\citenamefont {Simonis}(2018)}]{mgntmt47}%
  \BibitemOpen
  \bibfield  {author} {\bibinfo {author} {\bibfnamefont {V.}~\bibnamefont {Simonis}},\ }\href {https://arxiv.org/abs/1803.01809} {\bibinfo {title} {Improved predictions for magnetic moments and m1 decay widths of heavy hadrons}} (\bibinfo {year} {2018}),\ \Eprint {https://arxiv.org/abs/1803.01809} {arXiv:1803.01809 [hep-ph]} \BibitemShut {NoStop}%
\bibitem [{\citenamefont {Meng}\ \emph {et~al.}(2018)\citenamefont {Meng}, \citenamefont {Wang}, \citenamefont {Leng}, \citenamefont {Liu},\ and\ \citenamefont {Zhu}}]{mgntmt50}%
  \BibitemOpen
  \bibfield  {author} {\bibinfo {author} {\bibfnamefont {L.}~\bibnamefont {Meng}}, \bibinfo {author} {\bibfnamefont {G.-J.}\ \bibnamefont {Wang}}, \bibinfo {author} {\bibfnamefont {C.-Z.}\ \bibnamefont {Leng}}, \bibinfo {author} {\bibfnamefont {Z.-W.}\ \bibnamefont {Liu}},\ and\ \bibinfo {author} {\bibfnamefont {S.-L.}\ \bibnamefont {Zhu}},\ }\bibfield  {title} {\bibinfo {title} {Magnetic moments of the spin-$\frac{3}{2}$ singly heavy baryons},\ }\href {https://doi.org/10.1103/PhysRevD.98.094013} {\bibfield  {journal} {\bibinfo  {journal} {Phys. Rev. D}\ }\textbf {\bibinfo {volume} {98}},\ \bibinfo {pages} {094013} (\bibinfo {year} {2018})}\BibitemShut {NoStop}%
\bibitem [{\citenamefont {Liu}\ \emph {et~al.}(2017)\citenamefont {Liu}, \citenamefont {Wang}, \citenamefont {Liu},\ and\ \citenamefont {Guo}}]{EMFFlambdab}%
  \BibitemOpen
  \bibfield  {author} {\bibinfo {author} {\bibfnamefont {L.-L.}\ \bibnamefont {Liu}}, \bibinfo {author} {\bibfnamefont {C.}~\bibnamefont {Wang}}, \bibinfo {author} {\bibfnamefont {Y.}~\bibnamefont {Liu}},\ and\ \bibinfo {author} {\bibfnamefont {X.-H.}\ \bibnamefont {Guo}},\ }\bibfield  {title} {\bibinfo {title} {Electromagnetic form factors of ${\mathrm{\ensuremath{\Lambda}}}_{b}$ in the bethe-salpeter equation approach},\ }\href {https://doi.org/10.1103/PhysRevD.95.054001} {\bibfield  {journal} {\bibinfo  {journal} {Phys. Rev. D}\ }\textbf {\bibinfo {volume} {95}},\ \bibinfo {pages} {054001} (\bibinfo {year} {2017})}\BibitemShut {NoStop}%
\bibitem [{\citenamefont {Aaltonen}\ \emph {et~al.}(2007)\citenamefont {Aaltonen} \emph {et~al.}}]{sigmabfirst}%
  \BibitemOpen
  \bibfield  {author} {\bibinfo {author} {\bibfnamefont {T.}~\bibnamefont {Aaltonen}} \emph {et~al.} (\bibinfo {collaboration} {CDF}),\ }\bibfield  {title} {\bibinfo {title} {{First observation of heavy baryons $\Sigma_{b}$ and $\Sigma_{b}^*$}},\ }\href {https://doi.org/10.1103/PhysRevLett.99.202001} {\bibfield  {journal} {\bibinfo  {journal} {Phys. Rev. Lett.}\ }\textbf {\bibinfo {volume} {99}},\ \bibinfo {pages} {202001} (\bibinfo {year} {2007})},\ \Eprint {https://arxiv.org/abs/0706.3868} {arXiv:0706.3868 [hep-ex]} \BibitemShut {NoStop}%
\bibitem [{\citenamefont {Anderle}\ \emph {et~al.}(2021)\citenamefont {Anderle} \emph {et~al.}}]{eicc}%
  \BibitemOpen
  \bibfield  {author} {\bibinfo {author} {\bibfnamefont {D.~P.}\ \bibnamefont {Anderle}} \emph {et~al.},\ }\bibfield  {title} {\bibinfo {title} {{Electron-ion collider in China}},\ }\href {https://doi.org/10.1007/s11467-021-1062-0} {\bibfield  {journal} {\bibinfo  {journal} {Front. Phys. (Beijing)}\ }\textbf {\bibinfo {volume} {16}},\ \bibinfo {pages} {64701} (\bibinfo {year} {2021})},\ \Eprint {https://arxiv.org/abs/2102.09222} {arXiv:2102.09222 [nucl-ex]} \BibitemShut {NoStop}%
\bibitem [{\citenamefont {Accardi}\ \emph {et~al.}(2016)\citenamefont {Accardi} \emph {et~al.}}]{Accardi:2012qut}%
  \BibitemOpen
  \bibfield  {author} {\bibinfo {author} {\bibfnamefont {A.}~\bibnamefont {Accardi}} \emph {et~al.},\ }\bibfield  {title} {\bibinfo {title} {{Electron Ion Collider: The Next QCD Frontier}: {Understanding the glue that binds us all}},\ }\href {https://doi.org/10.1140/epja/i2016-16268-9} {\bibfield  {journal} {\bibinfo  {journal} {Eur. Phys. J. A}\ }\textbf {\bibinfo {volume} {52}},\ \bibinfo {pages} {268} (\bibinfo {year} {2016})},\ \Eprint {https://arxiv.org/abs/1212.1701} {arXiv:1212.1701 [nucl-ex]} \BibitemShut {NoStop}%
\bibitem [{\citenamefont {Aschenauer}\ \emph {et~al.}(2014)\citenamefont {Aschenauer} \emph {et~al.}}]{eRHIC}%
  \BibitemOpen
  \bibfield  {author} {\bibinfo {author} {\bibfnamefont {E.~C.}\ \bibnamefont {Aschenauer}} \emph {et~al.},\ }\href {https://arxiv.org/abs/1409.1633} {\bibinfo {title} {erhic design study: An electron-ion collider at bnl}} (\bibinfo {year} {2014}),\ \Eprint {https://arxiv.org/abs/1409.1633} {arXiv:1409.1633 [physics.acc-ph]} \BibitemShut {NoStop}%
\bibitem [{\citenamefont {Brodsky}\ \emph {et~al.}(1998)\citenamefont {Brodsky}, \citenamefont {Pauli},\ and\ \citenamefont {Pinsky}}]{BRODSKY1998299}%
  \BibitemOpen
  \bibfield  {author} {\bibinfo {author} {\bibfnamefont {S.~J.}\ \bibnamefont {Brodsky}}, \bibinfo {author} {\bibfnamefont {H.-C.}\ \bibnamefont {Pauli}}\ and\ \bibinfo {author} {\bibfnamefont {S.~S.}\ \bibnamefont {Pinsky}},\ }\bibfield  {title} {\bibinfo {title} {Quantum chromodynamics and other field theories on the light cone},\ }\href {https://doi.org/https://doi.org/10.1016/S0370-1573(97)00089-6} {\bibfield  {journal} {\bibinfo  {journal} {Physics Reports}\ }\textbf {\bibinfo {volume} {301}},\ \bibinfo {pages} {299 } (\bibinfo {year} {1998})}\BibitemShut {NoStop}%
\bibitem [{\citenamefont {Hu}\ \emph {et~al.}(2021)\citenamefont {Hu}, \citenamefont {Xu}, \citenamefont {Mondal}, \citenamefont {Zhao},\ and\ \citenamefont {Vary}}]{huzhi2021}%
  \BibitemOpen
  \bibfield  {author} {\bibinfo {author} {\bibfnamefont {Z.}~\bibnamefont {Hu}}, \bibinfo {author} {\bibfnamefont {S.}~\bibnamefont {Xu}}, \bibinfo {author} {\bibfnamefont {C.}~\bibnamefont {Mondal}}, \bibinfo {author} {\bibfnamefont {X.}~\bibnamefont {Zhao}},\ and\ \bibinfo {author} {\bibfnamefont {J.~P.}\ \bibnamefont {Vary}} (\bibinfo {collaboration} {BLFQ Collaboration}),\ }\bibfield  {title} {\bibinfo {title} {Transverse structure of electron in momentum space in basis light-front quantization},\ }\href {https://doi.org/10.1103/PhysRevD.103.036005} {\bibfield  {journal} {\bibinfo  {journal} {Phys. Rev. D}\ }\textbf {\bibinfo {volume} {103}},\ \bibinfo {pages} {036005} (\bibinfo {year} {2021})}\BibitemShut {NoStop}%
\bibitem [{\citenamefont {Brodsky}\ \emph {et~al.}(2006)\citenamefont {Brodsky}, \citenamefont {Gardner},\ and\ \citenamefont {Hwang}}]{Brodsky2006}%
  \BibitemOpen
  \bibfield  {author} {\bibinfo {author} {\bibfnamefont {S.~J.}\ \bibnamefont {Brodsky}}, \bibinfo {author} {\bibfnamefont {S.}~\bibnamefont {Gardner}}\ and\ \bibinfo {author} {\bibfnamefont {D.~S.}\ \bibnamefont {Hwang}},\ }\bibfield  {title} {\bibinfo {title} {Discrete symmetries on the light front and a general relation connecting the nucleon electric dipole and anomalous magnetic moments},\ }\href {https://doi.org/10.1103/PhysRevD.73.036007} {\bibfield  {journal} {\bibinfo  {journal} {Phys. Rev. D}\ }\textbf {\bibinfo {volume} {73}},\ \bibinfo {pages} {036007} (\bibinfo {year} {2006})}\BibitemShut {NoStop}%
\bibitem [{\citenamefont {Brisudova}\ and\ \citenamefont {Glazek}(1994)}]{Brisudova:1994it}%
  \BibitemOpen
  \bibfield  {author} {\bibinfo {author} {\bibfnamefont {M.~M.}\ \bibnamefont {Brisudova}}\ and\ \bibinfo {author} {\bibfnamefont {S.~D.}\ \bibnamefont {Glazek}},\ }\bibfield  {title} {\bibinfo {title} {{Relativistic scattering and bound state properties in a special Hamiltonian model}},\ }\href {https://doi.org/10.1103/PhysRevD.50.971} {\bibfield  {journal} {\bibinfo  {journal} {Phys. Rev. D}\ }\textbf {\bibinfo {volume} {50}},\ \bibinfo {pages} {971} (\bibinfo {year} {1994})}\BibitemShut {NoStop}%
\bibitem [{\citenamefont {Burkardt}(1998)}]{Burkardt:1998dd}%
  \BibitemOpen
  \bibfield  {author} {\bibinfo {author} {\bibfnamefont {M.}~\bibnamefont {Burkardt}},\ }\bibfield  {title} {\bibinfo {title} {{Dynamical vertex mass generation and chiral symmetry breaking on the light front}},\ }\href {https://doi.org/10.1103/PhysRevD.58.096015} {\bibfield  {journal} {\bibinfo  {journal} {Phys. Rev. D}\ }\textbf {\bibinfo {volume} {58}},\ \bibinfo {pages} {096015} (\bibinfo {year} {1998})},\ \Eprint {https://arxiv.org/abs/hep-th/9805088} {arXiv:hep-th/9805088} \BibitemShut {NoStop}%
\bibitem [{\citenamefont {Burkardt}\ and\ \citenamefont {Langnau}(1991)}]{Burkardt:1991tj}%
  \BibitemOpen
  \bibfield  {author} {\bibinfo {author} {\bibfnamefont {M.}~\bibnamefont {Burkardt}}\ and\ \bibinfo {author} {\bibfnamefont {A.}~\bibnamefont {Langnau}},\ }\bibfield  {title} {\bibinfo {title} {{Rotational invariance in light cone quantization}},\ }\href {https://doi.org/10.1103/PhysRevD.44.3857} {\bibfield  {journal} {\bibinfo  {journal} {Phys. Rev. D}\ }\textbf {\bibinfo {volume} {44}},\ \bibinfo {pages} {3857} (\bibinfo {year} {1991})}\BibitemShut {NoStop}%
\bibitem [{\citenamefont {Brodsky}\ \emph {et~al.}(2001)\citenamefont {Brodsky}, \citenamefont {Diehl},\ and\ \citenamefont {Hwang}}]{BRODSKY200199}%
  \BibitemOpen
  \bibfield  {author} {\bibinfo {author} {\bibfnamefont {S.~J.}\ \bibnamefont {Brodsky}}, \bibinfo {author} {\bibfnamefont {M.}~\bibnamefont {Diehl}}\ and\ \bibinfo {author} {\bibfnamefont {D.~S.}\ \bibnamefont {Hwang}},\ }\bibfield  {title} {\bibinfo {title} {Light-cone wavefunction representation of deeply virtual compton scattering},\ }\href {https://doi.org/https://doi.org/10.1016/S0550-3213(00)00695-7} {\bibfield  {journal} {\bibinfo  {journal} {Nuclear Physics B}\ }\textbf {\bibinfo {volume} {596}},\ \bibinfo {pages} {99 } (\bibinfo {year} {2001})}\BibitemShut {NoStop}%
\bibitem [{\citenamefont {Dokshitzer}(1977)}]{dglap1}%
  \BibitemOpen
  \bibfield  {author} {\bibinfo {author} {\bibfnamefont {Y.~L.}\ \bibnamefont {Dokshitzer}},\ }\bibfield  {title} {\bibinfo {title} {{Calculation of the Structure Functions for Deep Inelastic Scattering and e+ e- Annihilation by Perturbation Theory in Quantum Chromodynamics.}},\ }\href@noop {} {\bibfield  {journal} {\bibinfo  {journal} {Sov. Phys. JETP}\ }\textbf {\bibinfo {volume} {46}},\ \bibinfo {pages} {641} (\bibinfo {year} {1977})}\BibitemShut {NoStop}%
\bibitem [{\citenamefont {Gribov}\ and\ \citenamefont {Lipatov}(1972)}]{dglap2}%
  \BibitemOpen
  \bibfield  {author} {\bibinfo {author} {\bibfnamefont {V.~N.}\ \bibnamefont {Gribov}}\ and\ \bibinfo {author} {\bibfnamefont {L.~N.}\ \bibnamefont {Lipatov}},\ }\bibfield  {title} {\bibinfo {title} {{Deep inelastic e p scattering in perturbation theory}},\ }\href@noop {} {\bibfield  {journal} {\bibinfo  {journal} {Sov. J. Nucl. Phys.}\ }\textbf {\bibinfo {volume} {15}},\ \bibinfo {pages} {438} (\bibinfo {year} {1972})}\BibitemShut {NoStop}%
\bibitem [{\citenamefont {Altarelli}\ and\ \citenamefont {Parisi}(1977)}]{dglap3}%
  \BibitemOpen
  \bibfield  {author} {\bibinfo {author} {\bibfnamefont {G.}~\bibnamefont {Altarelli}}\ and\ \bibinfo {author} {\bibfnamefont {G.}~\bibnamefont {Parisi}},\ }\bibfield  {title} {\bibinfo {title} {{Asymptotic Freedom in Parton Language}},\ }\href {https://doi.org/10.1016/0550-3213(77)90384-4} {\bibfield  {journal} {\bibinfo  {journal} {Nucl. Phys. B}\ }\textbf {\bibinfo {volume} {126}},\ \bibinfo {pages} {298} (\bibinfo {year} {1977})}\BibitemShut {NoStop}%
\bibitem [{\citenamefont {Salam}\ and\ \citenamefont {Rojo}(2009)}]{hoppet}%
  \BibitemOpen
  \bibfield  {author} {\bibinfo {author} {\bibfnamefont {G.}~\bibnamefont {Salam}}\ and\ \bibinfo {author} {\bibfnamefont {J.}~\bibnamefont {Rojo}},\ }\bibfield  {title} {\bibinfo {title} {A higher order perturbative parton evolution toolkit (hoppet)},\ }\href {https://doi.org/https://doi.org/10.1016/j.cpc.2008.08.010} {\bibfield  {journal} {\bibinfo  {journal} {Computer Physics Communications}\ }\textbf {\bibinfo {volume} {180}},\ \bibinfo {pages} {120} (\bibinfo {year} {2009})}\BibitemShut {NoStop}%
\bibitem [{\citenamefont {Soffer}(1995)}]{Sofferbound}%
  \BibitemOpen
  \bibfield  {author} {\bibinfo {author} {\bibfnamefont {J.}~\bibnamefont {Soffer}},\ }\bibfield  {title} {\bibinfo {title} {Positivity constraints for spin-dependent parton distributions},\ }\href {https://doi.org/10.1103/PhysRevLett.74.1292} {\bibfield  {journal} {\bibinfo  {journal} {Phys. Rev. Lett.}\ }\textbf {\bibinfo {volume} {74}},\ \bibinfo {pages} {1292} (\bibinfo {year} {1995})}\BibitemShut {NoStop}%
\end{thebibliography}%

\end{document}